\documentclass{article}

\usepackage{arxiv}

\usepackage[utf8]{inputenc} 
\usepackage[T1]{fontenc}    
\usepackage{hyperref}       
\usepackage{url}            
\usepackage{booktabs}       
\usepackage{amsmath}        
\usepackage{amssymb}        
\usepackage{nicefrac}       
\usepackage{microtype}      
\usepackage{graphicx}       
\usepackage{natbib}         
\usepackage{doi}            
\usepackage{xcolor}         
\usepackage{CJK}            
\usepackage{enumerate}      
\usepackage{bm}             
\usepackage{indentfirst} 

\hypersetup{
  colorlinks=true,
  linkcolor=red,
  citecolor=orange,
  urlcolor=blue,
}

\graphicspath{{../figure}}

\usepackage{multicol} 

\usepackage{framed}
\usepackage{multirow}

\def\be{\bm{e}}
\def\bq{\bm{q}}
\def\bu{\bm{u}}
\def\bx{\bm{x}}
\def\bD{\bm{D}}
\def\bH{\bm{H}}
\def\bI{\bm{I}}
\def\btau{\bm{\tau}}
\def\dt{\mbox{${\mathit \Delta t}$}}
\def\dx{\mbox{${\mathit \Delta x}$}}

\makeatletter
\@addtoreset{equation}{section}
\makeatother

\title{A numerical method for low Mach number compressible flows by simultaneous relaxation of dependent variables
}

\author{\href{https://orcid.org/0000-0002-4875-8174}{\includegraphics[scale=0.06]{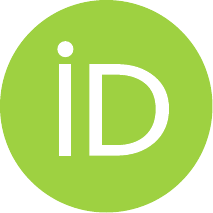}\hspace{1mm}Hideki Yanaoka}
\thanks{Email address for correspondence: yanaoka@iwate-u.ac.jp} \\
    Faculty of Science and Engineering, Iwate University, \\
    4-3-5 Ueda, Morioka, Iwate 020-8551, Japan \\
	\And
	Yuji Sato \\
	Mechanical and Aerospace Engineering, 
    Department of Science and Engineering, \\
    Graduate School of Integrated Arts and Sciences, Iwate University, \\
    4-3-5 Ueda, Morioka, Iwate 020-8551, Japan
	\texttt{} \\
}

\renewcommand{\shorttitle}{A numerical method for low Mach number compressible flows by simultaneous relaxation of dependent variables}

\makeatletter
\hypersetup{
pdfusetitle,
bookmarksnumbered,
pdftitle={\@title},
pdfsubject={},
pdfauthor={Hideki Yanaoka, Yuji Sato},
pdfkeywords={A numerical method for low Mach number compressible flows by simultaneous relaxation of dependent variables}
}
\makeatother

\begin{document}

\maketitle

\begin{abstract}
Density varies spatiotemporally in low Mach number flows. 
Hence, incompressibility cannot be assumed, 
and the density must be accurately solved. 
Various methods have been proposed to analyze low Mach number flows, 
but their energy conservation properties have not been investigated in detail. 
This study proposes a new method for simultaneously relaxing velocity, 
pressure, density, and internal energy 
using a conservative finite difference scheme with excellent energy conservation properties 
to analyze low Mach number flows. 
In the analysis for sound wave propagation in an inviscid compressible flow, 
the amplitude amplification ratio and frequency of sound wave obtained by this numerical method 
agree well with the theoretical values. 
In the analysis for a three-dimensional periodic inviscid compressible flow, 
each total amount for the momentum, total energy, and entropy are discretely conserved. 
When no approximation, such as low Mach number approximations, is applied 
to the fundamental equations, 
the excellent conservation properties of momentum, total energy, and entropy are achieved. 
In decaying compressible isotropic turbulence, 
this computational method can capture turbulence fluctuations. 
Analyzing the Taylor--Green decaying vortex, 
we confirmed the validity of this computational scheme 
for compressible viscous flows. 
In the calculation for the natural convection in a cavity, 
the validity of this numerical method was presented 
even in incompressible flows considering density variation. 
In a three-dimensional Taylor decaying vortex problem, 
it was shown that this numerical method can accurately calculate incompressible flows. 
We clarified the accuracy and validity of the present numerical method 
by analyzing various flow models 
and demonstrated the possibility of applying this method to complex flow fields.
\end{abstract}


\section{Introduction}

Incompressibility is assumed for computational simplicity when analyzing low Mach number flows. 
However, if the temperature difference between a heat source and fluid increases, 
the temperature dependence of density increases, 
and the incompressibility assumption does not hold. 
Compressible flows at low Mach numbers occur in flow fields with combustion 
or high-temperature heat sources and near walls in supersonic boundary layer flows. 
Additionally, compressibility must be considered in sound wave analysis. 
Therefore, it is necessary to construct a numerical method that can analyze 
low Mach number flows, but difficult problems arise.

In the analysis of compressible fluids, 
it is necessary to capture convective velocity and sound wave propagation. 
The Courant number is defined as $\nu = |V \pm c_0|\dt/\dx$, 
where $\dx$, $\dt$, $c_0$, and $V$ are a grid width, time step, sound speed, 
and convective velocity, respectively. 
If the stability condition for calculation is $\nu < 1$, 
the time step must satisfy the condition of $\dt < \dx/|V \pm c_0|$. 
As the sound wave is faster than the convective velocity, 
the Courant number is severely restricted 
when solving the fundamental equations of compressible fluids. 
Hence, compared with the analysis using the governing equations of incompressible fluids, 
we cannot set a large time increment, 
and a long timescale is required until the flow field reaches a statistically steady state.

There are two numerical methods for solving the momentum conservation equation governing flow: 
density- and pressure-based methods. 
The density-based scheme solves the density from the mass conservation equation 
and finds the pressure from the equation of state. 
This numerical method is generally used for compressible flow analyses 
but is unsuitable for incompressible flow analyses with low Mach numbers. 
At low Mach numbers, 
as the speed of sound is faster than convective velocity, 
the problem of stiffness in a convective term arises and deteriorates convergence. 
Using the quasi-compressible method \citep{Chorin_1967} based on the density-based method, 
incompressible fluid flows can also be analyzed. 
However, it is necessary to improve this method for applications to unsteady flows 
and the flow field with mixed incompressible and compressible characteristics. 
As another computational method, 
a preconditioning method that modifies the eigenvalues of the equation has been proposed 
\citep{Turkel_1987, Choi&Merkle_1993}.

The pressure-based method has been used in the analysis for incompressible flows 
and has been extended to compressible flow analyses 
\citep{Patnaik_et_al_1987, Rhie_1989, Karki&Patankar_1989, Chen&Pletcher_1991, Demirdzic_et_al_1993}. 
For this method, the Poisson equation for pressure or pressure correction value is derived 
from the mass and momentum conservation equations. 
After obtaining the pressure, the density is calculated from the equation of state.

In the analysis for incompressible flows at high Reynolds numbers, 
computational instability can be suppressed 
if transport quantities such as mass, momentum, and kinetic energy are discretely preserved in inviscid flows. 
However, when analyzing compressible flows, 
even if discrete conservation of transport quantities is established, 
the calculation becomes unstable at high Reynolds numbers. 
Thus, instead of solving the energy conservation equation, 
a method for solving the entropy conservation equation was proposed \citep{Harten_1983}. 
Additionally, as using the total energy equation is known to lead to unstable computations, 
Honein and Moin \citep{Honein&Moin_2004} used the internal energy equation. 
For the analysis of high Reynolds number flows, 
the setting of the dependent variable in the energy equation affects the calculation stability.

When analyzing compressible flows using the density-based method, 
the pressure in the momentum conservation equation is determined by the equation of state. 
When let $\rho_0$ be the density of the fluid and $g$ be the gravitational acceleration, 
the pressure is defined as the sum of thermodynamic pressure $P_\mathrm{th}$, 
hydrostatic pressure $p_\mathrm{hs} = \rho_0 g y$, 
and dynamic pressure $p_\mathrm{d}$ owing to fluid motion. 
At low Mach numbers, hydrostatic and dynamic pressures are lower than $P_\mathrm{th}$. 
However, as the hydrostatic and dynamic pressures significantly affect the momentum conservation equation, 
high accuracy is required when calculating the pressure using the equation of state. 
In the low Mach number approximation \citep{Rehm&Baum_1978, Quere_et_al_1992}, 
the pressure in the equation of state is replaced by $P_\mathrm{th}$, 
decoupling the link of the pressure between the equation of state and the momentum conservation equation. 
Therefore, density is affected by temperature only and does not change with pressure.

Numerical methods for analyzing low Mach number flows without using the low Mach approximation were also proposed 
\citep{Patnaik_et_al_1987, Wall_et_al_2002, Morinishi_2009, Morinishi_2010, Hou&Mahesh_2005, Bijl&Wesseling_1998, Kwatra_et_al_2009, Hennink_et_al_2021}. 
Patnaik et al. \citep{Patnaik_et_al_1987} proposed a density-based method (barely implicit correction: BIC) 
that transforms the total energy equation into an elliptic equation for pressure correction value 
and removes the time-step limitation caused by the speed of sound. 
After solving the pressure correction value, the BIC method modifies the velocity and total energy. 
Wall et al. \citep{Wall_et_al_2002} proposed a method to convert the Poisson equation 
of pressure correction into the Helmholtz equation, 
using the pressure-based method, and to avoid the Courant number limitation. 
The fully conservative finite difference scheme proposed by Morinishi \citep{Morinishi_2009, Morinishi_2010} 
is an improved version of the scheme by Wall et al. \citep{Wall_et_al_2002}, 
and if using the implicit midpoint rule, 
transport quantities and square quantities are preserved discretely in time and space directions. 
The method proposed by Hou and Mahesh \citep{Hou&Mahesh_2005} uses a collocated grid and, 
as with Wall et al. \citep{Wall_et_al_2002}, the time levels of density, pressure, 
and temperature are staggered from that of velocity. 
Similarly to Patnaik et al. \citep{Patnaik_et_al_1987}, 
the density is solved from the mass conservation equation, 
and the Poisson equation for the pressure correction value is derived from the total energy equation. 
They also used a nondimensionalization similar to Bijl and Wesseling \citep{Bijl&Wesseling_1998}, 
so that for low Mach numbers, the energy equation reduces to the form of the divergence-free condition for the velocity. 
Kwatra et al. \citep{Kwatra_et_al_2009} proposed a method for solving the pressure evolution equation. 
As the Mach number approaches zero, 
the Poisson equation for pressure reduces to the Poisson equation for incompressible flow. 
In this method, the density is solved from the mass conservation equation, 
and after solving the pressure, the velocity and total energy are corrected. 
Hennink et al. \citep{Hennink_et_al_2021} used the discontinuous Galerkin method 
to construct a pressure-based method for low Mach number flows. 
They adopted the method for solving mass flux instead of velocity. 
Boscheri and Tavelli \citep{Boscheri&Tavelli_2022} proposed a method to calculate low Mach number flows 
using a semi-implicit method. 
This method solves the density from the mass conservation equation. 
Furthermore, by substituting the momentum equation into the total energy equation, 
the Poisson equation for pressure (pressure wave equation) is derived, and the pressure is solved. 
As described above, various methods have been proposed to analyze low Mach number flows; 
however, energy conservation properties have not been investigated in the previous studies 
\citep{Hou&Mahesh_2005, Kwatra_et_al_2009, Hennink_et_al_2021}.

This study does not treat compressible flows with shock waves 
and analyzes low Mach number flows using the pressure-based method. 
This numerical method combines existing methods \citep{Wall_et_al_2002, Morinishi_2009, Morinishi_2010}, 
converts the Poisson equation for the pressure correction value 
into the Helmholtz equation, 
and discretizes the fundamental equation by a fully conservative finite difference scheme. 
To improve the stability, we propose a method to simultaneously relax the density and internal energy 
when modifying the velocity and pressure. 
We applied this numerical method to various flow models 
and investigated computational accuracy, convergence, and conservation properties.

The remainder of this paper is organized as follows: 
Section \ref{sec2} presents the fundamental equations and makes them dimensionless 
so that they can be applied to low Mach number flows. 
In Section \ref{sec3}, we construct a numerical method that can analyze low Mach number flows 
and propose a simultaneous relaxation method. 
In Section \ref{sec4}, we verify the computational accuracy, convergence, 
and conservation properties with the numerical method. 
Finally, Section \ref{sec5} summarizes the results.

\section{Modeling and fundamental equations}
\label{sec2}

\subsection{Fundamental equation}

In this study, we deal with compressible flows at low Mach numbers 
and analyze flows without shock waves. 
We also assume that the fluid is an ideal gas. 
The fundamental equations governing compressible flows are 
the conservation equations of mass, momentum, and internal energy 
and are expressed as
\begin{equation}
   \frac{Wo^2}{Re} \frac{\partial \rho}{\partial t} 
   + \nabla \cdot (\rho \bu) = 0,
   \label{mass}
\end{equation}
\begin{equation}
   \frac{Wo^2}{Re} \frac{\partial (\rho \bu)}{\partial t} 
   + \nabla \cdot (\rho \bu \otimes \bu) 
   = - \nabla p + \frac{1}{Re} \nabla \cdot \btau + \frac{1}{Fr^2} \rho \be_g,
   \label{momentum}
\end{equation}
\begin{equation}
   \frac{Wo^2}{Re} \frac{\partial (\rho e)}{\partial t} 
   + \nabla \cdot (\rho \bu e) 
   = - \frac{\kappa}{Re Pr} \nabla \cdot \bq 
   - (\kappa - 1) (\kappa Ma^2 p + 1) \nabla \cdot \bu 
   + \frac{\kappa (\kappa - 1) Ma^2}{Re} \btau : \nabla \bu,
   \label{energy_e}
\end{equation}
where $t$, $\rho$, $\bu$, $p$, and $e$ reprersent the time, density of fluid, 
velocity vector at the coordinate $\bx$,  pressure, 
and internal energy, respectively. 
$g$ represents the gravity acceleration, 
and $\be_g$ represents the unit vector in the direction of gravity. 
In this study, we do not apply the Boussinesq approximation 
to the momentum conservation equation so that we can analyze thermal convection 
with a high temperature difference. 
The second term on the right side of Eq. (\ref{energy_e}) 
represents work owing to volume change, 
and the third term represents the viscous dissipation term. 
We assume that the flow is a Newtonian fluid. 
The viscous stress tensor $\btau$ and heat flux vector $\bq$ are defined as
\begin{equation}
   \btau = \mu \left[ \nabla \bu + (\nabla \bu)^T 
         - \frac{2}{3} \bI \nabla \cdot \bu \right],
\end{equation}
\begin{equation}
   \bq = - k \nabla T,
\end{equation}
where the superscript $T$ represents the transpose and $\bI$ is the unit tensor. 
$\mu$, $k$, and $T$ represent the  viscosity coefficient of fluid, 
thermal conductivity, and temperature, respectively. 
Regarding the reference values used for nondimensionalization, 
the length, velocity, internal energy, temperature, time, viscosity coefficient, 
and thermal conductivity are $l_\mathrm{ref}$, $u_\mathrm{ref}$, $e_\mathrm{ref}$, 
$T_\mathrm{ref}$, $t_\mathrm{ref}$, $\mu_\mathrm{ref}$, 
and $k_\mathrm{ref}$, respectively. 
Using these reference values, the variables in the fundamental equations are 
nondimensionalized as follows:
\begin{subequations}
\begin{equation}
   \bx^* = \frac{\bx}{l_\mathrm{ref}}, \quad
   \bu^* = \frac{\bu}{u_\mathrm{ref}}, \quad
   \rho^* = \frac{\rho}{\rho_\mathrm{ref}}, \quad
   p^* = \frac{p - p_\mathrm{ref}}{\rho_\mathrm{ref} u_\mathrm{ref}^2}, \quad
   e^* = \frac{e}{e_\mathrm{ref}},
\end{equation}
\begin{equation}
   E^* = \frac{E}{e_\mathrm{ref}}, \quad
   T^* = \frac{T}{T_\mathrm{ref}}, \quad
   t^* = \frac{t}{t_\mathrm{ref}}, \quad 
   \mu^* = \frac{\mu}{\mu_\mathrm{ref}}, \quad 
   k^* = \frac{k}{k_\mathrm{ref}},
\end{equation}
\label{non-dimensionalize}
\end{subequations}
where the superscript $*$ represents the nondimensional variable, 
and is omitted in the fundamental equations. 
Addinally, $e_\mathrm{ref} = c_v T_\mathrm{ref}$, 
$c_v$ is the constant volume specific heat, 
and $p_\mathrm{ref}$ is the reference pressure. 
The nondimensional parameters in these fundamental equations are defined as follows: 
$Re$, $Wo$, $Fr$, $Pr$, and $Ma$ represent the Reynolds, Womersley, 
Froude,  Prandtl number, and Mach numbers, respectively:
\begin{equation}
   Re = \frac{u_\mathrm{ref} l_\mathrm{ref}}{\nu_\mathrm{ref}}, \quad 
   Wo = l_\mathrm{ref} \sqrt{\frac{1}{\nu_\mathrm{ref} t_\mathrm{ref}}}, \quad 
   Fr = \frac{u_\mathrm{ref}}{\sqrt{g l_\mathrm{ref}}}, \quad 
   Pr = \frac{\nu_\mathrm{ref}}{\alpha_\mathrm{ref}}, \quad 
   Ma = \frac{u_\mathrm{ref}}{c_\mathrm{ref}},
\end{equation}
where $\nu$ and $\alpha$ represent the dynamic viscosity 
and thermal diffusion coefficient, respectively.

The equation of state for an ideal gas is given as
\begin{equation}
   \kappa Ma^2 p + 1 = \rho e,
   \label{eqn_of_state}
\end{equation}
The total energy $E$ and its conservation equation are expressed as
\begin{equation}
   E = e + \frac{1}{2} \kappa (\kappa - 1) Ma^2 \bu \cdot \bu,
   \label{total_energy}
\end{equation}
\begin{align}
   \frac{Wo^2}{Re} \frac{\partial (\rho E)}{\partial t} 
   + \nabla \cdot (\rho \bu E) 
   &= - \frac{\kappa}{Re Pr} \nabla \cdot \bq 
   - (\kappa - 1) \nabla \cdot \left[ \bu (\kappa Ma^2 p + 1) \right] 
   \nonumber \\
   &
   + \frac{\kappa (\kappa - 1) Ma^2}{Re} \nabla \cdot (\btau \cdot \bu) 
   + \frac{1}{Fr^2} \rho g \be_g \cdot \bu,
   \label{energy_E}
\end{align}
As discontinuities such as shock waves occur in compressible flows, 
discontinuous variables such as density are accepted 
as solutions of partial differential equations in numerical calculations; 
hence, solutions are generally expressed in a weak form \citep{Morinishi_2009, Morinishi_2010}. 
Therefore, Eq. (\ref{energy_E}) is used instead of Eq. (\ref{energy_e}) 
for compressible flow analysis. 
For compressible flows at low Mach numbers, 
using the total energy equation is known to be computationally unstable \citep{Honein&Moin_2004}. 
Hence, suppression of nonlinear instability is more significant than capturing discontinuities. 
This study uses the internal energy equation (\ref{energy_e}) 
as with the previous studies \citep{Honein&Moin_2004, Morinishi_2009, Morinishi_2010}.

Using the thermodynamic relations, we get the following equation for entropy $s$:
\begin{equation}
   \frac{Wo^2}{Re} \frac{\partial (\rho s)}{\partial t} 
   + \nabla \cdot (\rho \bu s) 
   = \frac{1}{T} \left( - \frac{\kappa}{Re Pr} \nabla \cdot \bq 
   + \frac{Ec}{Re} \btau : \nabla \bu \right),
  \label{entropy_eq}
\end{equation}
where the entropy is dimensionless using $c_v$ and given as
\begin{equation}
  s = \ln \frac{\kappa Ma^2 p + 1}{\rho^{\kappa}}.
  \label{entropy}
\end{equation}
$Ec$ represents the Eckert number, defined as $Ec = u_\mathrm{ref}^2/(c_v T_\mathrm{ref})$. 
In inviscid flows, the total amount of entropy is conserved. 
Therefore, if the entropy changes with time 
when investigating the time variation of the total amount, 
we can find the occurrence of a nonphysical phenomenon.

\subsection{Low Mach number approximation model}

This study analyzed using the low Mach number approximation 
and compared calculation results obtained by the present numerical method 
with those with the approximation model. 
If a flow velocity is slower than a sound velocity, 
the low Mach number approximation model \citep{Rehm&Baum_1978} can be applied 
to the governing equations. 
Pressure can be decomposed into thermodynamic pressure $P_\mathrm{th}$, 
hydrostatic pressure $P_\mathrm{hs}$, and dynamic pressure $p'$ caused by fluid motion as follows:
\begin{equation}
   p = P_\mathrm{th}(t) + P_\mathrm{hs}(\bx) + p'(t, \bx).
\end{equation}
The hydrostatic pressure is defined as $P_\mathrm{hs}(\bx) = \rho_0 g \bx \cdot \be_g$, 
where $\rho_0$ is the density at the reference temperature $T_0$. 
Compared to the thermodynamic pressure $P_\mathrm{th}$, 
the dynamic and hydrostatic pressures are at low levels. 
When using the low Mach number approximation \citep{Rehm&Baum_1978}, 
the conservation equations of momentum and internal energy are expressed as
\begin{equation}
   \frac{Wo^2}{Re} \frac{\partial (\rho \bu)}{\partial t} 
   + \nabla \cdot (\rho \bu \otimes \bu) 
   = - \nabla p' + \frac{1}{Re} \nabla \cdot \btau 
   + \frac{1}{Fr^2} (\rho - \rho_0) \be_g,
   \label{momentum2}
\end{equation}
\begin{equation}
   \frac{Wo^2}{Re} \frac{\partial (\rho e)}{\partial t} 
   + \nabla \cdot (\rho \bu e) 
   = - \frac{\kappa}{Re Pr} \nabla \cdot \bq 
   - (\kappa - 1) (\kappa Ma^2 P_\mathrm{th} + 1) \nabla \cdot \bu 
   + \frac{\kappa (\kappa - 1) Ma^2}{Re} \btau : \nabla \bu,
   \label{energy_e2}
\end{equation}
where $\rho_0$ and $e_0$ are reference values. 
The pressure $p'$ in the above equation differs from that in Eq. (\ref{momentum}). 
In the following, $p'$ is replaced with $p$. 
The pressure in the equation of state considers only thermodynamic pressure, 
and the equation of state is given as
\begin{equation}
   P_\mathrm{th} 
   = \frac{1}{\kappa Ma^2} (\rho_0 e_0 - 1).
   \label{themodynamics_pressure.2}
\end{equation}
Consequently, as the pressures in the equation of state 
and the momentum conservation equation are independent and are not coupled, 
the analysis can be stably performed. 
This model does not consider density changes caused by local pressure changes, 
and density changes are affected only by temperature. 
The thermodynamic pressure $P_\mathrm{th}$ does not change over time in open space 
and can be assumed to be constant. 
Quere et al. \citep{Quere_et_al_1992} also proposed a method 
to define the thermodynamic pressure $P_\mathrm{th}$ considering time variation. 
From the mass conservation condition, 
the thermodynamic pressure is determined as follows:
\begin{equation}
   P_\mathrm{th} 
   = \frac{1}{\kappa Ma^2} \left( 
   \frac{M_0}{\frac{1}{\Omega} \int_{\Omega} \frac{1}{e} d \Omega} - 1 \right),
   \label{themodynamics_pressure.1}
\end{equation}
where $\Omega$ represents the computational domain. 
Additionally, $M_0$ is an initial average mass in the region and is given as
\begin{equation}
   M_0 = \frac{1}{\Omega} \int_{\Omega} \rho_0 d\Omega 
   = \frac{1}{\Omega} \int_{\Omega} \frac{\kappa Ma^2 p_0 + 1}{e_0} d \Omega.
\end{equation}
This study used the thermodynamic pressure $P_\mathrm{th}(t)$ that considers time variations. 
Generally, for low Mach number approximation models, 
the dependent variable in the energy equation is temperature, 
and the time variation $P_\mathrm{th}(t)/dt$ of the thermodynamic pressure should be calculated. 
In this study, to compare the results obtained by this numerical method 
with those using the low Mach number approximation, 
we used the internal energy as the dependent variable of the energy equation 
and calculated only the thermodynamic pressure $P_\mathrm{th}(t)$.

\subsection{Boussinesq approximation model}

This study analyzed using the Boussinesq approximation 
and compared calculation results obtained by the present numerical method 
with those with the approximation model. 
In the analysis of natural convection in the case of a small density change of fluid, 
the momentum equation with Boussinesq approximation is used. 
The momentum and energy equations are expressed as
\begin{equation}
   \frac{Wo^2}{Re} \frac{\partial \bu}{\partial t} 
   + \nabla \cdot (\bu \otimes \bu) 
   = - \nabla p + \frac{1}{Re} \nabla \cdot (2 \mu \bD) 
   - \frac{1}{Fr^2} (T - 1) \be_g,
   \label{momentum3}
\end{equation}
\begin{equation}
   \frac{Wo^2}{Re} \frac{\partial T}{\partial t} 
   + \nabla \cdot (\bu T) 
   = - \frac{1}{Re Pr} \nabla \cdot \bq 
   + \frac{(\kappa - 1) Ma^2}{Re} \btau : \nabla \bu.
   \label{energy_e3}
\end{equation}

The Boussinesq approximation cannot be applied to thermal convection fields 
where density and pressure fluctuations are great 
and high-temperature differences occur \citep{Spiegel&Veronis_1960}. 
For fluids such as air, the Boussinesq approximation is valid for flow fields 
with temperature differences lower than 30$^\circ$C \citep{Gray&Giorgini_1976}. 
In the case of water, the Boussinesq approximation can be applied 
for temperature differences below a few degrees. 
As the temperature difference increases, 
the temperature dependence of physical properties increases, 
making it hard to apply the Boussinesq approximation 
to the analysis of thermal convection using water.

\subsection{Incompressible limit}

In the incompressible limit where the Mach number is $Ma \rightarrow 0$, 
the terms of pressure work and viscous dissipation 
in Eqs. (\ref{energy_e}) and (\ref{energy_e2}) are eliminated, 
and the internal energy equation is transformed into the energy equation (\ref{energy_e3}) 
for temperature $T$. 
Furthermore, from the state equation (\ref{eqn_of_state}), $\rho e = 1$, 
and the flow field is isothermal. 
Thus, the above fundamental equations reduce to equations for incompressible fluids 
in the limit $Ma \rightarrow 0$. 
This is because $p^* = (p-p_\mathrm{ref})/(\rho u_\mathrm{ref}^2)$ is used 
for nondimensionalization of the pressure. 
Here, the reference pressure $p_\mathrm{ref}$ is obtained 
as $p_\mathrm{ref} = (\kappa-1)\rho e$ using the mainstream or initial densities and temperatures. 
A similar nondimensionalization was used in existing work \citep{Bijl&Wesseling_1998}.

\section{Numerical method}
\label{sec3}

\subsection{Definitions of finite difference and interpolation operations}

The Cartesian coordinates $(x, y, z)$ in the physical space are transformed 
into the computational space $(\xi, \eta, \zeta)$ for discretization in a nonuniform grid. 
We assume the relationship, $x = x(\xi)$, $y = y(\eta)$, and $z = z(\zeta)$, between both spaces. 
By letting dependent variables such as velocity and pressure be $\Phi$, 
the first derivative can be converted as follows:
\begin{equation}
   \frac{\partial \Phi}{\partial x} 
      = \frac{1}{J} \frac{\partial (J \xi_x \Phi)}{\partial \xi}, \quad 
   \frac{\partial \Phi}{\partial y} 
      = \frac{1}{J} \frac{\partial (J \eta_y \Phi)}{\partial \eta}, \quad 
   \frac{\partial \Phi}{\partial z} 
      = \frac{1}{J} \frac{\partial (J \zeta_z \Phi)}{\partial \zeta},
\end{equation}
where $J$ is the Jacobian defined as $J = x_{\xi} y_{\eta} z_{\zeta}$, 
and $\xi_x = y_{\eta} z_{\zeta}/J$, $\eta_y = z_{\zeta} x_{\xi}/J$, and 
$\zeta_z = x_{\xi} y_{\eta}/J$. 

The variable at a cell center $(i, j, k)$ is defined as $\phi_{i,j,k}$. 
The second-order central difference equation and interpolation 
in the $x$ ($\xi$)-direction for the variable $\Phi$ are given, respectively, as
\begin{equation}
   \left. \frac{\partial \Phi}{\partial x} \right|_{i,j,k} 
      = \frac{1}{J} \frac{(J \xi_x \bar{\Phi}^{\xi})_{i+1/2,j,k} 
                        - (J \xi_x \bar{\Phi}^{\xi})_{i-1/2,j,k}}{\Delta \xi},
\end{equation}
\begin{equation}
   \left. \bar{\Phi}^{\xi} \right|_{i+1/2,j,k} 
   = \frac{\Phi_{i,j,k} + \Phi_{i+1,j,k}}{2},
\end{equation}
where $\Delta \xi$ is the grid spacing in the computational space. 
The definitions of the $y$ ($\eta$)- and $z$ ($\zeta$)-directions are identical. 
The Jacobian is defined at cell centers. 
Derivative terms that are not directly related to conservation properties, 
such as momentum and total energy, are discretized 
without coordinate transformation, as follows:
\begin{equation}
   \left. \frac{\partial \Phi}{\partial x} \right|_{i,j,k} 
      = \frac{\bar{\Phi}^x_{i+1/2,j,k} - \bar{\Phi}^x_{i-1/2,j,k}}{\Delta x_i},
\end{equation}
where $\Delta x_i = x_{i+1/2}-x_{i-1/2}$ is the grid spacing. 
If a variable at time level $n$ is defined as $\Phi^n$, 
the derivative and interpolation of the variable for time 
are similarly expressed as follows:
\begin{equation}
   \left. \frac{\partial \Phi}{\partial t} \right|^{n+1/2} 
      = \frac{\Phi^{n+1} - \Phi^{n}}{\dt},
\end{equation}
\begin{equation}
   \Phi^{n+1/2} = \frac{\Phi^{n+1} + \Phi^{n}}{2},
\end{equation}
where $\Delta t$ is a time increment.

\subsection{Simultaneous relaxation method}

For periodic inviscid incompressible flows, 
the transport quantity, such as the kinetic energy, 
must be discretely conserved \citep{Morinishi_1996a,Morinishi_1998}. 
The generation of nonphysical kinetic energy leads to computational instability. 
Additionally, the transformation between the conservative and nonconservative forms 
of convection terms must be discretely satisfied 
\citep{Morinishi_1996a,Morinishi_1998}. 
A fully conservative finite difference scheme, 
in which the transport quantity is discretely conserved in the spatiotemporal direction, 
has been proposed. 
The transformation between conservative and nonconservative forms 
of the advection term is established \citep{Morinishi_2009,Morinishi_2010,Ham_et_al_2002}. 
We apply the fully conservative finite difference scheme 
to the analysis of low Mach number flows 
as with \citep{Morinishi_2009,Morinishi_2010,Ham_et_al_2002}. 
Furthermore, we adopt a spatiotemporal staggered grid, 
as with the existing studies \citep{Wall_et_al_2002, Morinishi_2009, Morinishi_2010}. 
The governing equation is discretized 
so that the transport quantity is discretely conserved in the spatiotemporal direction 
while maintaining second-order accuracy 
and compatibility between conservative and nonconservative forms is established. 
Velocity and internal energy are placed on the same time level, 
and density and pressure are placed on a time level that is half the time offset from the velocity. 
The Newton method is used to solve the unsteady solution. 
Using the implicit midpoint rule, Eqs. (\ref{mass}), (\ref{momentum}), 
and (\ref{energy_e}) are given as
\begin{subequations}
\begin{equation}
   \frac{Wo^2}{Re} \frac{\rho^{n+3/2,m+1} - \rho^{n+1/2}}{\dt} 
   = H_{\rho}^{n+1,m+1},
   \label{implicit.mass}
\end{equation}
\begin{equation}
   H_{\rho}^{n+1,m+1} = - \nabla \cdot (\rho \bu)^{n+1,m+1},
   \label{implicit.Hrho}
\end{equation}
\end{subequations}
\begin{subequations}
\begin{equation}
   \frac{Wo^2}{Re} \frac{(\rho \bu)^{n+1,m+1} - (\rho \bu)^n}{\dt} 
   = \bH_u^{n+1/2,m+1} 
   - \nabla \bar{p}^{n+1/2,m+1},
   \label{implicit.u}
\end{equation}
\begin{equation}
   \bH_u^{n+1/2,m+1} 
   = - \nabla \cdot [(\rho \bu)^{n+1/2,m+1} \otimes \hat{\bu}^{n+1/2,m+1}] 
   + \frac{1}{Re} \nabla \cdot \btau^{n+1/2,m+1} + \frac{1}{Fr^2} \rho^{n+1/2,m+1} \be_g,
   \label{implicit.Hu}
\end{equation}
\end{subequations}
\begin{subequations}
\begin{equation}
   \frac{Wo^2}{Re} \frac{(\rho e)^{n+1,m+1} - (\rho e)^n}{\dt} 
   = H_e^{n+1/2,m+1} 
   - (\kappa - 1) (\kappa Ma^2 \bar{p}^{n+1/2,m+1} + 1) 
   \nabla \cdot \hat{\bu}^{n+1/2,m+1},
   \label{implicit.e}
\end{equation}
\begin{align}
   H_e^{n+1/2,m+1} &= - \nabla \cdot [(\rho \bu)^{n+1/2,m+1} \hat{e}^{n+1/2,m+1}] 
   - \frac{\kappa}{Re Pr} \nabla \cdot \bq^{n+1/2,m+1} 
   \nonumber \\
   &
   + \frac{\kappa (\kappa - 1) Ma^2}{Re} \btau^{n+1/2,m+1} : \nabla \hat{\bu}^{n+1/2,m+1},
   \label{implicit.He}
\end{align}
\end{subequations}
where the superscripts $n$ and $m$ indicate the time and 
Newton iterative levels, respectively. 
The temporal levels are defined as the $n$ level for the velocity and internal energy 
and the $n+1/2$ level for the density and pressure. 
When using the Boussinesq approximation model, 
the scalar quantity is placed at the same time level as the velocity. 
Additionally, $\hat{\bu}$ and $\hat{e}$ in the above equation are 
the square-root density weighted interpolation, 
defined as follows \citep{Morinishi_2009, Morinishi_2010}:
\begin{equation}
   \hat{u}^{n+1/2,m+1} 
   = \frac{\overline{\sqrt{\overline{J \bar{\rho}^t}^{\xi}} u}^t}
          {\overline{\sqrt{\overline{J \bar{\rho}^t}^{\xi}}}^t}, \quad 
   \hat{v}^{n+1/2,m+1} 
   = \frac{\overline{\sqrt{\overline{J \bar{\rho}^t}^{\eta}} v}^t}
          {\overline{\sqrt{\overline{J \bar{\rho}^t}^{\eta}}}^t}, \quad 
   \hat{w}^{n+1/2,m+1} 
   = \frac{\overline{\sqrt{\overline{J \bar{\rho}^t}^{\zeta}} w}^t}
          {\overline{\sqrt{\overline{J \bar{\rho}^t}^{\zeta}}}^t},
\end{equation}
\begin{equation}
   \hat{e}^{n+1/2,m+1} 
   = \frac{\overline{\sqrt{\bar{\rho}^t} e}^t}
          {\overline{\sqrt{\bar{\rho}^t}}^t},
\end{equation}
\begin{equation}
   \bar{\rho}^{n+1,m+1} = \frac{\rho^{n+3/2,m+1} + \rho^{n+1/2,m+1}}{2}.
\end{equation}
This interpolated value is introduced 
to construct a fully conservative finite difference scheme \citep{Morinishi_2009,Morinishi_2010}. 
Wall et al. \citep{Wall_et_al_2002} have not used this interpolation. 
The density is obtained from the state equation (\ref{eqn_of_state}) as follows:
\begin{equation}
   \rho^{n+1,m+1} = \frac{\kappa Ma^2 p^{n+1,m+1} + 1}{e^{n+1,m+1}},
   \label{implicit.rho}
\end{equation}
To relax the Courant number limitation due to the speed of sound, 
we treat pressure implicitly 
and use the following double-time interpolation of pressure \citep{Wall_et_al_2002}:
\begin{equation}
   \bar{p}^{n+1/2} = \left( \frac{1}{4} - \varepsilon \right) p^{n-1/2} 
   + \frac{1}{2}  p^{n+1/2} 
   + \left( \frac{1}{4} + \varepsilon \right) p^{n+3/2},
   \label{implicit.weighted_average}
\end{equation}
where $\varepsilon$ is a parameter introduced to prevent numerical oscillations 
caused by high wavenumber acoustic modes. 
When nonphysical acoustic modes occur, 
they cannot be dissipated; hence, we need to prevent such vibrations. 
The previous study \citep{Wall_et_al_2002} used a value as small as $\varepsilon=0.005$. 
In this research, we basically set $\varepsilon = 0$.

By applying the simplified marker and cell (SMAC) method \citep{Amsden&Harlow_1970}, 
Eqs. (\ref{implicit.mass}), (\ref{implicit.u}), and (\ref{implicit.e}) are temporally split as follows:
\begin{subequations}
\begin{equation}
   \frac{Wo^2}{Re} \frac{\rho^{n+1,m+1} \tilde{\bu}^{n+1,m+1} 
                       - (\rho \bu)^n}{\dt} 
   = \bH_u^{n+1/2,m+1} - \nabla \bar{p}^{n+1/2,m},
   \label{predict.u}
\end{equation}
\begin{equation}
   \frac{Wo^2}{Re} 
   \frac{\rho^{n+1,m+1} \bu^{n+1,m+1} 
       - \rho^{n+1,m+1} \tilde{\bu}^{n+1,m+1}}{\dt} 
   = - \left( \frac{1}{4} + \varepsilon \right) \nabla (\Delta p^m),
   \label{correct.u.1}
\end{equation}
\begin{equation}
   \frac{Wo^2}{Re} \frac{\rho^{n+1,m} \tilde{e}^{n+1,m+1} 
                      - (\rho e)^n}{\dt} 
   = H_e^{n+1/2,m+1} 
   - (\kappa - 1) (\kappa Ma^2 \bar{p}^{n+1/2,m} + 1) 
   \nabla \cdot \hat{\bu}^{n+1/2,m},
   \label{predict.e}
\end{equation}
\begin{equation}
   \frac{Wo^2}{Re} \frac{\rho^{n+1,m+1} e^{n+1,m+1} 
                       - \rho^{n+1,m} \tilde{e}^{n+1,m+1}}{\dt} 
   = - \kappa (\kappa - 1) Ma^2 
   \left( \frac{1}{4} + \varepsilon \right) \Delta p^m 
   \nabla \cdot \hat{\bu}^{n+1/2,m},
   \label{correct.e.1}
\end{equation}
\begin{equation}
   p^{n+3/2,m+1} = p^{n+3/2,m} + \Delta p^m,
   \label{correct.p.1}
\end{equation}
\end{subequations}
where $\hat{\bu}^{n+1,m+1}$ and $\tilde{e}^{n+1,m+1}$ are the predicted values of velocity and internal energy, respectively, and $\Delta p^m$ is the pressure correction value. 
The velocity in $\bH_u^{n+1/2,m+1}$ on the right side of Eq. (\ref{predict.u}) 
is defined as $\bu^{n+1/2,m+1} = (\tilde{\bu}^{n+1,m+1} + \bu^n)/2$. 
The internal energy in $H_e^{n+1/2,m+1}$ on the right side of Eq. (\ref{predict.e}) 
is defined as $e^{n+1/2,m+1} = (\tilde{e}^{n+1,m+1} + e^n)/2$. 
When calculating the velocity $\tilde{\bu}^{n+1,m+1}$, 
the convective term is linearized as 
$\nabla \cdot [(\rho \bu)^{n+1/2,m} \otimes \hat{\bu}^{n+1/2,m+1}]$ 
using the $m$-level value. 
In Eqs. (\ref{predict.e}) and (\ref{correct.e.1}), 
the density and velocity are linearized as 
$\rho^{n+1,m+1}=\rho^{n+1,m}$ and $\hat{\bu}^{n+1,m+1}=\hat{\bu}^{n+1,m}$, respectively. 
Once the Newton iteration is completed, 
such a linearized approximation can be ignored, 
preserving second-order accuracy in the time integration. 
When using the Boussinesq approximation model, 
the internal energy correction process is not required, 
and $e^{n+1,m+1}$ or $T^{n+1,m+1}$ is obtained directly.

Substituting Eq. (\ref{correct.u.1}) 
into the mass conservation equation (\ref{implicit.mass}) yields 
the Poisson equation for the pressure correction value $\Delta p$ as follows:
\begin{equation}
   \left( \frac{1}{4} + \varepsilon \right) \nabla^2 (\Delta p^m) 
   = \frac{Wo^2}{\dt Re} \left[ 
   \frac{Wo^2}{Re} \frac{\rho^{n+3/2,m+1} - \rho^{n+1/2}}{\dt} 
   + \nabla \cdot (\rho^{n+1,m+1} \tilde{\bu}^{n+1,m+1}) \right].
   \label{poisson.1}
\end{equation}
To stabilize calculations, 
we incorporate the effect of the pressure correction value on the density 
in the time derivative of the density of Eq. (\ref{implicit.mass}) 
as follows \citep{Wall_et_al_2002}:
\begin{equation}
   \left. \frac{\partial \rho}{\partial t} \right|^{n+1,m+1} 
   \approx \frac{\rho^{n+3/2,m+1} - \rho^{n+1/2}}{\dt} 
   + \frac{1}{\dt} \left. \frac{\partial \rho}{\partial p} \right|_e \Delta p.
\end{equation}
$\partial \rho/\partial p|_e$ represents the derivative when the internal energy is constant. 
This derivative term is obtained using the equation of state as follows:
\begin{equation}
   \left. \frac{\partial \rho}{\partial p} \right|_e 
   = \frac{\kappa Ma^2}{e}.
\end{equation}
Using the above equation, 
the Poisson equation (\ref{poisson.1}) can be rewritten into the following Helmholtz equation:
\begin{align}
   &
   \left( \frac{1}{4} + \varepsilon \right) \nabla^2 (\Delta p^m) 
   - \frac{Wo^4}{\dt^2 Re^2} 
   \left. \frac{\partial \rho}{\partial p} \right|_e \Delta p^m 
   \nonumber \\
   & \qquad 
   = \frac{Wo^2}{\dt Re} \left[ 
   \frac{Wo^2}{Re} \frac{\rho^{n+3/2,m+1} - \rho^{n+1/2}}{\dt} 
   + \nabla \cdot (\rho^{n+1,m+1} \tilde{\bu}^{n+1,m+1}) \right].
   \label{poisson.2}
\end{align}

When using the low Mach number approximation model, 
the pressure $\bar{p}$ in Eq. (\ref{implicit.e}) is replaced with $P_\mathrm{th}$. 
Then, the energy equation is temporally split as follows:
\begin{subequations}
\begin{equation}
   \frac{Wo^2}{Re} \frac{\rho^{n+1,m} \tilde{e}^{n+1,m+1} 
                      - (\rho e)^n}{\dt} 
   = H_e^{n+1/2,m+1} 
   - (\kappa - 1) (\kappa Ma^2 P_\mathrm{th}^{n+1/2,m} + 1) 
   \nabla \cdot \hat{\bu}^{n+1/2,m},
   \label{predict.e.lm}
\end{equation}
\begin{equation}
   \frac{Wo^2}{Re} \frac{\rho^{n+1,m+1} e^{n+1,m+1} 
                       - \rho^{n+1,m} \tilde{e}^{n+1,m+1}}{\dt} 
   = - \kappa (\kappa - 1) Ma^2 
   \frac{1}{2} \Delta P_\mathrm{th}^m 
   \nabla \cdot \hat{\bu}^{n+1/2,m},
   \label{correct.e.lm.1}
\end{equation}
\begin{equation}
   P_\mathrm{th}^{n+1,m+1} = P_\mathrm{th}^{n+1,m} + \Delta P_\mathrm{th}^m,
   \label{correct.p.lm.1}
\end{equation}
\begin{equation}
   P_\mathrm{th}^{n+1,m} 
   = \frac{1}{\kappa Ma^2} \left( 
   \frac{M_0}{\frac{1}{\Omega} \int_{\Omega} \frac{1}{e^{n+1,m}} d \Omega} - 1 \right),
\end{equation}
\end{subequations}
where $\Delta P_\mathrm{th}$ represents the thermodynamic pressure correction value. 
The average mass at time $n+1$ can be obtained as
\begin{equation}
   M^{n+1,m+1} = \frac{1}{\Omega} \int_{\Omega} \rho^{n+1,m+1} d \Omega 
   = \frac{1}{\Omega} \int_{\Omega} 
   \frac{\kappa Ma^2 (P_\mathrm{th}^{n+1,m} + \Delta P_\mathrm{th}^m) + 1}{e^{n+1,m+1}} d \Omega,
\end{equation}
where $\Omega$ represents the computational domain. 
As the average mass $M^{n+1,m+1}$ is equal to the initial value $M_0$, 
$\Delta P_\mathrm{th}^m$ is obtained as follows:
\begin{equation}
   \Delta P_\mathrm{th}^m = \frac{M_0 - \frac{1}{\Omega} \int_{\Omega} 
   \frac{\kappa Ma^2 P_\mathrm{th}^{n+1,m} + 1}{e^{n+1,m}} d \Omega}
   {\frac{1}{\Omega} \int_{\Omega} \frac{\kappa Ma^2}{e^{n+1,m}} d \Omega}.
\end{equation}

Bijl and Wesseling \citep{Bijl&Wesseling_1998} and Kwatra et al. \citep{Kwatra_et_al_2009} 
proposed a method to solve the Poisson equation of pressure as a pressure-based method. 
The Poisson equation for pressure used in these existing studies is a complex form. 
In contrast, the Laplacian operator in the Poisson equation (\ref{poisson.2}) 
for the pressure correction value used in this study is linear and has a simple form.

The authors analyzed various flow fields 
using a simultaneous velocity and pressure relaxation method 
\citep{Yanaoka&Inafune_2023,Yanaoka_2023}. 
The same method is applied to the analysis of compressible fluids. 
In this study, the concept of the hihgly SMAC method \citep{Hirt_et_al_1975} is introduced 
to correct the velocity, pressure, internal energy, and density, 
and simultaneous relaxation is performed as follows: 
\begin{subequations}
\begin{align}
   &
   \left( \frac{1}{4} + \varepsilon \right) \nabla^2 (\Delta p^{m,l}) 
   - \frac{Wo^4}{\dt^2 Re^2} 
   \left. \frac{\partial \rho}{\partial p} \right|_e \Delta p^{m,l} 
   \nonumber \\
   & \qquad 
   = \frac{Wo^2}{\dt Re} \left[ 
   \frac{Wo^2}{Re} \frac{\rho^{n+3/2,m+1,l} - \rho^{n+1/2}}{\dt} 
   + \nabla \cdot (\rho^{n+1,m+1,l} \bu^{n+1,m+1,l}) \right],
   \label{poisson.3}
\end{align}
\begin{equation}
   \frac{Wo^2}{Re} 
   \frac{\rho^{n+1,m+1,l} \bu^{n+1,m+1,l+1} 
       - \rho^{n+1,m+1,l} \bu^{n+1,m+1,l}}{\dt} 
   = - \left( \frac{1}{4} + \varepsilon \right) \nabla (\Delta p^{m,l}),
   \label{correct.u.2}
\end{equation}
\begin{align}
   &
   \frac{Wo^2}{Re} \frac{\rho^{n+1,m+1,l} e^{n+1,m+1,l+1} 
                       - \rho^{n+1,m+1,l} e^{n+1,m+1,l}}{\dt} 
   \nonumber \\
   & \qquad 
   = - \kappa (\kappa - 1) Ma^2 
   \left( \frac{1}{4} + \varepsilon \right) \Delta p^{m,l} 
   \nabla \cdot \hat{\bu}^{n+1/2,m},
   \label{correct.e.2}
\end{align}
\begin{equation}
   p^{n+3/2,m+1,l+1} = p^{n+3/2,m+1,l} + \Delta p^{m,l},
   \label{correct.p.2}
\end{equation}
\begin{equation}
   \rho^{n+1,m+1,l+1} = \frac{\kappa Ma^2 p^{n+1,m+1,l+1} + 1}{e^{n+1,m+1,l+1}},
   \label{density}
\end{equation}
\end{subequations}
where the superscript $l$ represents the number of iterations. 
When $l = 1$, let $\bu^{n+1,m+1,l} = \tilde{\bu}^{n+1,m+1}$, 
$p^{n+3/2,m+1,l} = p^{n+3/2,m}$, $e^{n+1,m+1,l} = \tilde{e}^{n+1,m+1}$, and $\rho^{n+3/2,m+1,l} = \rho^{n+3/2,m}$, 
the velocity, pressure, internal energy, and density are simultaneously relaxed. 
We repeat the calculation up to a predetermined iteration number. 
After the simultaneous relaxation, 
let $\bu^{n+1,m+1} = \bu^{n+1,m+1,l+1}$, $p^{n+3/2,m+1} = p^{n+3/2,m+1,l+1}$, $e^{n+1,m+1} = e^{n+1,m+1,l+1}$, 
and $\rho^{n+3/2,m+1} = \rho^{n+3/2,m+1,l+1}$. 
Takemitsu \citep{Takemitsu_1985} proposed a similar method 
that simultaneously iterates the velocity correction equation 
and Poisson equation of the pressure correction. 
However, the Poisson equation for pressure should be solved after correcting the velocity. 
The present numerical method does not require the Poisson equation for the pressure to be solved. 
The velocity, pressure, internal energy, and density are simultaneously relaxed 
for the low Mach number model as well. 
In the Boussinesq approximation, 
the velocity and pressure are simultaneously relaxed.

We use the Euler implicit method and implicit midpoint rule 
to analyze steady and unsteady flows, respectively. 
For spatial differentiation, we use second-order accuracy central difference methods. 
The biconjugate gradient stabilized method \citep{Vorst_1992} is applied 
to solve simultaneous linear equations. 
These discretized equations are solved by following the subsequent procedure.
\begin{enumerate}[Step 1 :]
\setlength{\leftskip}{1.0em}
\setlength{\itemsep}{0mm}
\item At $m = 1$, 
      let $\rho^{n+3/2,m} = \rho^{n+1/2}$, $\bu^{n+1,m} = \bu^{n}$, 
      $p^{n+3/2,m} = p^{n+1/2}$, $e^{n+1,m} = e^{n}$, and $T^{n+1,m} = T^{n}$.
\item Solve Eq. (\ref{predict.u}), and predict the velocity $\hat{\bu}^{n+1,m+1}$.
\item Solve Eq. (\ref{predict.e}), and predict the internal energy $\hat{e}^{n+1,m+1}$.
\item Calculate the density $\rho^{n+3/2,m+1}$ by Eq. (\ref{implicit.rho}).
\item Solve the pressure correction value $\Delta p^{m,l}$ using the Helmholtz equation (\ref{poisson.2}).
\item Correct the velocity $\bu^{n+1,m+1,l+1}$, pressure $p^{n+3/2,m+1,l+1}$, 
      internal energy $e^{n+1,m+1,l+1}$, and density $\rho^{n+3/2,m+1,l+1}$ 
      using Eqs. (\ref{correct.u.2}), (\ref{correct.p.2}), (\ref{correct.e.2}), and (\ref{density}), respectively. 
      At the end of simultaneous relaxation, 
      set $\bu^{n+1,m+1} = \bu^{n+1,m+1,l+1}$, $p^{n+3/2,m+1} = p^{n+3/2,m+1,l+1}$, 
      $e^{n+1,m+1} = e^{n+1,m+1,l+1}$, and $\rho^{n+3/2,m+1} = \rho^{n+3/2,m+1,l+1}$.
\item Repeat Steps 2 to 7. 
      After the Newton iteration is completed, 
      set $\rho^{n+3/2} = \rho^{n+3/2,m+1}$, $\bu^{n+1} = \bu^{n+1,m+1}$, 
      $p^{n+3/2} = p^{n+3/2,m+1}$, $e^{n+1} = e^{n+1,m+1}$, 
      and $T^{n+1} = T^{n+1,m+1}$.
\item Advance the time step and return to Step 1.
\end{enumerate}

\subsection{Correction of pressure in cavity flow}

For the flow in a cavity, 
the Neumann condition is used as the pressure boundary condition; 
hence, the representative pressure $p_\mathrm{ref}$ cannot be fixed in the flow field 
when the calculation is performed without the low Mach number approximation. 
As for an initial condition, we give the initial pressure at an initial temperature. 
As time progresses, the reference pressure shifts and the mass error in the container occurs. 
Vierendeel et al. \citep{Vierendeel_et_al_2003} corrected the pressure so that mass is conserved. 
The mass $M$ is calculated as
\begin{equation}
   M = \int_V \rho^{n+1,m+1} dV 
   = \int_V \frac{\kappa Ma^2 
   \left(p^{n+3/2,m+1} + p^{n+1/2} \right)/2 +1}
   {e^{n+1,m+1}} dV.
\end{equation}
If the pressure correction amount is $\delta_p$, 
the corrected mass $M'$ becomes $M' = 1$ as follows:
\begin{align}
   M' &= \int_V \frac{\kappa Ma^2 
   \left[ (p^{n+3/2,m+1} + \delta_p) + p^{n+1/2} \right]/2 + 1}
   {e^{n+1,m+1}} dV 
   \nonumber \\
   &= \int_V \frac{\kappa Ma^2 p^{n+1,m+1}}{e^{n+1,m+1}} dV 
   + \frac{1}{2} \int_V \frac{\kappa Ma^2 \delta_p}{e^{n+1,m+1}} dV 
   \nonumber \\
   &= M + \delta_p \frac{1}{2} \int_V \frac{\kappa Ma^2}{e^{n+1,m+1}} dV = 1.
\end{align}
Therefore, the pressure is corrected as
\begin{subequations}
\begin{align}
   & \hat{p}^{n+3/2,m+1} = p^{n+3/2,m+1} + \delta_p, \\
   & \delta_p = \frac{1 - M}{1/2 \int_V \frac{\kappa Ma^2}{e^{n+1,m+1}} dV}.
\end{align}
\end{subequations}
In the analysis using low Mach number approximation or Boussinesq approximation, 
such pressure correction is not necessary.

\subsection{Evaluation of transport quantity}

We analyzed a periodic inviscid flow to evaluate conservation properties 
for the momentum and energy. 
In this flow field, each total amount of transport quantity is conserved. 
By observing the time evolution of the total amount for the momentum, total energy, and entropy, 
we can evaluate the properties of the conservative finite difference scheme. 
The total amount $\langle \rho \bu \rangle$ of momentum is obtained 
by the volume integral of momentum and is defined as
\begin{equation}
   \langle \rho u \rangle 
   = \sum_{i,j,k} \left. \frac{\overline{J \rho}^x u}{\bar{J}^x} 
   \Delta V \right|_{i+1/2,j,k}, \quad 
   \langle \rho v \rangle 
   = \sum_{i,j,k} \left. \frac{\overline{J \rho}^y v}{\bar{J}^y} 
   \Delta V \right|_{i,j+1/2,k}, \quad 
   \langle \rho w \rangle 
   = \sum_{i,j,k} \left. \frac{\overline{J \rho}^z w}{\bar{J}^z} 
   \Delta V \right|_{i,j,k+1/2},
\end{equation}
where $V$ represents the computational domain 
and $\Delta V$ is the volume of a cell.

The total amounts, $\langle \rho E \rangle$ and $\langle \rho s \rangle$, 
for total energy and entropy are given as
\begin{equation}
   \langle \rho E \rangle 
   = \sum_{i,j,k} \left. \rho E \Delta V \right|_{i,j,k}, \quad 
   \rho E = \rho e + \frac{1}{2 J} \kappa (\kappa - 1) Ma^2 \left( 
     \overline{\overline{J \rho}^x u^2}^x 
   + \overline{\overline{J \rho}^y v^2}^y 
   + \overline{\overline{J \rho}^z w^2}^z \right),
\end{equation}
\begin{equation}
   \langle \rho s \rangle 
   = \sum_{i,j,k} \left. \rho s \Delta V \right|_{i,j,k}.
\end{equation}

Subsequently. we define error to evaluate preservation property. 
We confirm the temporal change of the error 
and the variation of the error with the number of grid points 
and verify the conservation property and convergence of the finite difference scheme. 
The momentum errors $\varepsilon_{\rho u}$, $\varepsilon_{\rho v}$, 
and $\varepsilon_{\rho w}$ are defined as
\begin{equation}
   \varepsilon_{\rho u} 
   = \frac{\langle \rho u \rangle - \langle \rho u \rangle_0}
          {\langle \rho u \rangle_0}, \quad 
   \varepsilon_{\rho v} 
   = \frac{\langle \rho v \rangle - \langle \rho v \rangle_0}
          {\langle \rho v \rangle_0}, \quad 
   \varepsilon_{\rho u} 
   = \frac{\langle \rho w \rangle - \langle \rho w \rangle_0}
          {\langle \rho w \rangle_0},
\end{equation}
where the subscript $0$ represents the total amount of an initial value. 
For a periodic flow, the sum of the initial values of momentum may be zero. 
In that case, the error is defined as
\begin{equation}
   \varepsilon_{\rho u} 
   = \frac{\langle \rho u \rangle - \langle \rho u \rangle_0}
          {\langle (\rho u)^2 \rangle_0^{1/2}}, \quad 
   \varepsilon_{\rho v} 
   = \frac{\langle \rho v \rangle - \langle \rho v \rangle_0}
          {\langle (\rho v)^2 \rangle_0^{1/2}}, \quad 
   \varepsilon_{\rho u} 
   = \frac{\langle \rho w \rangle - \langle \rho w \rangle_0}
          {\langle (\rho w)^2 \rangle_0^{1/2}}.
\end{equation}
The total energy and entropy errors $\varepsilon_{\rho E}$ and $\varepsilon_{\rho s}$ are defined as follows:
\begin{equation}
   \varepsilon_{\rho E} 
   = \frac{\langle \rho E \rangle - \langle \rho E \rangle_0}
          {\langle \rho E \rangle_0}, \quad 
   \varepsilon_{\rho s} 
   = \frac{\langle \rho s \rangle - \langle \rho s \rangle_0}
          {\langle \rho s \rangle_0}.
\end{equation}
If the initial values of the dimensionless pressure and density are constant, 
the entropy is zero. 
Then, the entropy error is defined as the absolute error as follows:
\begin{equation}
   \varepsilon_{\rho s} 
   =\langle \rho s \rangle - \langle \rho s \rangle_0.
\end{equation}

\section{Verification of numerical method}
\label{sec4}

\subsection{Sound wave propagation}

To verify the validity of the present numerical method for the analysis of compressible flows, 
we analyzed the propagation of an acoustic wave in a one-dimensional periodic inviscid compressible flow. 
In this flow field, no energy attenuation owing to viscosity occurs; 
thus, the amplitude of the sound wave remains constant. 
Additionally, a theoretical value is given by the one-dimensional linear theory 
for sound wave propagation. 
By investigating the variation rate of the amplitude and frequency of the sound wave to the theoretical values, 
we can verify the validity of this numerical scheme.

The initial values for the density and temperature are uniform and defined 
as $\rho_0$ and $T_0$, respectively. 
The initial value of sound speed is given as $c_0 = \sqrt{\kappa (\kappa-1) c_v T_0}$. 
As in the existing research \citep{Wall_et_al_2002}, 
the following velocity distribution is imposed as an initial disturbance, 
and the sound wave is propagated:
\begin{equation}
  u(x) = \Delta u_\mathrm{max} \cos \left( \frac{2 \pi x}{\lambda} \right),
\end{equation}
where $\lambda$ and $\Delta u_\mathrm{max}$ are the wavelength and amplitude 
of the disturbance, respectively, 
and we set $\Delta u_\mathrm{max} = 10^{-6} c_0$. 
The sound wave returns to its original position after the time $t = \lambda /c_0$. 
The computational domain is $\lambda$ in the $x$-direction 
and $\lambda/(N-1)$ in the $y$- and $z$-directions. 
Here, $N$ is the number of grid points in the $x$-direction. 
As for the boundary condition, periodic boundary conditions are imposed for velocity, 
pressure, and internal energy. 
The reference values in this calculation are as follows: 
the length, velocity, time, density, pressure, temperature, and internal energy are set to 
$l_\mathrm{ref} = \lambda$, $u_\mathrm{ref} = c_0$, 
$t_\mathrm{ref} = l_\mathrm{ref}/u_\mathrm{ref}$, 
$\rho_\mathrm{ref} = \rho_0$, 
$p_\mathrm{ref} = (\kappa -1) \rho_\mathrm{ref} c_v T_\mathrm{ref}$, 
$T_\mathrm{ref} = T_0$, 
and $e_\mathrm{ref} = c_v T_\mathrm{ref}$, respectively. 
In this calculation, the specific heat ratio is $\kappa = 1.4$, 
and the Mach number is $Ma = u_\mathrm{ref}/c_0 = 1.0$. 
To compare the calculated result with the existing result \citep{Wall_et_al_2002}, 
we used a uniform grid with $N \times 2 \times 2$ grid points, 
where $N = 3$, 5, 7, 11, 21, 31, and 41. 
The time step was set so that the Courant number 
$\mathrm{CFL} = \Delta t c_0 / \Delta x$ was CFL = 0.5, 2.0, and 8.0 
for all grids.

\begin{figure}[!t]
\begin{minipage}[b]{0.49\hsize}
\begin{center}
\includegraphics[trim=0mm 0mm 0mm 0mm, clip, width=75mm]{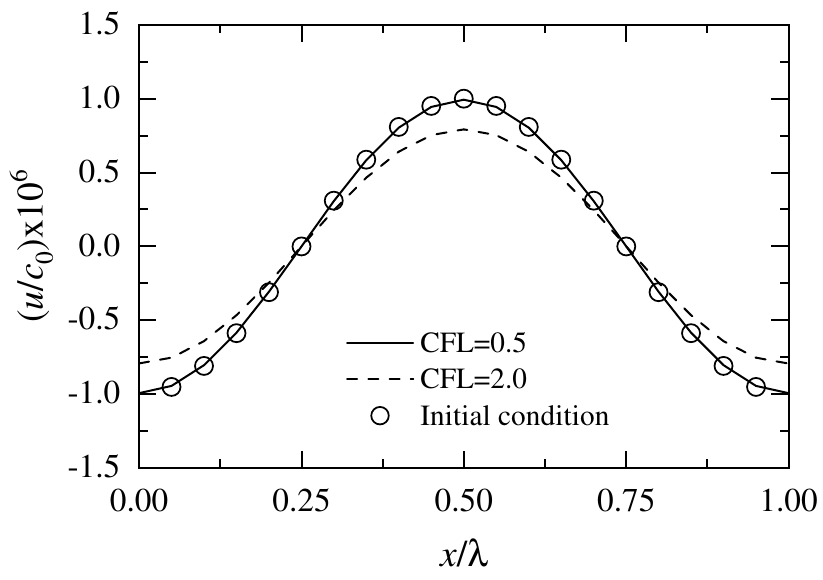} \\
{\small (a) $t/(\lambda /c_0) = 3$}
\end{center}
\end{minipage}
%
\begin{minipage}[b]{0.49\hsize}
\begin{center}
\includegraphics[trim=0mm 0mm 0mm 0mm, clip, width=75mm]{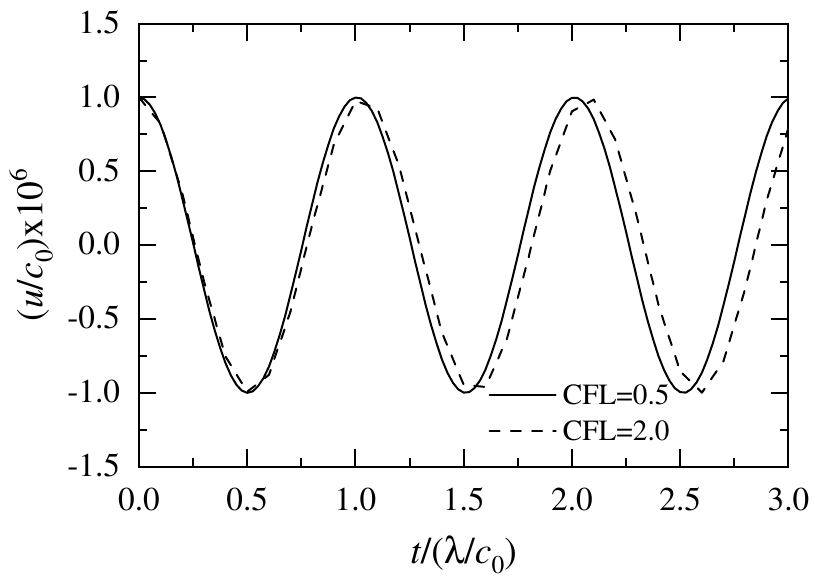} \\
{\small (b) $x/\lambda = 0.5$}
\end{center}
\end{minipage}
\caption{Velocity distribution: $N = 21$}
\label{sound_fig01}
\end{figure}

Figure \ref{sound_fig01}(a) shows the velocity distribution 
obtained using the number of grid points $N = 21$ 
at dimensionless time $t/(\lambda /c_0) = 3$. 
For CFL = 0.5, the velocity distribution shows the same distribution 
as the initial velocity distribution, 
and the periodicity of the sound wave is maintained in time. 
In contrast, at CFL = 2.0, 
the velocity distribution differs from the initial one, 
and the phase of the sound wave is shifted. 
Figure \ref{sound_fig01}(b) shows the time variation of the velocity at $x/\lambda = 0.5$. 
For CFL = 0.5, the velocity becomes maximum at $t/(\lambda/c_0) = 1.0$, 2.0, and 3.0, 
and no phase shift of the sound wave occurs. 
For CFL = 2.0, the velocity becomes maximum at $t/(\lambda/c_0) = 1.036$ and 2.073, 
causing a phase shift of the sound wave.

\begin{figure}[!t]
\centering
\includegraphics[trim=0mm 0mm 0mm 0mm, clip, width=75mm]{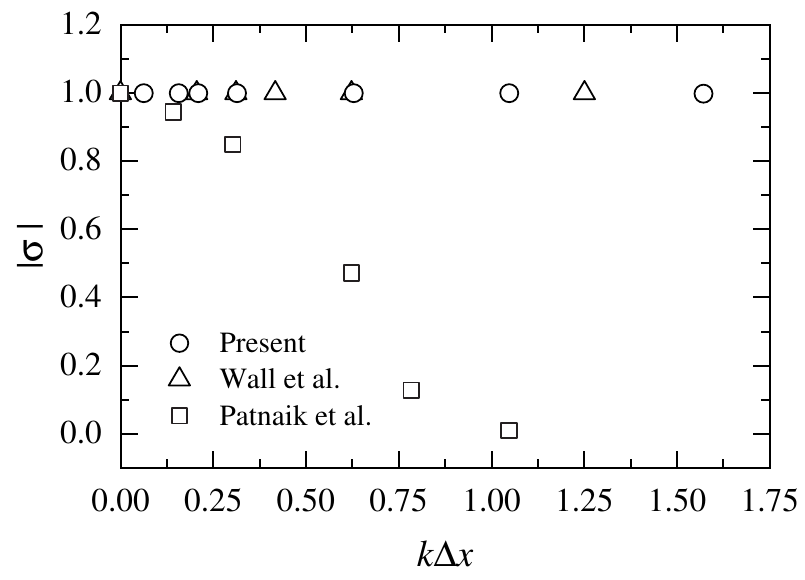}
\caption{Comparison of acoustic wave amplitude: CFL = 0.5}
\label{sound_fig02}
\end{figure}

By checking the error of the sound wave amplitude, 
we can evaluate the error of the finite difference scheme. 
Figure \ref{sound_fig02} shows the amplification ratio $|\sigma|$ of the sound wave amplitude 
at each grid resolution $k \Delta x$ for CFL = 0.5. 
Here, $k = 2\pi/\lambda$ represents the wavenumber of the sound wave, 
and $\Delta x$ represents the grid width. 
Analysis by Wall et al. \citep{Wall_et_al_2002} gives a theoretical value of $|\sigma| = 1$. 
In this calculation, 
$|\sigma|$ is obtained as the ratio of the amplitude of the sound wave 
after one cycle to the initial amplitude. 
This definition is the same as Patnaik et al. \citep{Patnaik_et_al_1987}. 
From the figure, it can be seen that this calculation result does not change 
with the grid resolution and is kept constant. 
In the method (Barely Implicit Correction method) by Patnaik et al. \citep{Patnaik_et_al_1987}, 
the amplitude of the sound wave is attenuated as the grid resolution becomes coarse. 
An increase in $k\Delta x$ means an increase in wavenumber. 
Hence, the higher the wavenumber, the greater the attenuation of the sound wave. 
Additionally, this calculated result agrees well with the theoretical value 
as with the result of Wall et al. \citep{Wall_et_al_2002}.

\begin{figure}[!t]
\begin{minipage}[t]{0.49\hsize}
\begin{center}
\includegraphics[trim=0mm 0mm 0mm 0mm, clip, width=75mm]{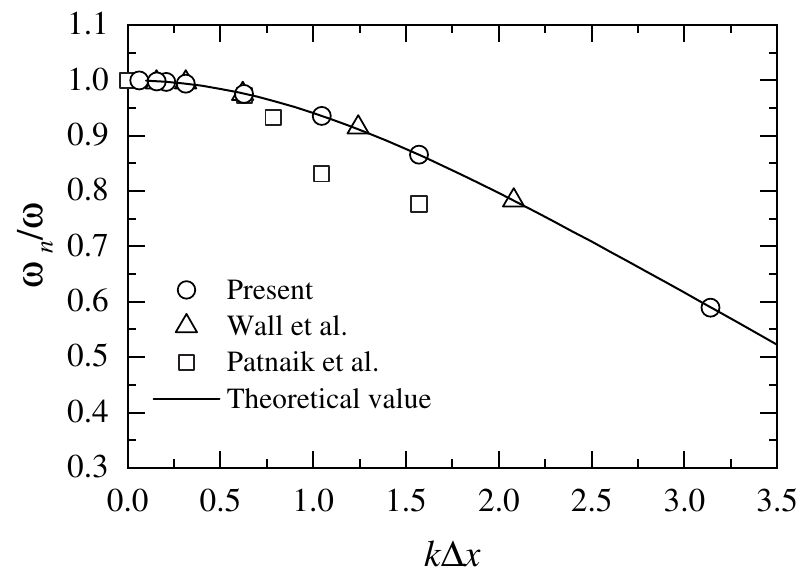} \\
{\small (a) CFL = 0.5}
\end{center}
\end{minipage}
\begin{minipage}[t]{0.49\hsize}
\begin{center}
\includegraphics[trim=0mm 0mm 0mm 0mm, clip, width=75mm]{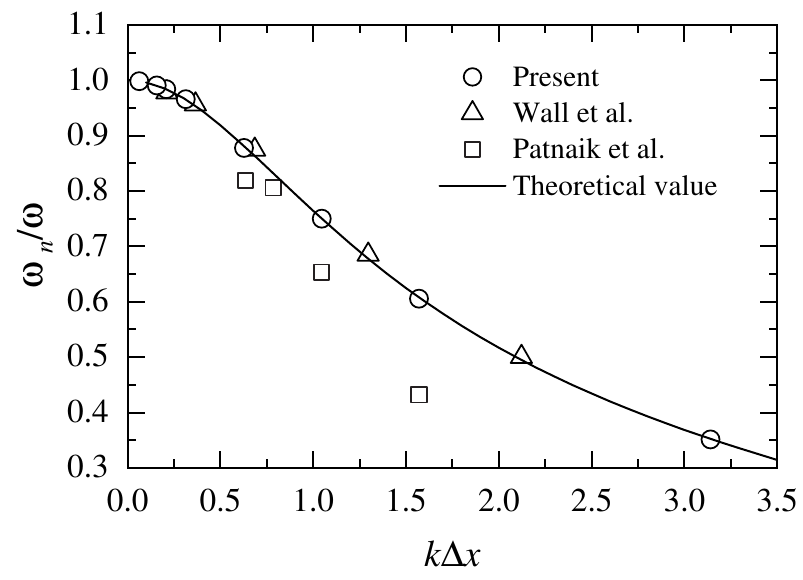} \\
{\small (b) CFL = 2.0}
\end{center}
\end{minipage}
\caption{Comparison of acoustic wave frequency}
\label{sound_fig03}
\end{figure}

Figure \ref{sound_fig03} shows the ratio $\omega _n/\omega$ of sound frequency 
at each grid resolution for CFL = 0.5 and 2.0. 
Here, $\omega _n = \Delta \theta/\Delta t$ represents the numerical solution of the sound wave frequency, 
and $\omega = k c_0$ represents the theoretical value. 
$\Delta \theta$ is the phase difference. 
$\omega _n$ is calculated as the average frequency from the wavelength of three periods. 
The validity of the method for finding the frequency will be discussed later. 
The theoretical value of the sound wave frequency ratio $\omega _n/\omega$ 
calculated by Wall et al. \citep{Wall_et_al_2002} is shown below:
\begin{equation}
  \frac{\omega _n}{\omega} = \frac{1}{(\mathrm{CFL}) k \Delta x} {\tan}^{-1}
  \left[ \frac{2 (\mathrm{CFL}) (k' \Delta x)}
              {2 - \frac{1}{2} [(\mathrm{CFL}) (k' \Delta x)]^2} \right],
\end{equation}
where $k'$ represents the modified wavenumber. 
Using the second-order central difference method, 
the relationship between $k\Delta x$ and $k' \Delta x$ is given as
\begin{equation}
  k' \Delta x = \sqrt{2 - 2 \cos (k \Delta x)}.
\end{equation}
Regardless of the CFL, 
the present calculation result agrees well with the previous result \citep{Wall_et_al_2002} 
and theoretical value. 
Additionally, it can be seen that this calculation result is more appropriate than 
the result of Patnaik et al. \citep{Patnaik_et_al_1987}.

\begin{figure}[!t]
\begin{minipage}[t]{0.49\hsize}
\begin{center}
\includegraphics[trim=0mm 0mm 0mm 0mm, clip, width=75mm]{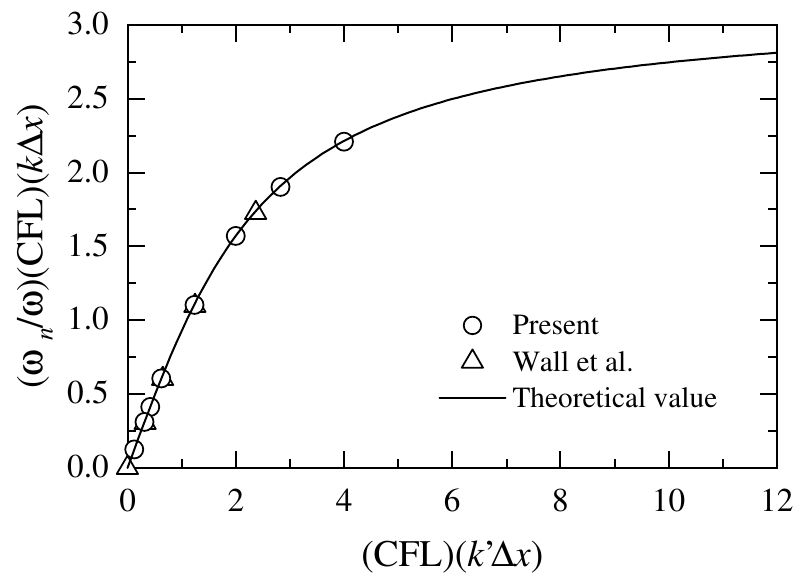} \\
{\small (a) CFL = 2.0}
\end{center}
\end{minipage}
\begin{minipage}[t]{0.49\hsize}
\begin{center}
\includegraphics[trim=0mm 0mm 0mm 0mm, clip, width=75mm]{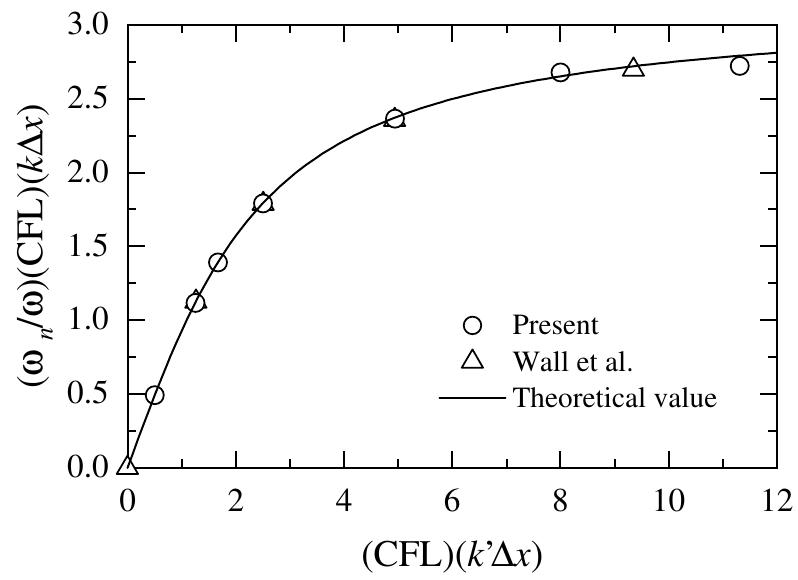} \\
{\small (b) CFL = 8.0}
\end{center}
\end{minipage}
\caption{Influences of space and time discretizations on acoustic wave frequency}
\label{sound_fig04}
\end{figure}

\begin{figure}[!t]
\begin{minipage}[t]{0.49\hsize}
\begin{center}
\includegraphics[trim=0mm 0mm 0mm 0mm, clip, width=75mm]{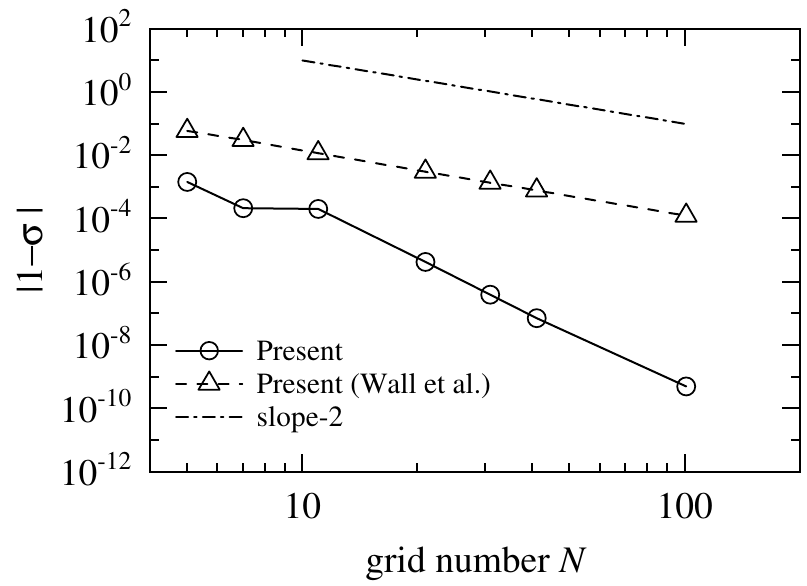} \\
{\small (a) Amplitude}
\end{center}
\end{minipage}
\begin{minipage}[t]{0.49\hsize}
\begin{center}
\includegraphics[trim=0mm 0mm 0mm 0mm, clip, width=75mm]{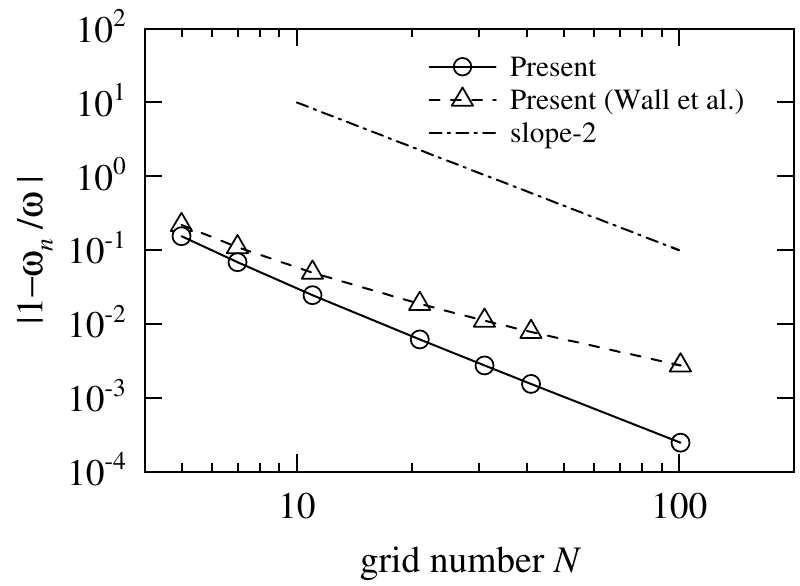} \\
{\small (b) Frequency}
\end{center}
\end{minipage}
\caption{Errors of amplitude and frequency of acoustic wave for CFL $= 0.5$}
\label{sound_fig05}
\end{figure}

We investigated the effect of discretization in time and space 
on the frequency of the sound wave. 
Figure \ref{sound_fig04} shows the frequency ratio of the sound wave 
at each grid resolution for CFL = 2.0 and 8.0. 
The theoretical value becomes one curve regardless of $k \Delta x$ and CFL. 
The calculated results agree well with existing results \citep{Wall_et_al_2002} 
and theoretical solutions, 
and valid results are obtained even at a high Courant number.

Figure \ref{sound_fig05} shows the amplitude and frequency errors of the sound wave 
at each number of grid points for CFL = 0.5. 
The figure shows the results of this numerical method with the time level of the internal energy 
as $n+1$, as with Morinishi \citep{Morinishi_2009}, 
and $n+3/2$, as with Wall et al. \citep{Wall_et_al_2002}. 
Regarding the amplitude of the sound wave, 
when the time level of the internal energy is set to $n+3/2$, 
the error $|1-\sigma|$ decreases with the slope $-2$ as the number of grid points increases, 
and the second-order accuracy is maintained. 
On the other hand, when the time level of the internal energy is set to $n+1$, 
the error significantly decreases compared to the result using the method of Wall et al. \citep{Wall_et_al_2002}. 
Additionally, when $N > 11$, the error transitions with a slope smaller than the slope of $-2$, 
indicating high convergence. 
As in the amplitude, the error $|1-\omega _n/\omega|$ 
obtained using the present method is lower than that obtained by the same method as Wall et al. \citep{Wall_et_al_2002}. 
The present numerical method provides high convergence to the number of grid points.

\begin{figure}[!t]
\begin{minipage}[t]{0.32\hsize}
\begin{center}
\includegraphics[trim=0mm 0mm 0mm 0mm, clip, width=50mm]{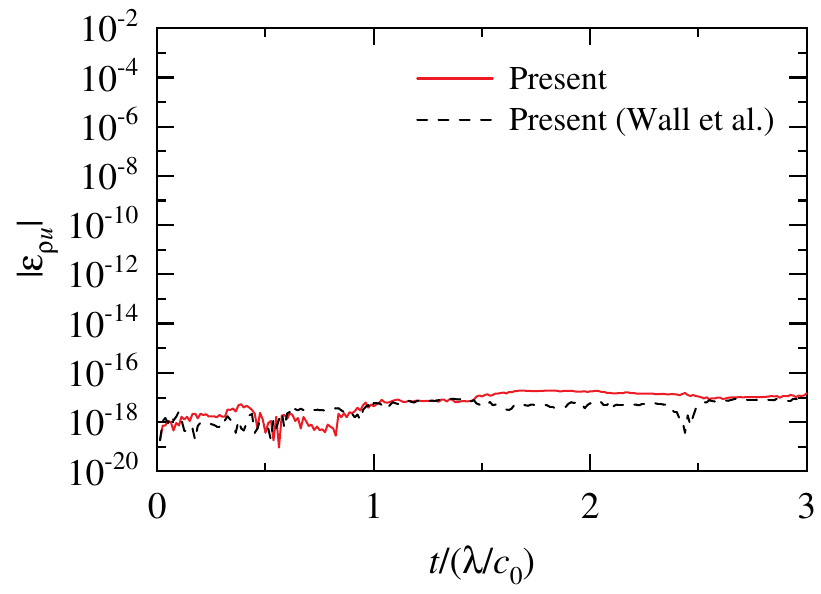} \\
{\small (a) Momentum}
\end{center}
\end{minipage}
\begin{minipage}[t]{0.32\hsize}
\begin{center}
\includegraphics[trim=0mm 0mm 0mm 0mm, clip, width=50mm]{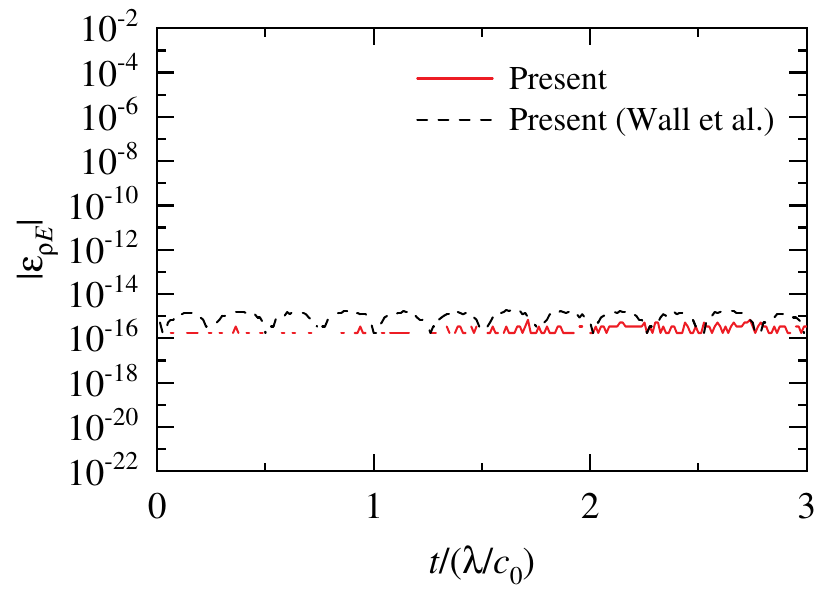} \\
{\small (b) Total energy}
\end{center}
\end{minipage}
\begin{minipage}[t]{0.32\hsize}
\begin{center}
\includegraphics[trim=0mm 0mm 0mm 0mm, clip, width=50mm]{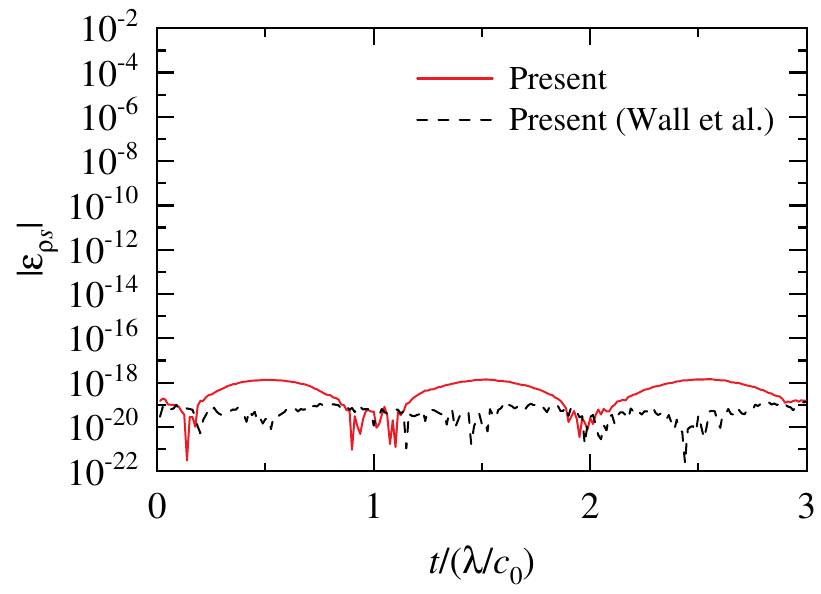} \\
{\small (c) Entropy}
\end{center}
\end{minipage}
\caption{Relative errors of momentum and total energy, and absolute error of entropy: $N = 41$, CFL $= 0.5$}
\label{sound_fig06}
\end{figure}

Figure \ref{sound_fig06} shows the relative errors, 
$\varepsilon_{\rho u} = (\langle \rho u \rangle - \langle \rho u \ rangle_0)/\langle (\rho u)^2 \rangle_0$ 
and 
$\varepsilon_{\rho E} = (\langle \rho E \rangle - \langle \rho E \rangle_0)/\langle \rho E \rangle_0 $, 
in the momentum and total energy, and absolute error, 
$\varepsilon _{\rho s} = \langle \rho s \rangle - \langle \rho s \rangle_0$, in the entropy, 
where $\langle \rho u \rangle$, $\langle \rho E \rangle$, 
and $\langle \rho s \rangle$ are the total amounts of momentum, total energy, 
and entropy, respectively, 
and the subscript 0 represents the initial value. 
As the initial value of entropy is zero, the absolute error in the entropy is shown. 
In Fig. \ref{sound_fig06}(b), the solid line is not shown 
because of $|\varepsilon_{\rho E}| = 0$ at where the solid line is interrupted. 
Regardless of the numerical method, 
the error is the rounding error level, and the momentum, total energy, 
and entropy are discretely conserved.

\begin{figure}[!t]
\begin{minipage}[t]{0.32\hsize}
\begin{center}
\includegraphics[trim=0mm 0mm 0mm 0mm, clip, width=50mm]{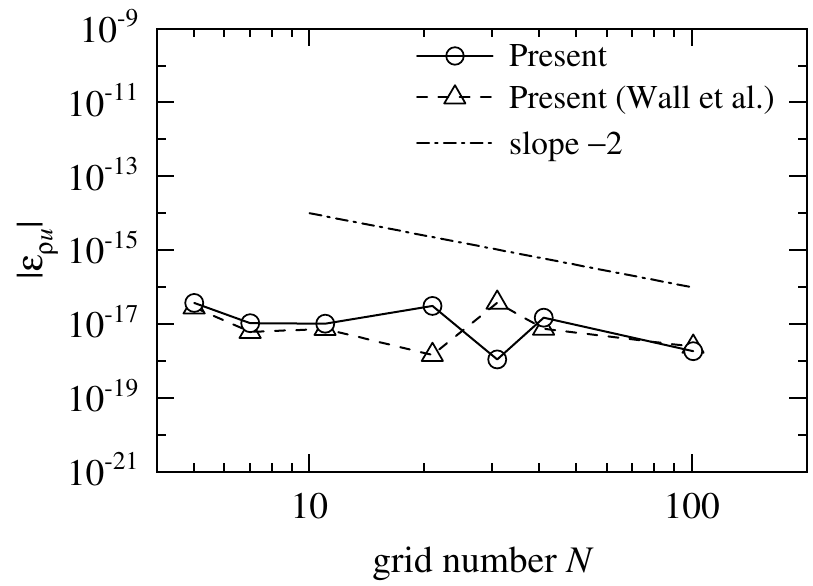} \\
{\small (a) Momentum}
\end{center}
\end{minipage}
\begin{minipage}[t]{0.32\hsize}
\begin{center}
\includegraphics[trim=0mm 0mm 0mm 0mm, clip, width=50mm]{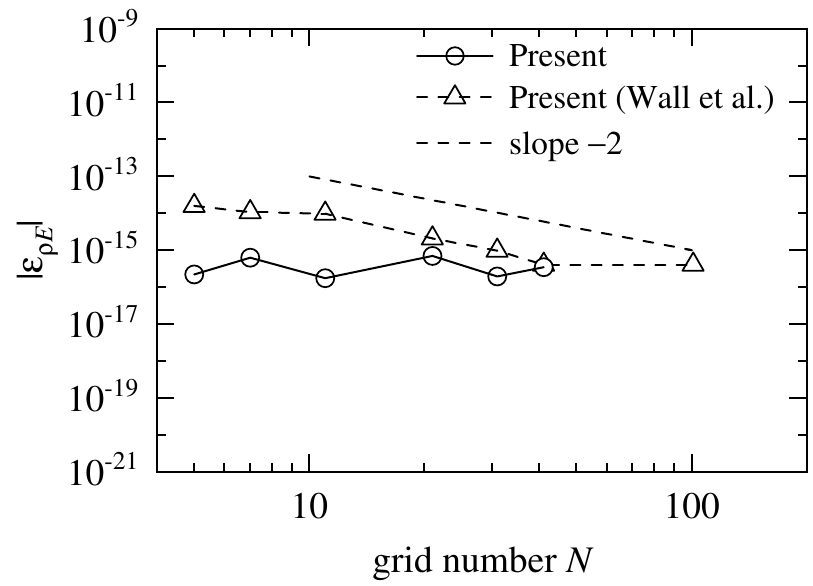} \\
{\small (b) Total energy}
\end{center}
\end{minipage}
\begin{minipage}[t]{0.32\hsize}
\begin{center}
\includegraphics[trim=0mm 0mm 0mm 0mm, clip, width=50mm]{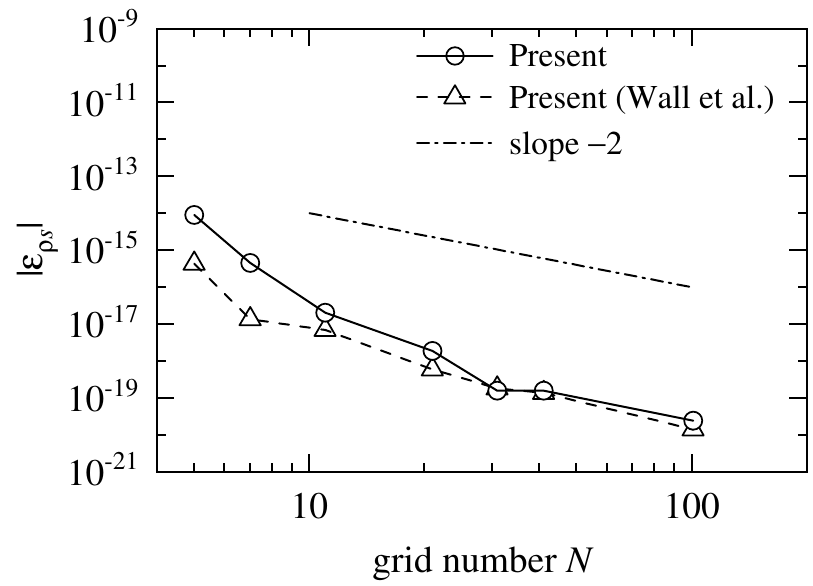} \\
{\small (c) Entropy}
\end{center}
\end{minipage}
\caption{Relative errors of momentum and total energy, and absolute error of entropy for CFL $= 0.5$}
\label{sound_fig07}
\end{figure}

Figure \ref{sound_fig07} shows the relative errors in the momentum 
and total energy and absolute error in the entropy for CFL = 0.5. 
Both numerical methods have low error regardless of the number of grid points 
and conserve momentum, total energy, and entropy. 
Compared with the method of Wall et al. \citep{Wall_et_al_2002}, 
the error in the total energy for this method is low, 
and the error is at the level of rounding error even for coarse grid resolution.

From the above results, 
it can be seen that the present numerical method has excellent calculation accuracy, 
convergence, and energy conservation in the analysis of inviscid compressible flows.

\subsection{Periodic inviscid compressible flow}
\label{subsec2}

We analyzed a three-dimensional periodic inviscid compressible flow 
to verify the validity of this numerical method. 
In the flow field, each total amount of momentum and total energy 
inside the computational domain is temporally conserved. 
If an inappropriate difference scheme is used, each total amount is not preserved 
because of the generation of nonphysical momentum and total energy. 
In the inviscid analysis, 
we can verify the discrete conservation of momentum and total energy 
as no energy attenuation caused by viscosity occurs.

The calculation region is a cube with a side $L$. 
Similarly to the previous study \citep{Morinishi_2009}, 
the initial condition for the velocity is three-dimensionally derived 
using the vector potential with uniform random numbers. 
The velocity is normalized to satisfy 
that the volume-averaged velocity $\langle \bm{u} \rangle$ becomes zero, 
and the volume-averaged velocity fluctuation $\frac{1}{3} \langle u'^2+v'^2+w'^2 \rangle$ is a constant value $U_0^2$. 
The initial values for density and temperature are $\rho_0$ and $T_0$, respectively, 
and are uniform. 
As for boundary conditions, periodic boundary conditions are given to the velocity, 
pressure, and internal energy.

The reference values in this calculation are as follows: 
the length, velocity, time, density, pressure, temperature, and internal energy are 
$l_\mathrm{ref} = L$, $u_\mathrm{ref} = U_0$, 
$t_\mathrm{ref} = l_\mathrm{ref}/u_\mathrm{ref}$, 
$\rho_\mathrm{ref} = \rho_0$, 
$p_\mathrm{ref} = (\kappa -1) \rho_\mathrm{ref} c_v T_\mathrm{ref}$, 
$T_\mathrm{ref} = T_0$, 
and $e_\mathrm{ref} = c_v T_\mathrm{ref}$, respectively. 
The specific heat ratio is set to $\kappa = 1.4$, 
and the initial fluctuating Mach number is given as $Ma = U_0/c_0 = 0.2$. 
Referring to the existing research \citep{Morinishi_2009}, 
the time interval $\Delta t$ is set to $\Delta t/(L/U_0)= 0.0001$, 0.0002, 0.0005, 0.001, 0.002, 0.005, 0.01, 0.02, 0.05, and 0.1. 
The calculation is performed until dimensionless time $t/(L/U_0) = 10$. 
For $\Delta t/(L/U_0) = 0.1$, the initial Courant number $\mathrm{CFL} = \Delta t (U_0+c_0)/\Delta x$ 
considering the speed of sound is CFL = 6.00, 
and the local Courant number is CFL = 7.60. 
Herein, $\Delta x$ is a grid width. 
In this analysis, we first use a uniform grid with $11 \times 11 \times 11$ grid points.

\begin{figure}[!t]
\begin{minipage}[t]{0.49\hsize}
\begin{center}
\includegraphics[trim=0mm 0mm 0mm 0mm, clip, width=75mm]{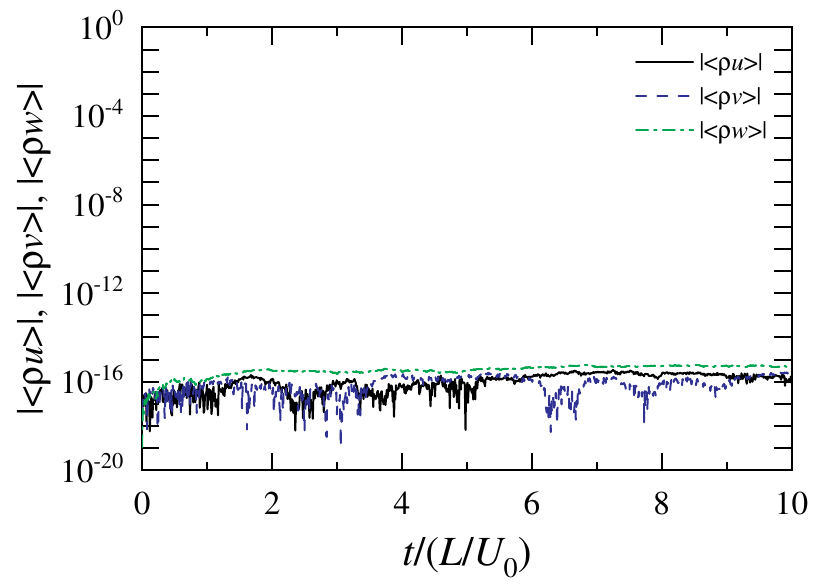} \\
{\small (a) Momentum: present method}
\end{center}
\end{minipage}
\begin{minipage}[t]{0.49\hsize}
\begin{center}
\includegraphics[trim=0mm 0mm 0mm 0mm, clip, width=75mm]{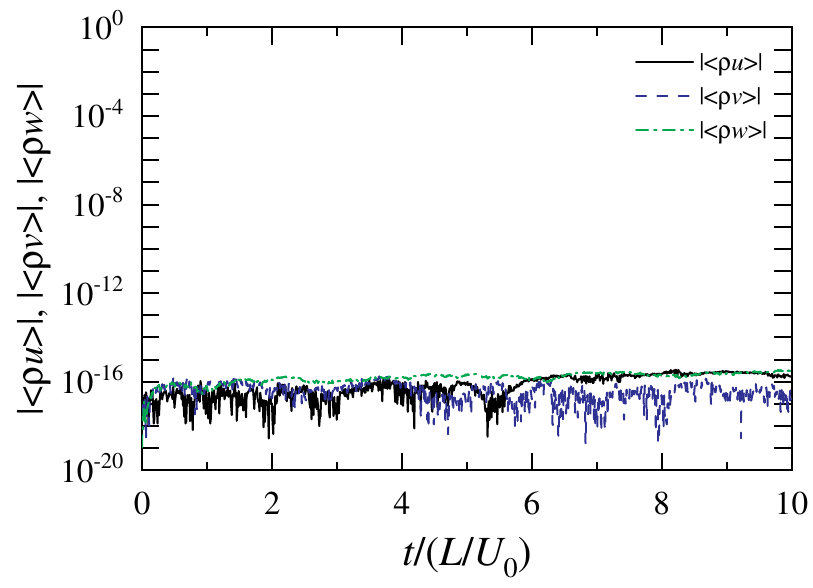} \\
{\small (b) Momentum: Wall et al.'method}
\end{center}
\end{minipage}
\begin{minipage}[t]{0.49\hsize}
\begin{center}
\includegraphics[trim=0mm 0mm 0mm 0mm, clip, width=75mm]{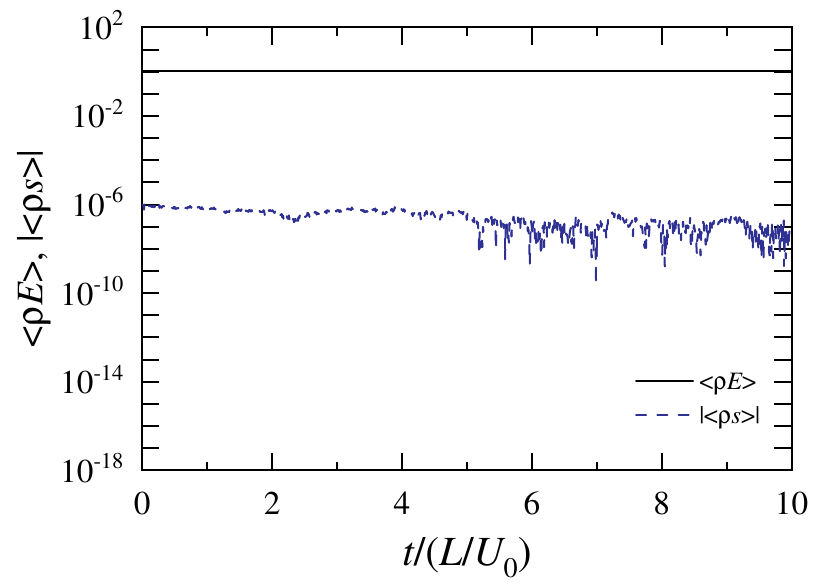} \\
{\small (c) Total energy and entropy: present method}
\end{center}
\end{minipage}
\begin{minipage}[t]{0.49\hsize}
\begin{center}
\includegraphics[trim=0mm 0mm 0mm 0mm, clip, width=75mm]{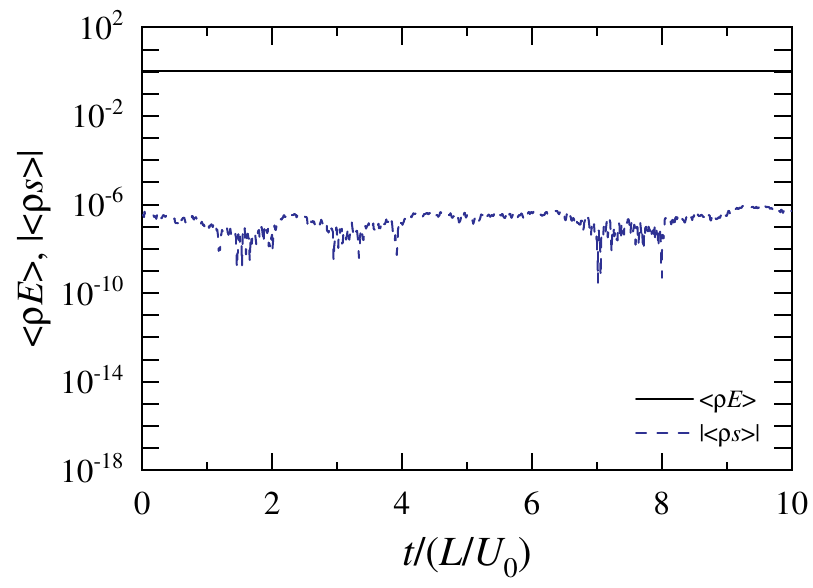} \\
{\small (d) Total energy and entropy: Wall et al.'s method}
\end{center}
\end{minipage}
\caption{Total amounts of momentum, total energy, and entropy: 
$\Delta t/(L/U_0) = 0.002$; No approximation}
\label{invis_fig01}
\end{figure}

In this analysis, 
the methods without approximation to the fundamental equation 
and using low Mach number approximation were compared 
to verify the effectiveness of the present numerical method. 
First, we show the results of the proposed method without approximation. 
Figures \ref{invis_fig01}(a) and (b) show the total amount 
$\langle \rho \bm{u} \rangle$ of the momentum for the time interval $\Delta t/(L/U_0)= 0.002$. 
Figures \ref{invis_fig01}(c) and (d) present the total amounts, 
$\langle \rho E \rangle$ and $\langle \rho s \rangle$, 
of the total energy and entropy. 
Figures \ref{invis_fig01}(a) and (c) show the results of this numerical method 
when the time level of the internal energy is set to $n+1$ 
in the discretization of the fundamental equation, as with Morinishi \citep{Morinishi_2009}. 
Figures \ref{invis_fig01}(b) and (d) show the results 
when the time level of the internal energy is $n+3/2$, as with Wall et al. \citep{Wall_et_al_2002}. 
Regardless of the numerical method, 
$|\langle \rho \bm{u} \rangle|$ changes on the order of $10^{-16}$, 
and the momentum is temporally conserved even at a discrete level. 
Additionally, $\langle \rho E \rangle$ and $|\langle \rho s \rangle|$ change 
on the order of 1.0336 and $10^{-6}$, respectively, 
and it is found that the initial state is maintained. 
From the above results, for each total amount, 
it was confirmed that there was no difference in the characteristics of the two numerical methods.

\begin{figure}[!t]
\begin{minipage}[t]{0.49\hsize}
\begin{center}
\includegraphics[trim=0mm 0mm 0mm 0mm, clip, width=75mm]{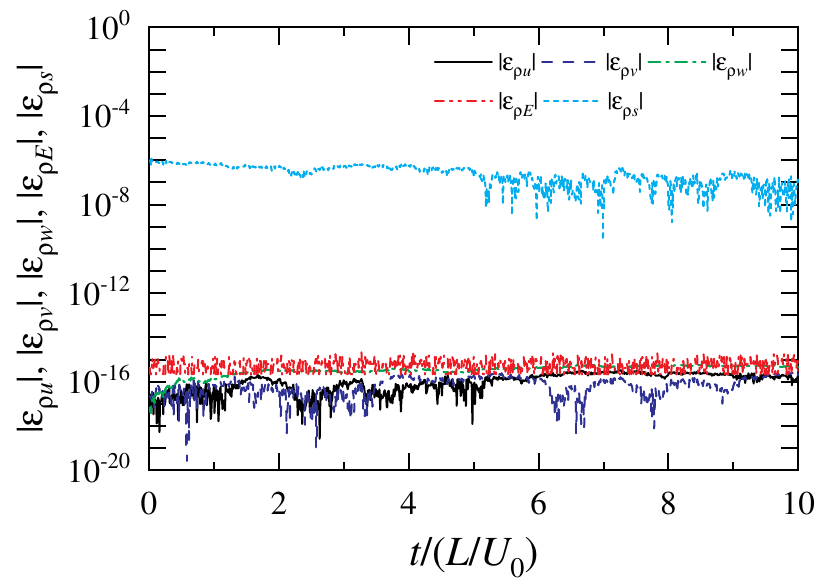} \\
{\small (a) Present method}
\end{center}
\end{minipage}
\begin{minipage}[t]{0.49\hsize}
\begin{center}
\includegraphics[trim=0mm 0mm 0mm 0mm, clip, width=75mm]{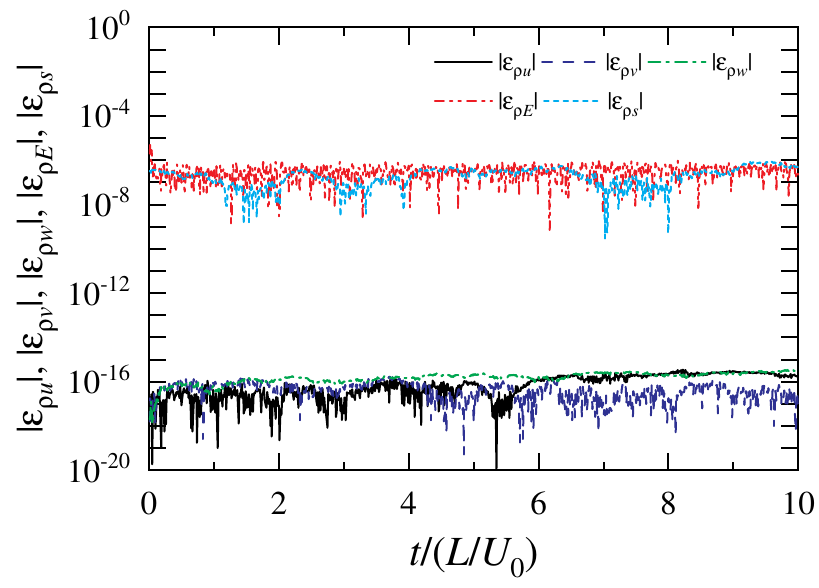} \\
{\small (b) Wall et al.'s method}
\end{center}
\end{minipage}
\caption{Relative errors of momentum and total energy, and absolute error of entropy: 
$\Delta t/(L/U_0)= 0.002$; No approximation}
\label{invis_fig02}
\end{figure}

Figure \ref{invis_fig02} shows the error for $\Delta t/(L/U_0) = 0.002$. 
The relative error in the momentum is 
$\varepsilon_{\rho \bm{u}} = (\langle \rho \bm{u} \rangle - \langle \rho \bm{u} \rangle_0)/\langle (\rho \ bm{u})^2 \rangle_0 ^{1/2}$, 
the relative error in the total energy is 
$\varepsilon_{\rho E} = (\langle \rho E \rangle - \langle \rho E \rangle_0)/ \langle \rho E \rangle_0$, 
and the absolute error in the entropy is 
$\varepsilon_{\rho s} = \langle \rho s \rangle - \langle \rho s \rangle_0$. 
The subscript 0 indicates the initial value. 
As $\langle \rho s \rangle_0$ is zero, we define the error as the absolute error. 
Additionally, as $\langle (\rho \bm{u})^2 \rangle_0^{1/2}$ and $\langle \rho E \rangle_0$ are on the order of $10^0$,  
we can compare the orders of $\varepsilon_{\rho \bm{u}}$, $\varepsilon_{\rho E}$, 
and $\varepsilon_{\rho s}$, although the definition of error is different. 
Figure \ref{invis_fig02}(a) shows the results of this analysis method 
when the time level of the internal energy is set to $n+1$ in the discretization of the fundamental equation 
as with Morinishi \citep{Morinishi_2009}. 
$|\varepsilon_{\rho \bm{u}}|$ and $|\varepsilon_{\rho E}|$ are at the orders of $10^{-17}$ and $10^{-15 }$, respectively, 
and the momentum and total energy are conserved in time. 
In contrast, $|\varepsilon_{\rho s}|$ changes on the order of $10^{-6}$ 
and shows a higher value than the momentum and total energy errors. 
The discrete conservation property for entropy worsens. 
This is because the entropy conservation equation cannot be derived discretely. 
Figure \ref{invis_fig02}(b) shows the result obtained using the same method as Wall et al. \citep{Wall_et_al_2002} 
when solving the fundamental equations. 
$|\varepsilon_{\rho \bm{u}}|$ changes on the order of $10^{-17}$, 
and the momentum is conserved in time. 
However, as $|\varepsilon_{\rho E}|$ fluctuates on the order of $10^{-7}$, 
the total energy conservation property deteriorates compared with the result of the present method. 
Furthermore, as with the result of this numerical method, 
$|\varepsilon_{\rho s}|$ changes in the order of $10^{-6}$ and shows a high value.

\begin{figure}[!t]
\centering
\includegraphics[trim=0mm 0mm 0mm 0mm, clip, width=75mm]{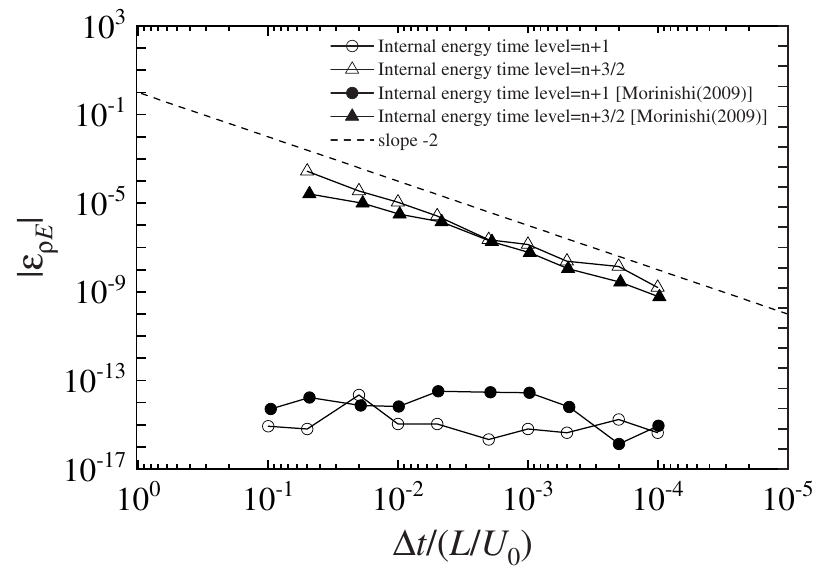} \\
\caption{Relative error of total energy: $t/(L/U_0) = 10$; No approximation}
\label{invis_fig03}
\end{figure}

Subsequently, at $t/(L/U_0) = 10$, 
the relative error $\varepsilon_{\rho E}$ in the total energy 
to $\Delta t/(L/U_0)$ is shown in Fig. \ref{invis_fig03}. 
In the figure, the results are compared 
when the time level of the internal energy is set to $n+3/2$ as in Wall et al. \citep{Wall_et_al_2002}, 
and $n+1$ as in Morinishi \citep{Morinishi_2009}. 
The results of Morinishi \citep{Morinishi_2009} are also compared. 
The dashed line indicates a straight line with a slope of $-2$. 
When the time level of the internal energy is placed at $n+3/2$, 
$|\varepsilon_{\rho E}|$ decreases as $\Delta t$ decreases. 
The slope is $-2$, and the second-order accuracy is maintained. 
The decreasing tendency of the error is the same as the result of Morinishi \citep{Morinishi_2009}. 
In contrast, when the time level of the internal energy is placed at $n+1$, 
$|\varepsilon_{\rho E}|$ is on the order of $10^{-15}$ at each time step 
and is kept to the roundoff error level as in Morinishi \citep{Morinishi_2009}. 
Evidently from the above results, 
if the kinetic and internal energies are obtained discretely at the same time level $n+1$, 
the total energy is discretely preserved at the rounding error level.

\begin{figure}[!t]
\begin{minipage}[t]{0.49\hsize}
\begin{center}
\includegraphics[trim=0mm 0mm 0mm 0mm, clip, width=75mm]{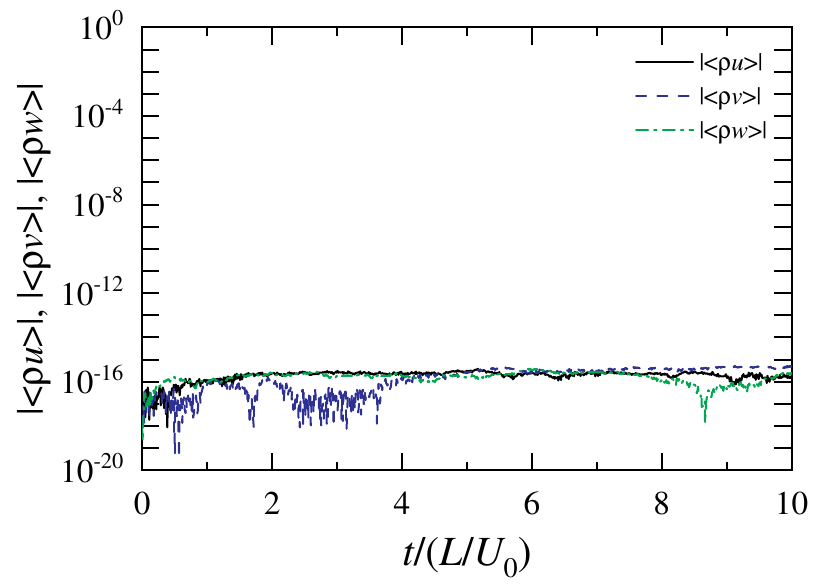} \\
{\small (a) Momentum}
\end{center}
\end{minipage}
\begin{minipage}[t]{0.49\hsize}
\begin{center}
\includegraphics[trim=0mm 0mm 0mm 0mm, clip, width=75mm]{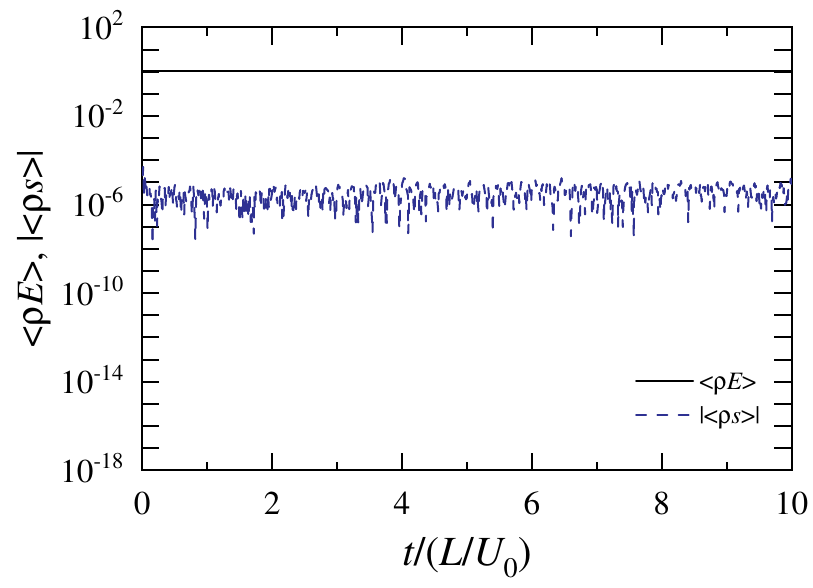} \\
{\small (b) Total energy and entropy}
\end{center}
\end{minipage}
\caption{Total amounts of momentum, total energy, and entropy for nonuniform grid: 
$\Delta t/(L/U_0)= 0.002$; No approximation}
\label{invis_fig04}
\end{figure}

Second, we show momentum and energy conservation properties 
in a nonuniform grid. 
As in the existing research \citep{Morinishi_2009}, 
we use the following nonuniform grid for the analysis:
\begin{equation}
   \alpha_i = L \frac{i-1}{N_{\alpha}-1} 
     + \gamma_{\alpha} \frac{L}{N_{\alpha}-1} 
     \sin \left[ \frac{2 \pi (i - 1)}{N_{\alpha}-1} \right], 
     \quad i=1, \cdots, N_{\alpha},
\end{equation}
where $\alpha = x$, $y$, and $z$, 
$\gamma_x = 0.1$, $\gamma_y = 0.2$, and $\gamma_z = 0.3$, 
and $N_x = N_y = N_z = 11$.

As in Fig. \ref{invis_fig01}, for $\Delta t/(L/U_0) = 0.002$, 
the total amounts, $\langle \rho \bm{u} \rangle$, $\langle \rho E \rangle$, 
and $\langle \rho s \rangle$, of the momentum, total energy, and entropy, respectively, 
are shown in Fig. \ref{invis_fig04}. 
Even in the nonuniform grid, 
the momentum and total energy are kept in the initial state 
when the finite difference scheme of this study is used, 
indicating excellent conservation property. 
Additionally, $|\langle \rho s \rangle|$ changes on the order of $10^{-6}$ 
and shows a value close to the initial value. 
This confirmed that the present numerical method discretely conserves 
momentum and energy even when using a nonuniform grid 
to the same extent as when using a uniform grid.

\begin{figure}[!t]
\centering
\includegraphics[trim=0mm 0mm 0mm 0mm, clip, width=75mm]{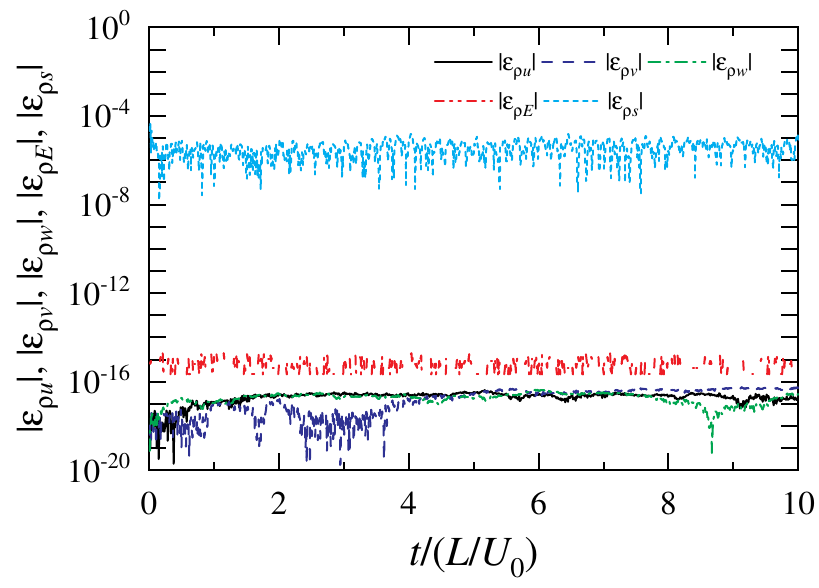}
\vspace*{-0.5\baselineskip}
\caption{Relative errors of momentum and total energy, 
and absolute error of entropy for nonuniform grid: 
$\Delta t/(L/U_0)= 0.002$; No approximation}
\label{invis_fig05}
\end{figure}

Similarly to Fig. \ref{invis_fig02}, for $\Delta t/(L/U_0) = 0.002$, 
the relative errors, $\varepsilon_{\rho \bm{u}}$ and $\varepsilon_{ \rho E}$, 
in the momentum and total energy, 
and absolute entropy error $\varepsilon_{\rho s}$ are shown in Fig. \ref{invis_fig05}. 
It can be seen that the present computational method obtains 
the excellent conservation of momentum and total energy even in the nonuniform grid. 
Additionally, $|\varepsilon_{\rho s}|$ changes on the order of $10^{-6}$, 
comparable to the error when using the uniform grid. 
It was clarified from the above results that even when using the nonuniform grid, 
this numerical method can obtain the same conservation property 
as when using the uniform grid.

\begin{figure}[!t]
\begin{minipage}[t]{0.49\hsize}
\begin{center}
\includegraphics[trim=0mm 0mm 0mm 0mm, clip, width=75mm]{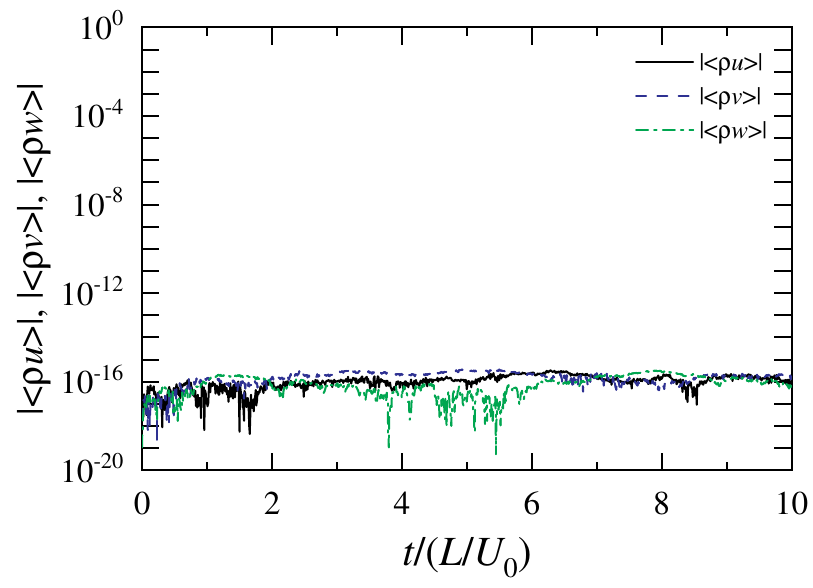} \\
{\small (a) Momentum: uniform grid}
\end{center}
\end{minipage}
\begin{minipage}[t]{0.49\hsize}
\begin{center}
\includegraphics[trim=0mm 0mm 0mm 0mm, clip, width=75mm]{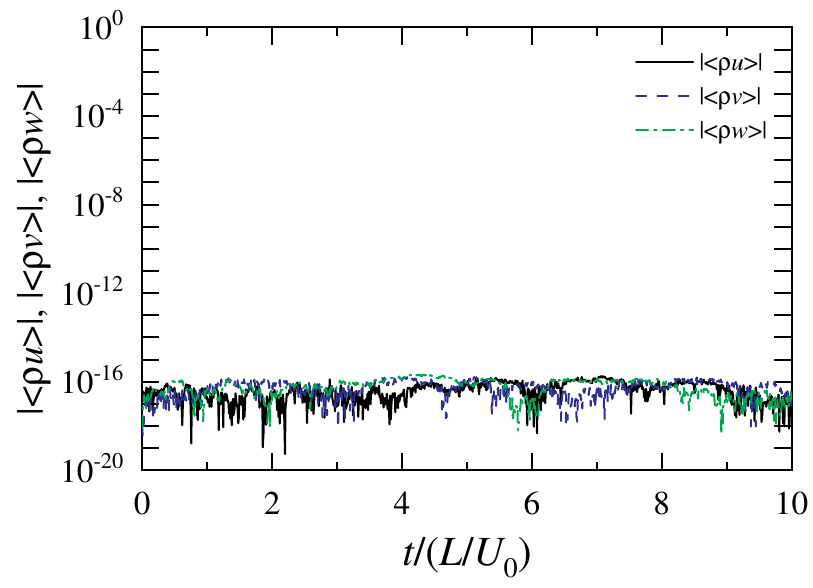} \\
{\small (b) Momentum: nonuniform grid}
\end{center}
\end{minipage}

\vspace*{0.5\baselineskip}
\begin{minipage}[t]{0.49\hsize}
\begin{center}
\includegraphics[trim=0mm 0mm 0mm 0mm, clip, width=75mm]{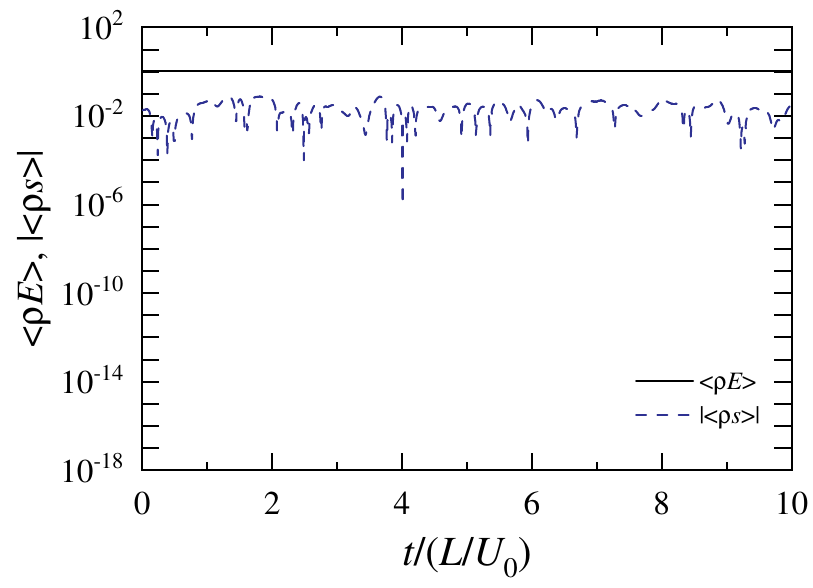} \\
{\small (c) Total energy and entropy: uniform grid}
\end{center}
\end{minipage}
\begin{minipage}[t]{0.49\hsize}
\begin{center}
\includegraphics[trim=0mm 0mm 0mm 0mm, clip, width=75mm]{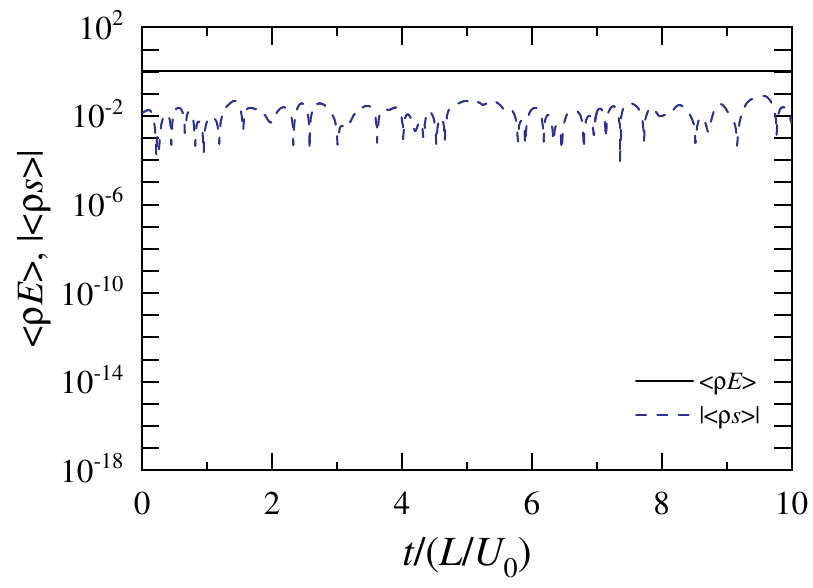} \\
{\small (d) Total energy and entropy: nonuniform grid}
\end{center}
\end{minipage}
\caption{Total amounts of momentum, total energy, and entropy: 
$\Delta t/(L/U_0) = 0.002$; Low Mach number approximation}
\label{invis_fig06}
\end{figure}

To demonstrate the effectiveness of this numerical method, 
which does not approximate the fundamental equations, 
we show the calculation results using the low Mach number approximation. 
Figure \ref{invis_fig06} shows the total amounts, 
$\langle \rho \bm{u} \rangle$, $\langle \rho E \rangle$, 
and $\langle \rho s \rangle$, of the momentum, total energy, and entropy 
for $\Delta t/(L/U_0) = 0.002$. 
$|\langle \rho \bm{u} \rangle |$ changes on the order of $10^{-16}$ 
for both the uniform and nonuniform grids and the momentum is conserved in time. 
$\langle \rho E \rangle$ is kept constant, 
as with the result without approximation. 
In contrast, $|\langle \rho s \rangle|$ changes on the order of $10^{-2}$, 
showing a higerh value than the results using the numerical method without approximation.

\begin{figure}[!t]
\begin{minipage}[t]{0.49\hsize}
\begin{center}
\includegraphics[trim=0mm 0mm 0mm 0mm, clip, width=75mm]{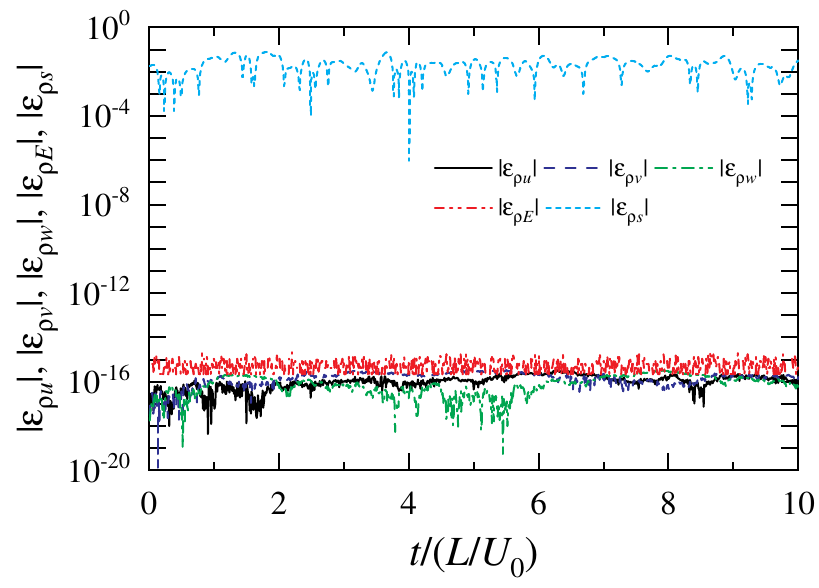} \\
{\small (a) Uniform grid}
\end{center}
\end{minipage}
\begin{minipage}[t]{0.49\hsize}
\begin{center}
\includegraphics[trim=0mm 0mm 0mm 0mm, clip, width=75mm]{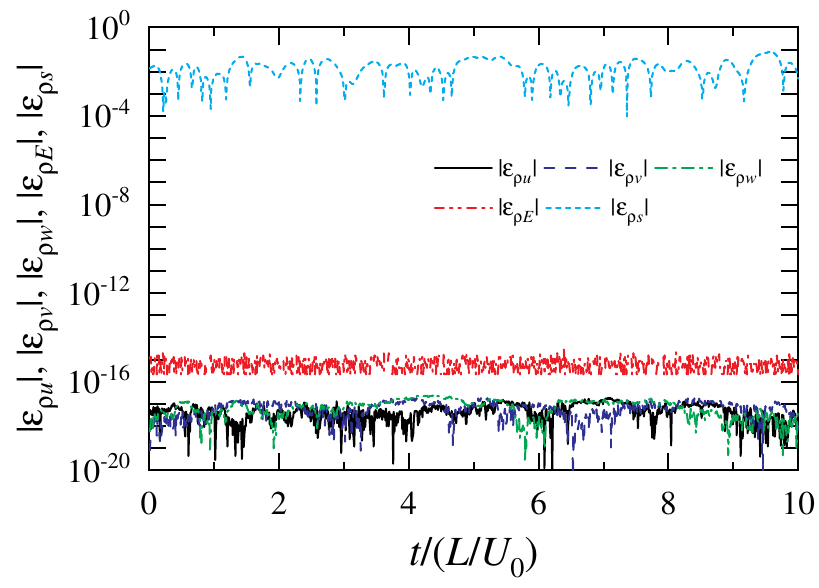} \\
{\small (b) Non-uniform grid}
\end{center}
\end{minipage}
\caption{Relative errors of momentum and total energy, and absolute error of entropy: 
$\Delta t/(L/U_0) = 0.002$; Low Mach number approximation}
\label{invis_fig07}
\end{figure}

Figure \ref{invis_fig07} shows the relative errors, 
$\varepsilon_{\rho \bm{u}}$ and $\varepsilon_{\rho E}$, in the momentum and total energy, 
and absolute error $\varepsilon_{\rho s}$ of the entropy for $\Delta t/(L/U_0) = 0.002$. 
Regardless of the grid, 
$|\varepsilon_{\rho \bm{u}}|$ and $|\varepsilon_{\rho E}|$ remain at low levels, 
and the momentum and total energy are conserved in time. 
$|\varepsilon_{\rho s}|$ changes on the order of $10^{-2}$, 
and the conservation performance worsens compared with the numerical method without approximation.

From the above results, 
it can be seen that the conservation properties of momentum and energy are excellent 
in the present numerical scheme that does not apply approximation to the fundamental equations. 
In the following, we will not use the same method as Wall et al. \citep{Wall_et_al_2002}. 
Furthermore, we will not use the low Mach number approximation and analyze 
using this numerical method that does not apply approximation to the fundamental equations.

\subsection{Decaying compressible isotropic turbulence}

Decaying compressible isotropic turbulence is a benchmark test used 
to verify the validity of computational methods 
\citep{Samtaney_et_al_2001, Honein&Moin_2004, Morinishi_2009, Morinishi_2010}. 
To confirm the accuracy of this numerical method, 
we analyzed compressible isotropic turbulence without using a turbulence model. 
As a condition for the divergence-free velocity field, the energy spectrum $E(k)$ 
for the initial velocity in an incompressible flow is expressed as follows \citep{Samtaney_et_al_2001}:
\begin{equation}
   E(k) = 16 \sqrt{\frac{2}{\pi}} \frac{u_\mathrm{rms}^2}{k_p} 
   \left( \frac{k}{k_p} \right)^4 e^{- 2 (k/k_p)^2},
   \label{enegy_spectrum}
\end{equation}
where $k$ represents the wavenumber, 
$k_p$ represents the peak wavenumber at which the spectrum is maximum, 
and $u_\mathrm{rms}$ represents the root mean square of the velocity fluctuation. 
The initial turbulence energy is given as $3u_\mathrm{rms}^2/2$, 
and the large eddy turnover time scale is $\tau = 2/(k_p u_\mathrm{rms})$. 
The phase of the wavenumber space is given using a uniform random number of $[0, 2\pi]$. 
Initial values of density, pressure, and internal energy are uniformly set to $\rho_0$, $p_0$, and $e_0$, respectively, 
where $p_0 = (\kappa-1)\rho_0 e_0$. 
Periodic boundary conditions are applied to all dependent variables. 
The computational domain is $[0, 2 \pi L]$ in each coordinate direction, 
and the wavenumber is set to $k_p = 4/L$.

The initial velocity fluctuation is defined as $q' = \langle (u'^2+v'^2+w'^2)/3 \rangle^{1/2}$, 
where $\langle \quad \rangle$ represents the volume average in the computational domain. 
As the flow is isotropic turbulence, $q' = \langle u'^2 \rangle^{1/2} = u_\mathrm{rms}$. 
The reference values in this calculation are as follows: 
the length, velocity, time, density, pressure, and internal energy are 
$l_\mathrm{ref} = L$, $u_\mathrm{ref} = u_\mathrm{rms}$, 
$t_\mathrm{ref} = l_\mathrm{ref}/u_\mathrm{ref}$, 
$\rho_\mathrm{ref} = \rho_0$, $p_\mathrm{ref} = p_0$, 
and $e_\mathrm{ref} = e_0$, respectively. 
Similarly to existing research \citep{Honein&Moin_2004,Morinishi_2010}, 
the Reynolds number based on Taylor microscale $\lambda$ is set to 
$Re_{\lambda} = \rho_0 u_\mathrm{rms} \lambda/\mu_0 = 30$. 
Additionally, we set the turbulent Mach number $M_{t0} = \sqrt{3} q'/c_0 = 0.3$, 
Prandtl number $Pr = 0.71$, and specific heat ratio $\kappa = 1.4$. 
The Taylor microscale is defined as 
$\lambda = [\langle u'^2 \rangle / \langle (\partial u'/\partial x)^2 \rangle]^{1/2}$. 
We use the Maxwell--Rayleigh power law $\mu/\mu_0 = (e/e_0)^{0.76}$ as the viscosity coefficient, 
where $\mu_0$ is the initial viscosity coefficient. 
In this analysis, inviscid flow was also analyzed 
to investigate the conservation property of this numerical method. 
As in the existing research \citep{Morinishi_2009,Morinishi_2010}, 
each coordinate direction is divided into 64 cells. 
This grid is the resolution that can be regarded as direct numerical simulation \citep{Honein&Moin_2004}. 
The time step is set to $\Delta t/\tau = 0.02$ ($\Delta t/(L/u_\mathrm{rms}) = 0.01$). 
The initial Courant number is CFL$ = \Delta t (u_\mathrm{rms}+c_0)/\Delta x = 0.690$, 
where $c_0$ is the speed of sound. 
The maximum initial Courant number using the local velocity is CFL$ = \Delta t(u +c_0 )/\Delta x = 0.971$.

\begin{figure}[!t]
\begin{minipage}[t]{0.49\hsize}
\begin{center}
\includegraphics[trim=0mm 0mm 0mm 0mm, clip, width=70mm]{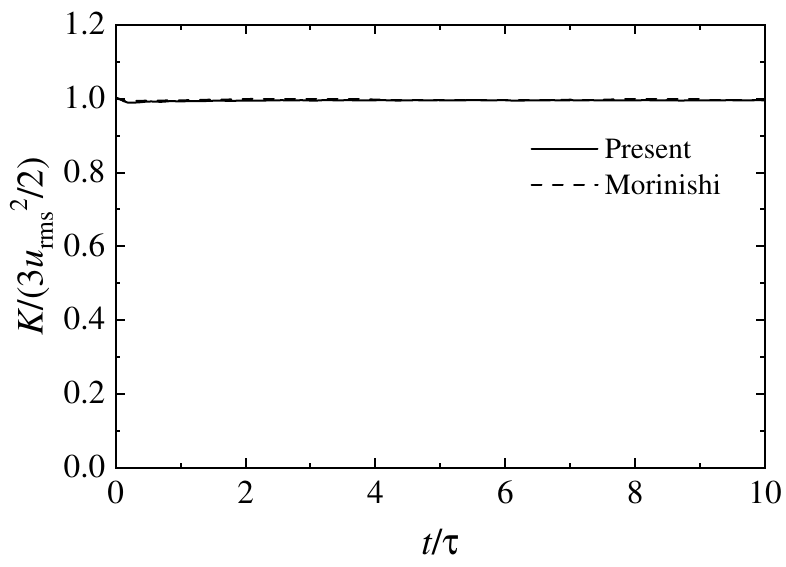} \\
{\small (a) Turbulence kinetic energy}
\end{center}
\end{minipage}
\begin{minipage}[t]{0.49\hsize}
\begin{center}
\includegraphics[trim=0mm 0mm 0mm 0mm, clip, width=70mm]{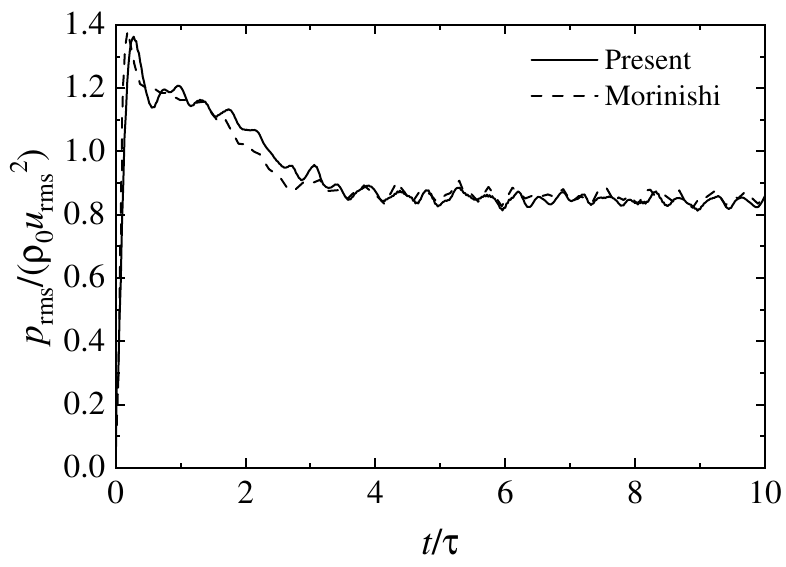} \\
{\small (b) Fluctuation intensity of pressure}
\end{center}
\end{minipage}

\vspace*{0.5\baselineskip}
\begin{minipage}[t]{0.49\hsize}
\begin{center}
\includegraphics[trim=0mm 0mm 0mm 0mm, clip, width=70mm]{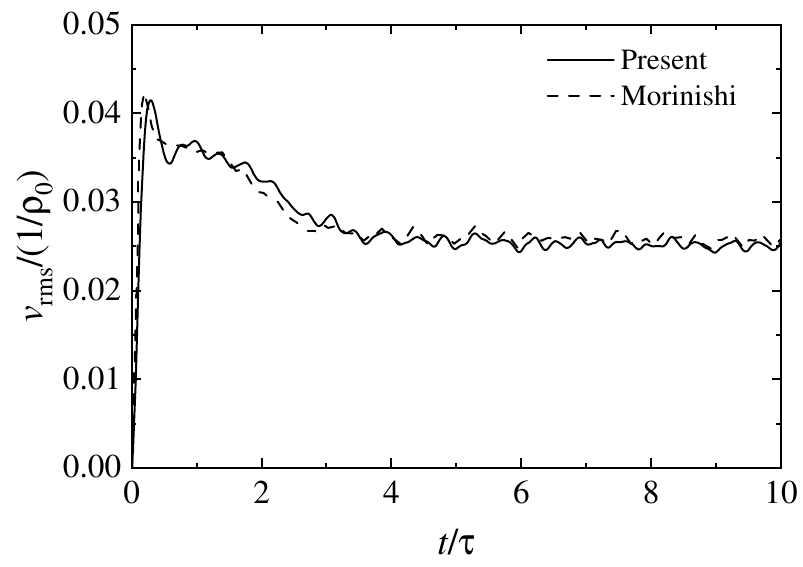} \\
{\small (c) Fluctuation intensity of specific volume}
\end{center}
\end{minipage}
\begin{minipage}[t]{0.49\hsize}
\begin{center}
\includegraphics[trim=0mm 0mm 0mm 0mm, clip, width=70mm]{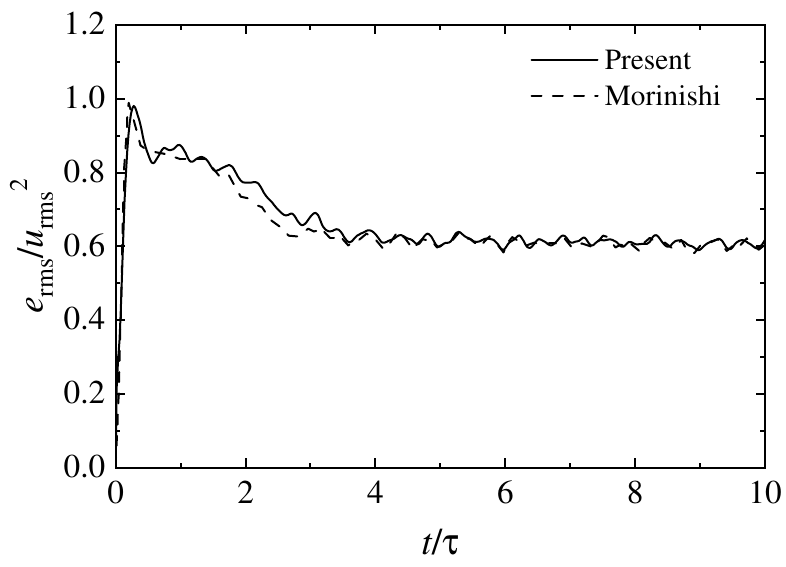} \\
{\small (d) Fluctuation intensity of internal energy}
\end{center}
\end{minipage}
\caption{Time variations of turbulence kinetic energy and fluctuation intensities of pressure, 
specific volume, and internal energy for $Re_{\lambda} = \infty$}
\label{isot_fig01}
\end{figure}

\begin{figure}[!t]
\begin{minipage}[t]{0.49\hsize}
\begin{center}
\includegraphics[trim=0mm 0mm 0mm 0mm, clip, width=70mm]{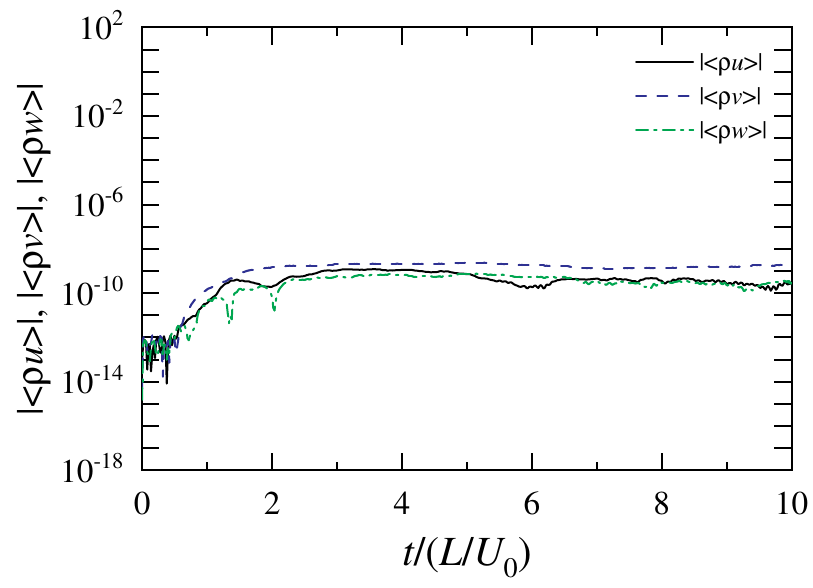} \\
{\small (a) Momentum}
\end{center}
\end{minipage}
\begin{minipage}[t]{0.49\hsize}
\begin{center}
\includegraphics[trim=0mm 0mm 0mm 0mm, clip, width=70mm]{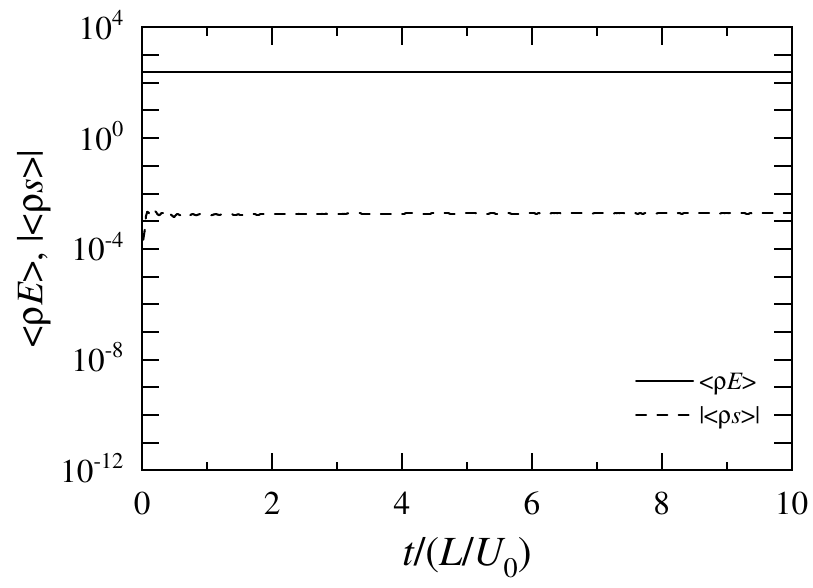} \\
{\small (b) Total energy and entropy}
\end{center}
\end{minipage}
\caption{Total amounts of momentum, total energy, and entropy for $Re_{\lambda} = \infty$}
\label{isot_fig02}
\end{figure}

\begin{figure}[!t]
\centering
\includegraphics[trim=0mm 0mm 0mm 0mm, clip, width=71mm]{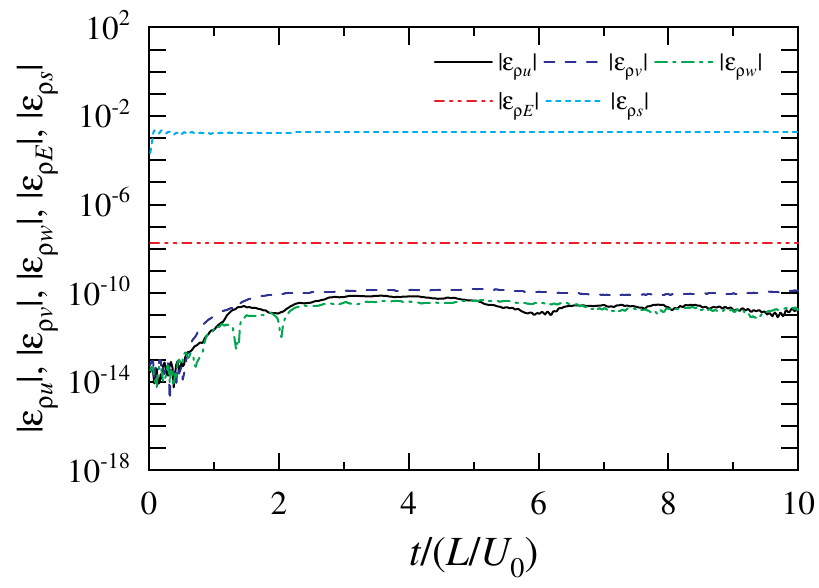} \\
\vspace*{-0.5\baselineskip}
\caption{Relative errors of momentum and total energy, and absolute error of entropy for $Re_{\lambda} = \infty$}
\label{isot_fig03}
\end{figure}

First, we investigated inviscid flow. 
Figure \ref{isot_fig01} shows turbulence energy 
$K = (u_\mathrm{rms}^2+v_\mathrm{rms}^2+w_\mathrm{rms}^2)/2$, 
pressure fluctuation $p_\mathrm{rms} = \langle p'^2 \rangle^{1/2}$, 
specific volume fluctuation $v_\mathrm{rms} = \langle (1/\rho)'^2 \rangle^{1/2}$, 
and internal energy fluctuation $e_\mathrm{rms} = \langle e'^2 \rangle^{1/2}$. 
$K$ remains constant and agrees well with an existing result \citep{Morinishi_2010}. 
$p_\mathrm{rms}$, $v_\mathrm{rms}$, and $e_\mathrm{rms}$ become maximum 
at $t/\tau \approx 0.30$, and decrease with time at $t/\tau > 0.30$. 
$p_\mathrm{rms}$, $v_\mathrm{rms}$, and $e_\mathrm{rms}$ asymptotically approaches constant values over time. 
These fluctuating distributions are similar to the previous result \citep{Morinishi_2010}. 
From the above results, the validity of this numerical method was confirmed 
in inviscid analysis considering compressibility.

Subsequently, we clarify the conservation property of this computational method 
for inviscid compressible isotropic turbulence. 
The total amount, $\langle \rho \bm{u} \rangle$, of the momentum is shown in Fig. \ref{isot_fig02}(a), 
and the total amounts, $\langle \rho E \rangle$ and $\langle \rho s \rangle$, 
of the total energy and entropy are shown in Fig. \ref{isot_fig02}(b). 
$|\langle \rho \bm{u} \rangle |$ changes on the order of $10^{-9}$, 
and the momentum is conserved in time. 
Additionally, $\langle \rho E \rangle$ and $|\langle \rho s \rangle|$ change 
around $254.30$ and $10^{-3}$, respectively, 
and the initial state is maintained. 
Figure \ref{isot_fig03} shows the relative error 
$\varepsilon_{\rho \bm{u}} = (\langle \rho \bm{u} \rangle - \langle \rho \bm{u} \rangle_0) /\langle (\rho \bm{u})^2 \rangle_0^{1/2}$ 
in the momentum, 
relative error $\varepsilon_{\rho E} = (\langle \rho E \rangle - \langle \ rho E \rangle_0)/\langle \rho E \rangle_0$ 
in the total energy, 
and absolute error $\varepsilon_{\rho s} = \langle \rho s \rangle - \langle \rho s \rangle_0$ 
in the entropy. 
As $\langle \rho s \rangle_0$ is zero, we defined the error as the absolute error. 
The subscript 0 indicates the initial value. 
$|\varepsilon_{\rho \bm{u}}|$ and $|\varepsilon_{\rho E}|$ variate 
on the order of $10^{-10}$ and $10^{-8}$, respectively, 
and the momentum and total energy are conserved in time. 
In contrast, $|\varepsilon_{\rho s}|$ changes on the order of $10^{-3}$, 
and although the definition of error is different, 
it shows a higher value than the errors of the momentum and total energy. 
This trend is similar to the results obtained in Subsection \ref{subsec2}. 
It was confirmed from the above results that this numerical method has good conservation properties 
for momentum and total energy in the inviscid compressible flow.

\begin{figure}[!t]
\begin{minipage}[t]{0.49\hsize}
\begin{center}
\includegraphics[trim=0mm 0mm 0mm 0mm, clip, width=70mm]{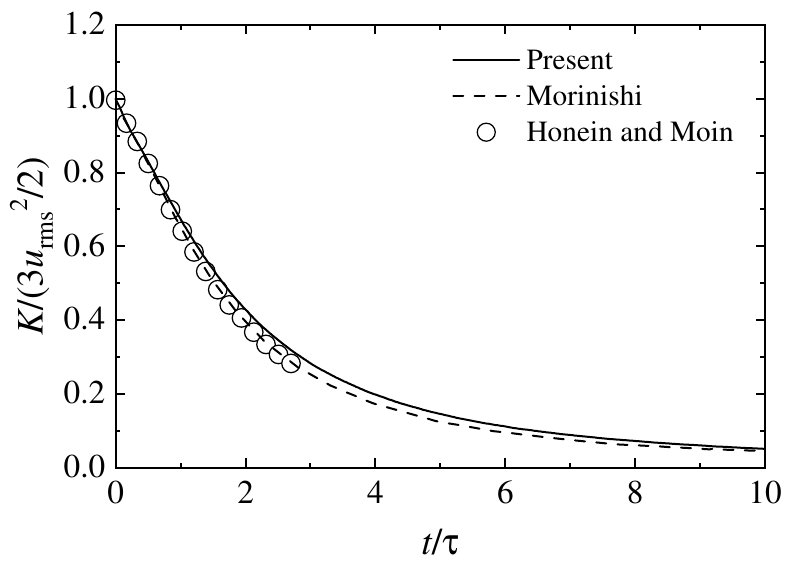} \\
{\small (a) Turbulence kinetic energy}
\end{center}
\end{minipage}
\begin{minipage}[t]{0.49\hsize}
\begin{center}
\includegraphics[trim=0mm 0mm 0mm 0mm, clip, width=70mm]{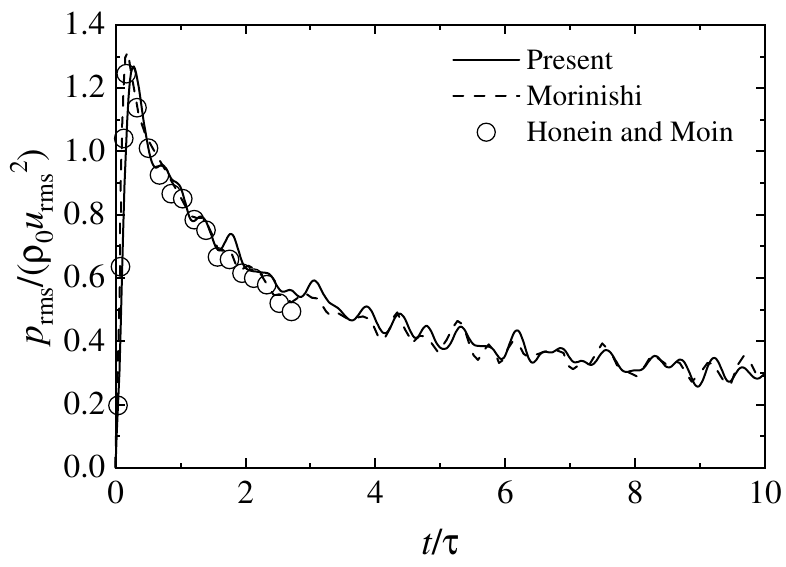} \\
{\small (b) Fluctuation intensity of pressure}
\end{center}
\end{minipage}

\vspace*{0.5\baselineskip}
\begin{minipage}[t]{0.49\hsize}
\begin{center}
\includegraphics[trim=0mm 0mm 0mm 0mm, clip, width=70mm]{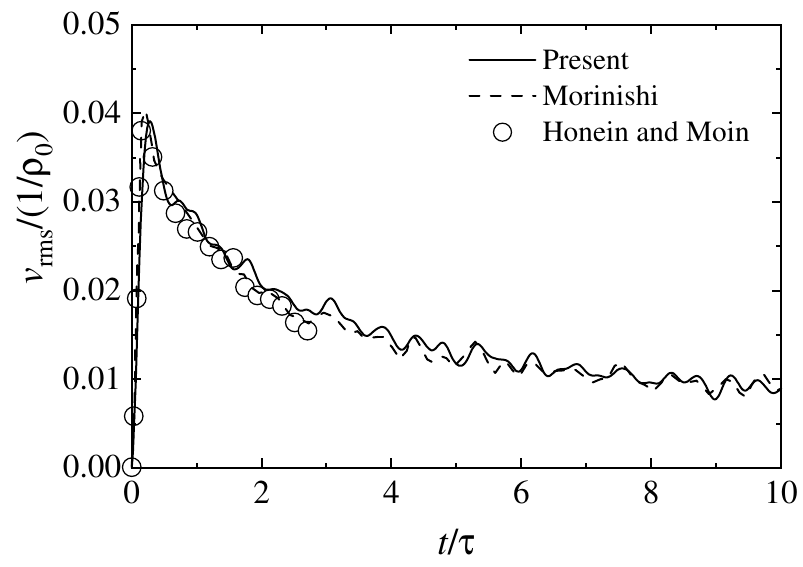} \\
{\small (c) Fluctuation intensity of specific volume}
\end{center}
\end{minipage}
\begin{minipage}[t]{0.49\hsize}
\begin{center}
\includegraphics[trim=0mm 0mm 0mm 0mm, clip, width=70mm]{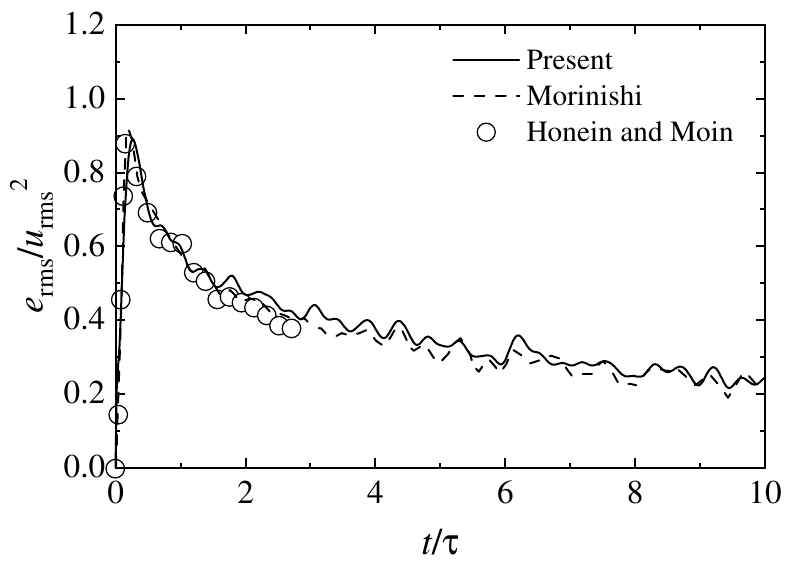} \\
{\small (d) Fluctuation intensity of internal energy}
\end{center}
\end{minipage}
\caption{Time variations of turbulence kinetic energy and fluctuation intensities of pressure, 
specific volume, and internal energy for $Re_{\lambda} = 30$}
\label{isot_fig04}
\end{figure}

Finally, we investigated the validity of this numerical method in the analysis 
considering compressibility and viscosity. 
Figure \ref{isot_fig04} shows the time variation of each fluctuation. 
The existing result \citep{Honein&Moin_2004} in the figure was obtained 
by direct numerical simulation using a spectral method. 
$K$ decreases over time. 
Although the turbulence kinetic energy is slightly overestimated, 
the calculated results agree well with the previous results 
\citep{Honein&Moin_2004, Morinishi_2010}. 
The distributions of $p_\mathrm{rms}$, $v_\mathrm{rms}$, and $e_\mathrm{rms}$ have extreme values 
at $t/\tau \approx 0.27$, and decrease with time at $t/\tau > 0.27$. 
These fluctuating distributions are similar to existing results \citep{Honein&Moin_2004, Morinishi_2010}. 
From the above results, the validity of the present numerical method was confirmed 
in the analysis considering compressibility and viscosity.

\subsection{Taylor--Green decaying vortex}

Taylor and Green \citep{Taylor&Green_1937} proposed a flow field with the initial condition 
for a decaying vortex in three-dimensional flow. 
No exact solution is given. 
Large-scale vortices generated by a smooth initial condition interact, 
and the flow transitions to turbulence, forming small-scale vortex structures. 
Ultimately, the vortex decays due to viscosity. 
Analysis of the Taylor--Green decaying vortex using the Navier--Stokes equation 
for compressible fluid has been performed 
to compare high-order computational methods \citep{Wang_et_al_2013}. 
We compared the present and existing results 
\citep{Wang_et_al_2013, Shirokov&Elizarova_2014, Kulikov&Son_2018, Orszag_1974, Shirokov&Elizarova_2014}, 
and verified the validity of this numerical method for the analysis 
considering compressibility and viscosity.

The Taylor--Green decaying vortex flow is a periodic flow field 
with a wavelength $2 \pi L$. 
The computational domain is [$-\pi L$, $\pi L$] in each coordinate. 
The wavenumber is $k = 1/L$. The maximum velocity of the vortex is $V_0$, 
and the density and pressure at the initial uniform temperature $T _0$ are $\rho _0$ and $p_0$, respectively. 
The viscosity coefficient is $\mu_0$, and the thermal conductivity is $k_0$. 
The reference values in this calculation are as follows: 
$l_\mathrm{ref} = L$, $u_\mathrm{ref} = V_0$, $\rho _\mathrm{ref} = \rho _0$, 
$p_\mathrm{ref} = p_0$, $e_\mathrm{ref} = c_v T _0$, 
$\mu _\mathrm{ref} = \mu _0$, $k _\mathrm{ref} = k_0$. 
The dimensionless initial value is given as
\begin{align}
   u &=   \sin(k x) \cos(k y) \cos(k z), \\
   v &= - \cos(k x) \sin(k y) \cos(k z), \\
   w &= 0, \\
   p &= \frac{1}{16} 
        \left[ \cos(2 k x) + \cos(2 k y) \right] 
        \left[ \cos(2 k z) + 2 \right],
\end{align}
where the dimensionless wavenumber is $k = 1$. 
The initial density is obtained from the equation of state. 
As for boundary conditions, periodic boundary conditions are given 
to the velocity, pressure, and internal energy.

To verify the conservation property of the finite difference scheme, 
we investigated the conservation properties of momentum and total energy 
by inviscid analysis. 
As the flow of the Taylor--Green decaying vortex is periodic, 
the volume integral of the initial value of the momentum is zero as follows:
\begin{equation}
   \langle \rho u \rangle = \int_V \rho u dV = 0, \quad 
   \langle \rho v \rangle = \int_V \rho v dV = 0, \quad 
   \langle \rho w \rangle = \int_V \rho w dV = 0,
\end{equation}
where $V$ represents the computational domain. 
The kinetic energy is given by $K = \rho |\bm{u}|^2/2$. 
Using the initial value, 
the volume-integrated kinetic energy $\langle K \rangle$ is given as
\begin{equation}
   \langle K \rangle = \int_V K dV = \pi^3.
\end{equation}
The total energy is $\rho E = \rho e+\kappa(\kappa-1)Ma^2K$. 
As the initial temperature is constant in the domain, 
$\langle \rho e \rangle = (2\pi)^3$. 
The volume-integrated total energy $\langle \rho E \rangle$ is given as
\begin{equation}
   \langle \rho E \rangle = \pi^3 \left[ 8 + \kappa (\kappa - 1) Ma^2 \right].
\end{equation}

In periodic inviscid flow, 
as momentum and total energy are kept in an initial state, 
we can clarify the conservation property of a computational method. 
In this analysis, the physical properties at the initial temperature $T_0$ are constant, 
the specific heat ratio is given as $\kappa = 1.4$, 
and the Mach number is set to $Ma = V_0/c_0 = 0.1$, 
where $c_0$ is the speed of sound. 
For the calculation, we use a uniform grid $N \times N \times N$ with $N = 41$, 
which has almost the same resolution as the existing research 
\citep{Shirokov&Elizarova_2014, Kulikov&Son_2018}. 
Additionally, the time step $\Delta t$ is set to $\Delta t/(L/V_0) = 0.01$ 
as with the condition for the viscous analysis at the high Reynolds number shown later. 
The initial Courant number in this analysis is $\mathrm{CFL} = V_0 \Delta t / \Delta x = 0.0637$, 
and the initial Courant number considering the speed of sound is 
$\mathrm{CFL} = (V_0+c_0) \Delta t/\Delta x = 0.700$, 
where $\Delta x$ is a grid width.

\begin{figure}[!t]
\begin{minipage}[t]{0.49\hsize}
\begin{center}
\includegraphics[trim=0mm 0mm 0mm 0mm, clip, width=70mm]{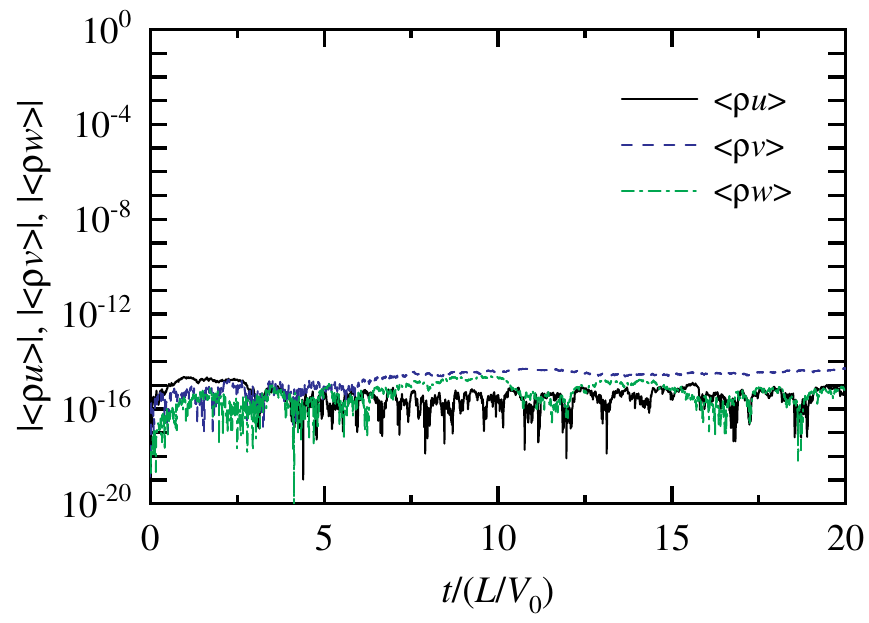} \\
{\small (a) Momentum}
\end{center}
\end{minipage}
\begin{minipage}[t]{0.49\hsize}
\begin{center}
\includegraphics[trim=0mm 0mm 0mm 0mm, clip, width=70mm]{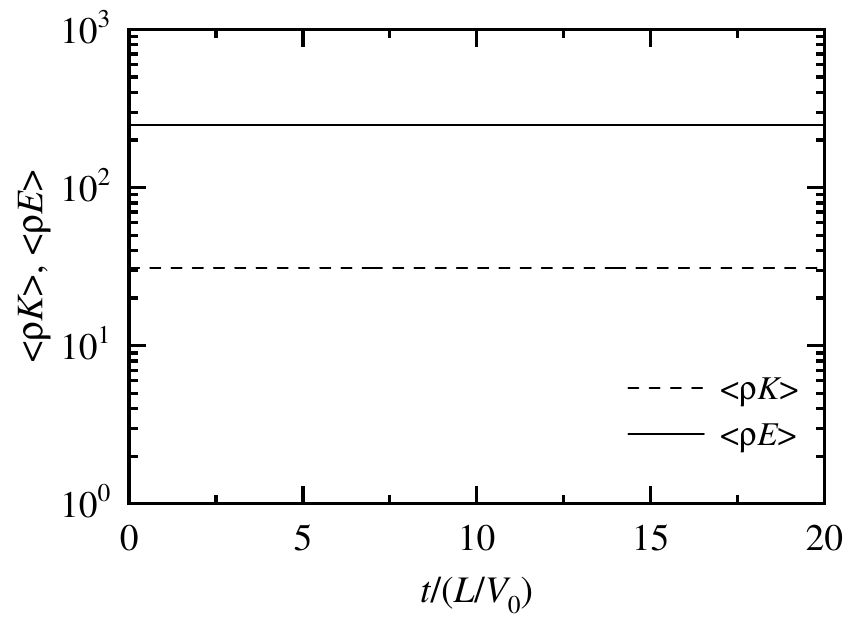} \\
{\small (b) Kinetic energy and total energy}
\end{center}
\end{minipage}
\caption{Total amounts of momentum, kinetic energy, and total energy for $Re = \infty$: $N = 41$}
\label{tg_fig01}
\end{figure}

\begin{figure}[!t]
\centering
\includegraphics[trim=0mm 0mm 0mm 0mm, clip, width=68mm]{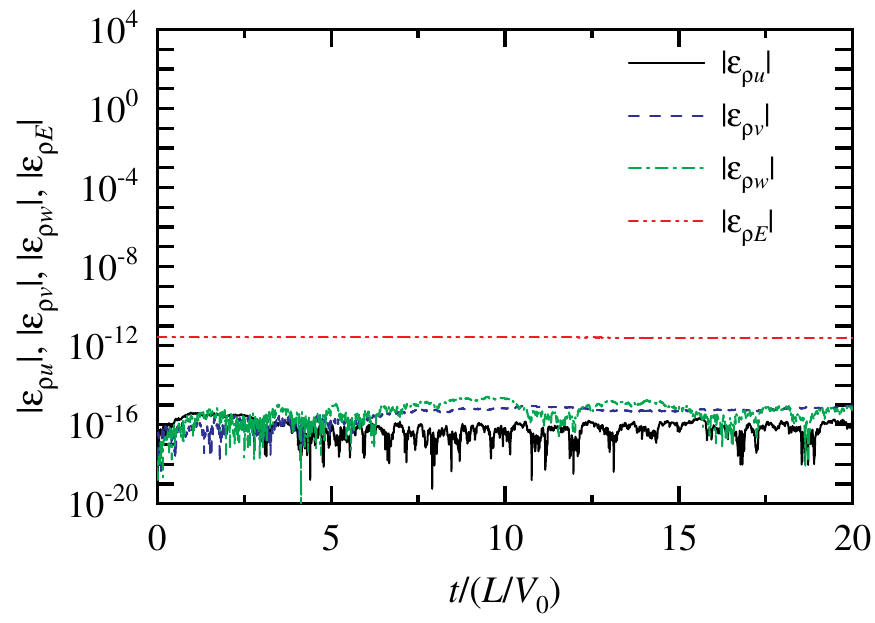}
\vspace*{-0.5\baselineskip}
\caption{ Errors of momentum and total energy for $ Re = \infty$: $N = 41$}
\label{tg_fig02}
\end{figure}

Figure \ref{tg_fig01}(a) shows the total amount $\langle \rho \bm{u} \rangle$ of the momentum. 
$|\langle \rho \bm{u} \rangle |$ changes on the order of $10^{-16}$, 
and the momentum is temporally conserved. 
Figure \ref{tg_fig01}(b) shows the total amounts, 
$\langle \rho K \rangle$ and $\langle \rho E \rangle$, respectively, 
of the kinetic and total energies. 
Each total amount is kept constant. 
The relative error of each value from the exact solution is $6.00 \times 10^{-2} \%$ and $1.96 \times 10^{-5} \%$, 
and the kinetic and total energies are conserved in time. 
Figure \ref{tg_fig02} shows the relative errors, 
$\varepsilon_{\rho u} = (\langle \rho u \rangle - \langle \rho u \rangle_0)/\langle (\rho u)^2 \ rangle_0^{1/2}$ 
and $\varepsilon_{\rho v} = (\langle \rho v \rangle - \langle \rho v \rangle_0)/\langle (\rho v)^2 \rangle_0^{1 /2}$, 
of the momentums in $x$- and $y$-directions, 
absolute error $\varepsilon_{\rho w} = \langle \rho w \rangle - \langle \rho w \rangle_0$ in the momentum in $z$-direction, 
and relative error $\varepsilon_{\ rho E} = (\langle \rho E \rangle - \langle \rho E \rangle_0)/\langle \rho E \rangle_0$ in the total energy. 
Here, as $\langle \rho w \rangle_0$ is zero, 
the error of the momentum $\rho w$ is defined as the absolute error. 
The subscript 0 represents the initial value. 
The errors, $|\varepsilon_{\rho u}|$, $|\varepsilon_{\rho v}|$, 
and $|\varepsilon_{\rho w}|$, change on the order of $10^{-16}$, 
and the momentum is temporally preserved. 
Additionally, $|\varepsilon_{\rho E}|$ changes on the order of $10^{-12}$, 
and the total energy is also conserved. 
Evidently from the above results, 
this computational method achieves excellent conservation properties 
of the momentum and total energy.

\begin{figure}[!t]
\begin{minipage}[t]{0.49\hsize}
\begin{center}
\includegraphics[trim=0mm 0mm 0mm 0mm, clip, width=70mm]{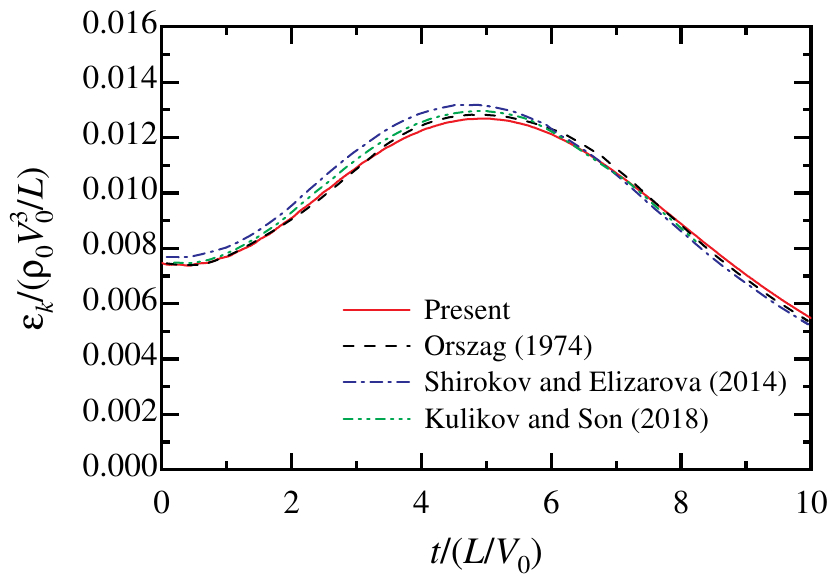} \\
{\small (a) $Re = 100$}
\end{center}
\end{minipage}
\begin{minipage}[t]{0.49\hsize}
\begin{center}
\includegraphics[trim=0mm 0mm 0mm 0mm, clip, width=70mm]{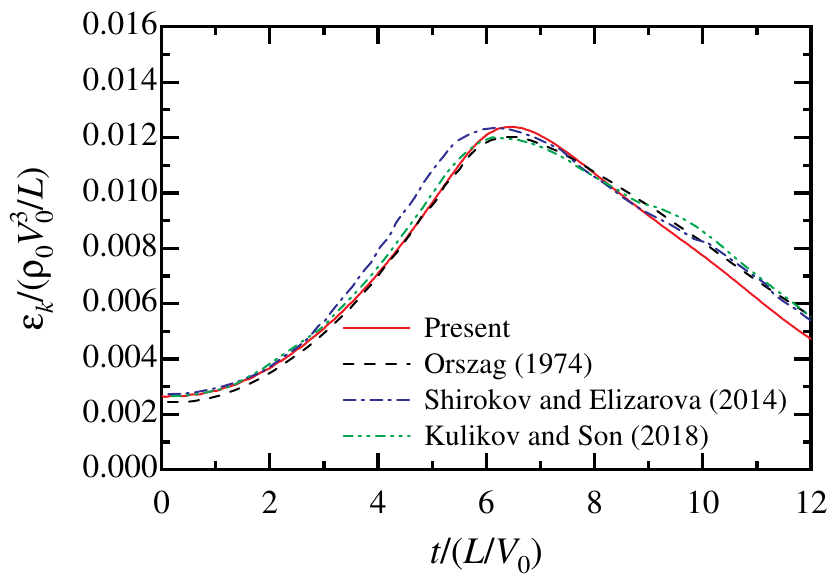} \\
{\small (b) $Re = 280$}
\end{center}
\end{minipage}
\caption{Comparisons of dissipation rate of kinetic energy for $Re = 100$ and $280$: $N=41$}
\label{tg_fig06}
\end{figure}

Subsequently, referring to existing studies 
\citep{Orszag_1974, Shirokov&Elizarova_2014, Kulikov&Son_2018}, 
we analyzed the conditions of the Reynolds number $Re = \rho_0 V_0 L /\mu _0 = 100$ 
and $280$, Prandtl number $Pr = \mu _0 c_p/k_0 = 0.71$, 
and Mach number $Ma = V_0/c_0 = 0.1$. 
A uniform grid with $N = 41$ grid points is used for the calculation. 
The time step $\Delta t$ is set to $\Delta t/(L/V_0) = 0.01$ 
such that the initial Courant number CFL is $\mathrm{CFL} < 1$. 

To evaluate the properties of the Taylor--Green decaying vortex, 
we investigated the following kinetic energy dissipation rate $\varepsilon_k$:
\begin{equation}
   \varepsilon_k = - \frac{dK}{dt}, \quad 
   K = \frac{1}{V} \int_V \rho \frac{|\bu|^2}{2} dV.
\end{equation}
The kinetic energy dissipation rate $\varepsilon_k$ is shown in Fig. \ref{tg_fig06}. 
The results of Shirokov and Elizarova \citep{Shirokov&Elizarova_2014} and Kulikov and Son \citep{Kulikov&Son_2018} 
in Fig. \ref{tg_fig06} were obtained with the $N = 64$ grid points. 
Orszag \citep{Orszag_1974} reported the results obtained using the spectral method 
for the conditions $N = 32$ and $Re = 100$, 200, 300, and 400. 
Figures \ref{tg_fig06}(a) and \ref{tg_fig06}(b) compare the results of Orszag \citep{Orszag_1974} 
at $Re = 100$ and 300, respectively. 
Although this calculation uses a smaller number of grid points than the existing studies 
\citep{Shirokov&Elizarova_2014, Kulikov&Son_2018}, 
the dissipation rates for $Re = 100$ and 280 agree well with the existing results 
\citep{Shirokov&Elizarova_2014, Kulikov&Son_2018}. 
Moreover, the dissipation rate agrees well with the existing result 
\citep{Orszag_1974} obtained using the spectral method.

Subsequently, as with the workshop \citep{Wang_et_al_2013} on the analysis of the Taylor--Green decaying vortex 
by high-order accurate computational methods, 
we analyzed the condition of the Reynolds number $Re = \rho_0 V_0 L/\mu_0 = 1600$, 
Prandtl number $Pr = \mu _0 c_p/k_0 = 0.71$, and Mach number $Ma = V_0/c_0 = 0.1$. 
The physical properties at the initial temperature $T_0$ are constant. 
Uniform grids with $N = 41$, 161, and 256 grid points are used for the calculations. 
The time step $\Delta t$ is set to $\Delta t/(L/V_0) = 0.01$ 
so that the initial Courant number CFL is $\mathrm{CFL} = V_0 \Delta t/\Delta x < 1$. 
In the grid of $N = 256$, the initial Courant number is CFL= 0.406, 
and the initial Courant number considering the speed of sound is $\mathrm{CFL} = (V_0+c_0) \Delta t/\Delta x = 4.46$. 
The existing analysis \citep{Wang_et_al_2013} investigated the trend of the dissipation rate up to $t/(L/V_0) = 20$. 
The dissipation rate is maximum at $t/(L/V_0) \approx 8$, 
where minimum turbulent structure occurs.

\begin{figure}[!t]
\begin{minipage}[t]{0.325\hsize}
\begin{center}
\includegraphics[trim=0mm 0mm 0mm 0mm, clip, width=50mm]{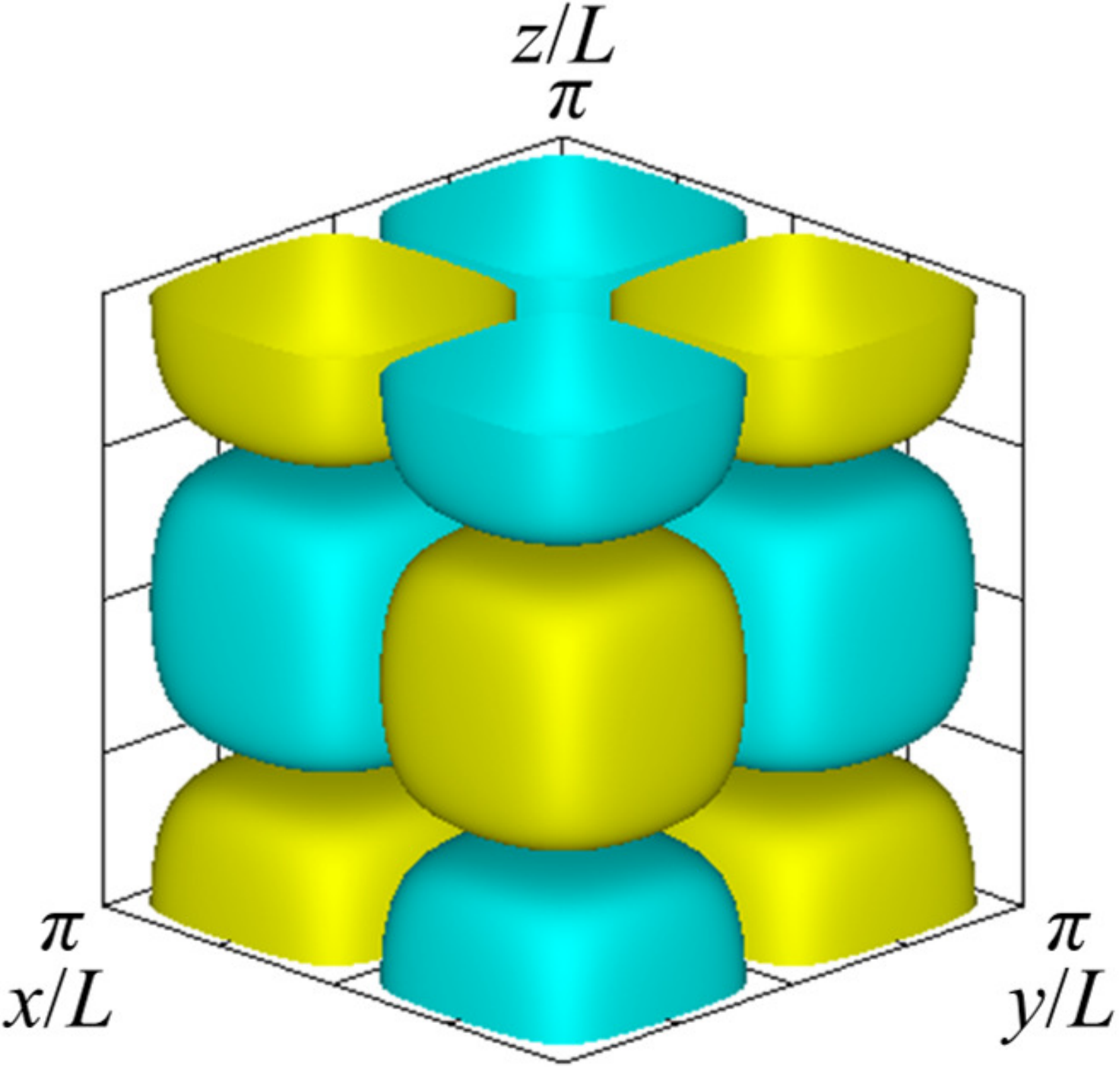} \\
{\small (a) $t/(L/V_0) = 0$}
\end{center}
\end{minipage}
\centering
\begin{minipage}[t]{0.325\hsize}
\begin{center}
\includegraphics[trim=0mm 0mm 0mm 0mm, clip, width=50mm]{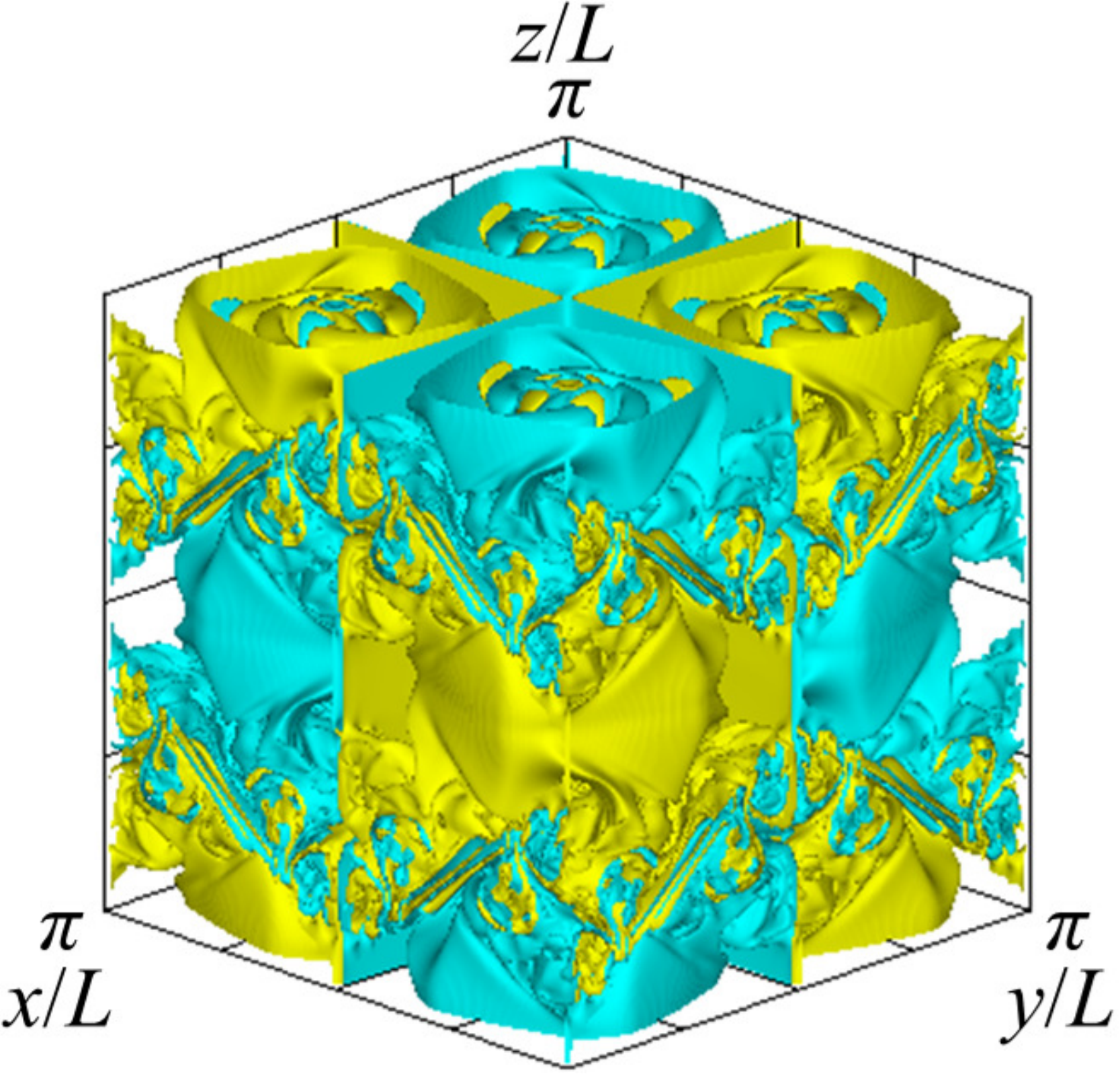} \\
{\small (b) $t/(L/V_0) = 10$}
\end{center}
\end{minipage}
\centering
\begin{minipage}[t]{0.325\hsize}
\begin{center}
\includegraphics[trim=0mm 0mm 0mm 0mm, clip, width=50mm]{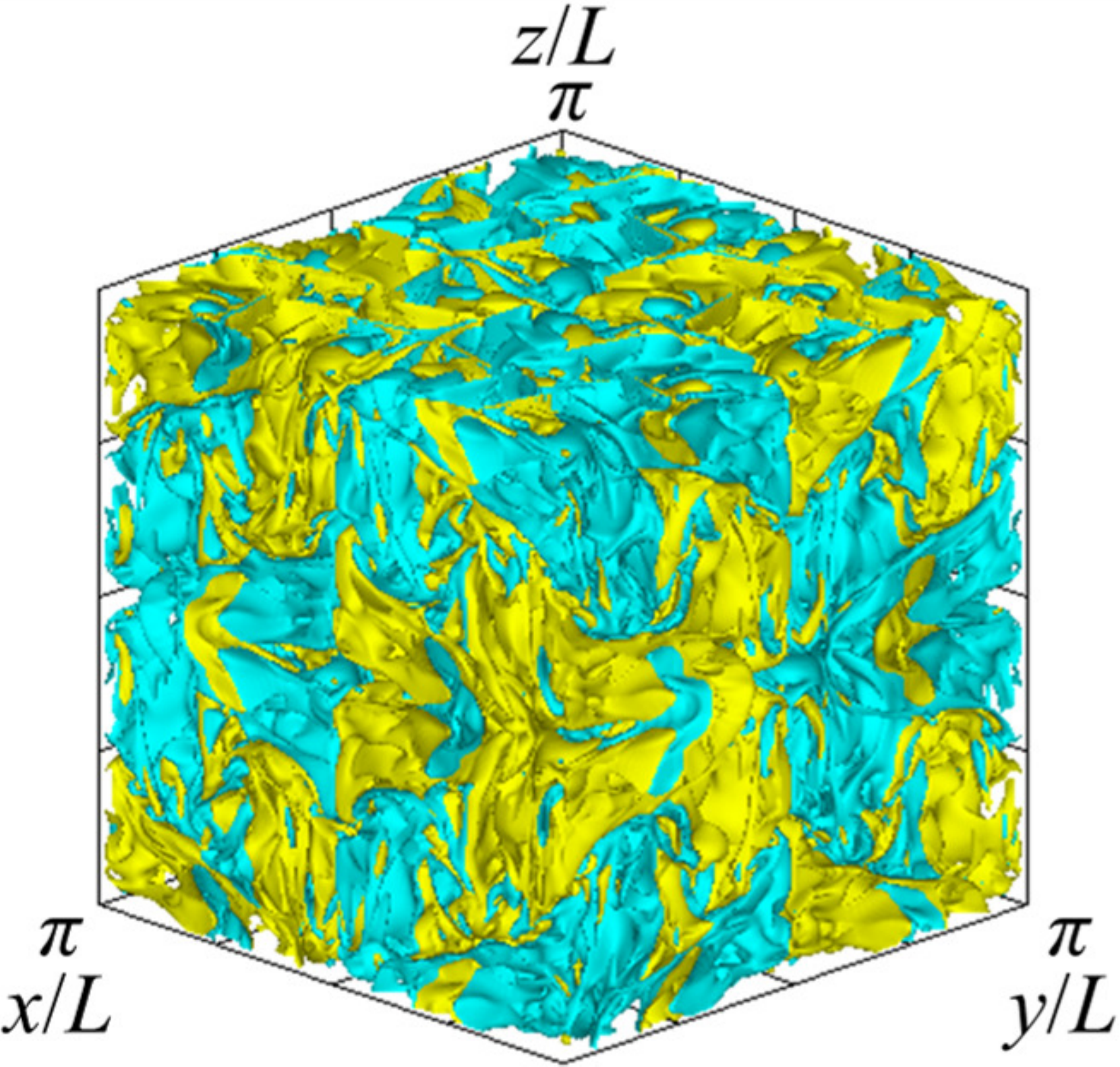} \\
{\small (c) $t/(L/V_0) = 20$}
\end{center}
\end{minipage}
\caption{Isosurfaces of $z$-direction vorticity for $Re = 1600$: $N = 256$}
\label{tg_fig12}
\end{figure}

\begin{figure}[!t]
\begin{minipage}[t]{0.49\hsize}
\begin{center}
\includegraphics[trim=0mm 0mm 0mm 0mm, clip, width=70mm]{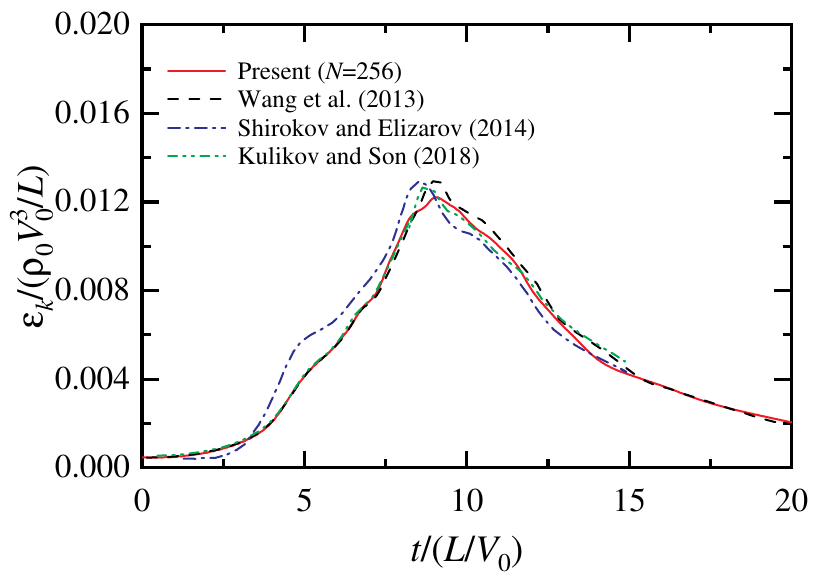} \\
{\small (a) Comparison with existing value}
\end{center}
\end{minipage}
\begin{minipage}[t]{0.49\hsize}
\begin{center}
\includegraphics[trim=0mm 0mm 0mm 0mm, clip, width=70mm]{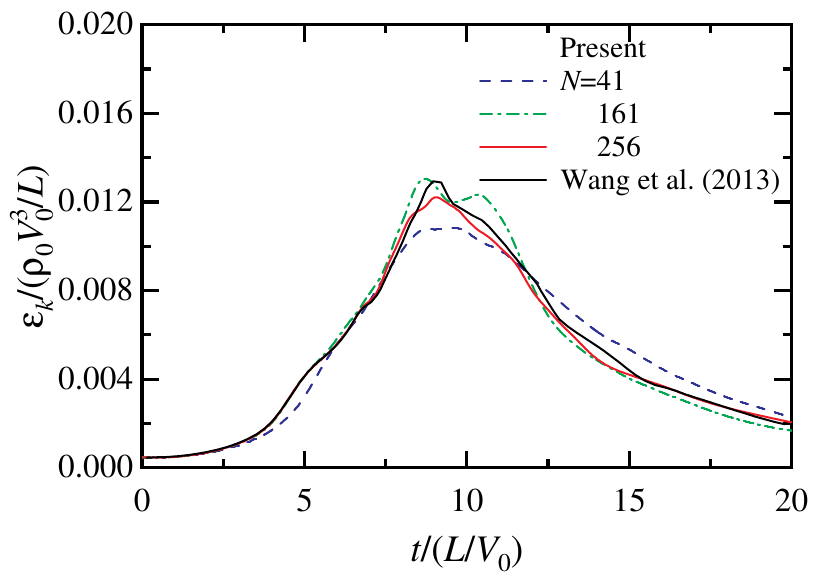} \\
{\small (b) Confirmation of grid dependence}
\end{center}
\end{minipage}
\caption{Dissipation rate of kinetic energy for $Re = 1600$: $N = 256$}
\label{tg_fig04}
\end{figure}

Figure \ref{tg_fig12} shows isosurfaces of the vorticity, $\omega_z$, 
in the $z$-direction for $t/(L/V_0) = 0$, 10, and 20. 
In the figure, the isosurfaces for $\omega_z = 0.2$ and $\omega_z = -0.2$ are shown 
in yellow and blue, respectively. 
The vortex scale decreases with time, 
and at $t/(L/V_0) = 10$, the vortex structure becomes anisotropic. 
The vortex grows isotropically at $t/(L/V_0) = 20$. 
This flow tendency is similar to the existing result \citep{Shirokov&Elizarova_2014}.

The dissipation rate $\varepsilon_k$ of the kinetic energy 
for the $N = 256$ grid points is shown in Fig. \ref{tg_fig04}(a). 
Existing results by Wang et al. \citep{Wang_et_al_2013}, Shirokov and Elizarova \citep{Shirokov&Elizarova_2014}, 
and Kulikov and Son \citep{Kulikov&Son_2018} were obtained using $N = 512$, 128, and 256 grid points, respectively. 
The maximum value of the dissipation rate exists around almost the same time 
in the results of this study and Wang et al. \citep{Wang_et_al_2013}. 
Furthermore, the present result agrees well with those of Wang et al. \citep{Wang_et_al_2013} and Kulikov and Son \citep{Kulikov&Son_2018} 
up to $t/(L/V_0) = 7$. 
In contrast, there is a difference between the result of Shirokov and Elizarova \citep{Shirokov&Elizarova_2014} and this result. 
Shirokov and Elizarova \citep{Shirokov&Elizarova_2014} obtained the result using the $N = 128$ grid points. 
It is considered that the difference in the results occurred owing to the grid dependency. 
Figure \ref{tg_fig04}(b) shows the dissipation rate $\varepsilon_k$ for $N = 41$, 161, and 256. 
It is found that with an increase in the number of grid points, 
this calculation result asymptotically approaches the result of Wang et al. \citep{Wang_et_al_2013}. 
It was clarified from the above results that this computational method is valid 
in the analysis considering compressibility and viscosity.

\begin{figure}[!t]
\centering
\includegraphics[trim=0mm 0mm 0mm 0mm, clip, width=70mm]{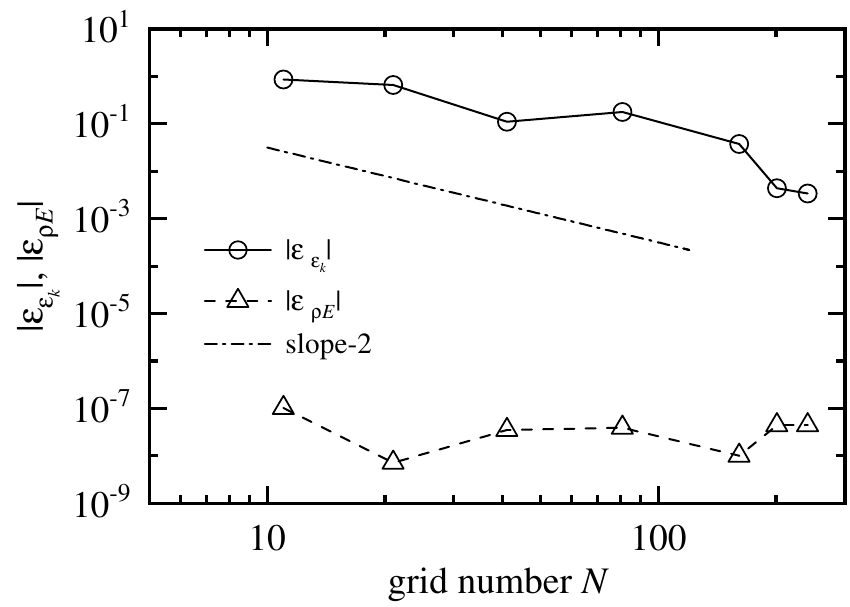}
\caption{Relative errors of dissipation rate of kinetic energy and total amount of total energy}
\label{tg_fig10}
\end{figure}

To investigate the convergence of the calculation results for this analysis, 
we use $N = 11$, 21, 41, 81, 161, 201, 241, and 256 grid points 
for $Re = 1600$, $Pr = 0.71$, and $Ma = 0.1$. 
Figure \ref{tg_fig10} shows the relative errors, 
$\varepsilon_{\varepsilon_k}$ and $\varepsilon_{\rho E}$, 
of the dissipation rate of kinetic energy and the total amount of total energy. 
The error is defined as the relative difference between the result obtained using $N=256$ 
and that using each grid. 
$|\varepsilon_{\varepsilon_k}|$ decreases as the number of grid points increases 
and remains a slope of approximately $-2$. 
$|\varepsilon_{\rho E}|$ is maintained on the order of $10^{-7}$ 
regardless of the number of grid points. 
It can be seen from the above results 
that this computational method has good convergence to the number of grid points.

\subsection{Natural convection in the cavity}

We analyzed natural convection in a cavity 
to investigate the validity of this numerical method 
for incompressible fluid with density variation. 
The origin is placed at the bottom of the container, 
the $x$ and $y$ axes are in the horizontal and vertical directions, respectively, 
and the $z$ axis is perpendicular to the $x$--$y$ plane. 
The size of the cavity on the $x$--$y$ plane is $L \times L$, 
and the computational region in the $z$-direction is set to the minimum grid width $\Delta x_\mathrm{min}$. 
As initial conditions, the fluid is stationary and isothermal. 
As for the boundary condition at the wall, 
a no-slip condition is imposed for the velocity, 
and a zero gradient condition is given for the pressure. 
The left and right walls of the cavity are heated and cooled 
at uniform temperatures $T_{H}$ and $T_{C}$, respectively. 
Other walls are insulated. 
A periodic boundary condition is given in the spanwise direction. 
Uniform initial values for the density and temperature are $\rho_0$ and $T_0 = (T_H+T_C)/2$, respectively. 
The viscosity coefficient and thermal conductivity at temperature $T _0$ 
are $\mu _0$ and $k_0$, respectively. 
The reference values used in this calculation are as follows: 
the length, velocity, time,density, temperature, internal energy, and pressure are 
$l_\mathrm{ref} = L$, $u_\mathrm{ref} = \sqrt{g \beta \Delta T L}$, 
$t_\mathrm {ref} = L^{2}/ \alpha $, $\rho_\mathrm{ref} = \rho_0$, 
$T_\mathrm{ref} = T_0$, $e_\mathrm{ ref} = c_v T_\mathrm{ref}$, 
and $p_\mathrm{ref} = (\kappa -1) \rho _\mathrm{ref} e_\mathrm{ref}$, respectviely. 
Here, $\beta$ is the volume expansion coefficient. 
Additionally, the viscosity coefficient and thermal conductivity are calculated 
using Sutherland's equation as with the previous study \citep{Vierendeel_et_al_2003}.

We investigated the heat transfer characteristics 
at a high-temperature difference $\Delta T = 720^\circ$C. 
The Rayleigh number $Ra$ is defined as $Ra = g \beta \Delta T L^3/(\nu_0 \alpha_0)$, 
where $\Delta T = T_H-T_C$. 
We set $Ra = 10^2$, $10^3$, $10^4$, $10^5$, and $10^6$, and $Pr = 0.71$, 
referring to the previous study \citep{Vierendeel_et_al_2003}. 
The Mach numbers $Ma = \sqrt{g \beta \Delta T L}/c_0$ 
at $Ra=10^2$, $10^3$, $10^4$, $10^5$, and $10^6$ are 
$3.84 \ times 10^{-4}$, $5.63 \times 10^{-4}$, $8.27 \times 10^{-4}$, $1,21 \times 10^{-3}$, and $1.78 \times 10^{ -3}$, respectively, 
where $c_0 = \sqrt{\kappa(\kappa-1)c_ v T_0}$. 
The Froude number is $Fr = u_\mathrm{ref}/\sqrt{g l_\mathrm{ref}} = 1.095$. 
A grid with $61 \times 61 \times 2$ is used for the calculation. 
The grid is set to be dense near the wall, 
and the minimum grid width is $\Delta x_\mathrm{min} = 3.04 \times 10^{-3}$. 
In this analysis, the time step is adjusted 
such that the Courant number CFL is $\mathrm{CFL} = (\sqrt{g \beta \Delta T L} \Delta t/\Delta x_\mathrm{min} ) < 1$. 
At $Ra = 10^6$, the maximum initial Courant number is $\mathrm{CFL} = 0.277$, 
and the maximum Courant number CFL considering the speed of sound is $\mathrm{CFL} = \Delta t (\sqrt{g \beta \Delta T L} + c_0)/\Delta x_\mathrm{min} = 16.17$. 
The existing result \citep{Vierendeel_et_al_2003} was obtained 
from a grid with $1024 \times 1024$ grid points 
without using any approximation to the fundamental equation.

\begin{figure}[!t]
\begin{minipage}[t]{0.24\hsize}
\begin{center}
\includegraphics[trim=0mm 0mm 0mm 0mm, clip, width=41mm]{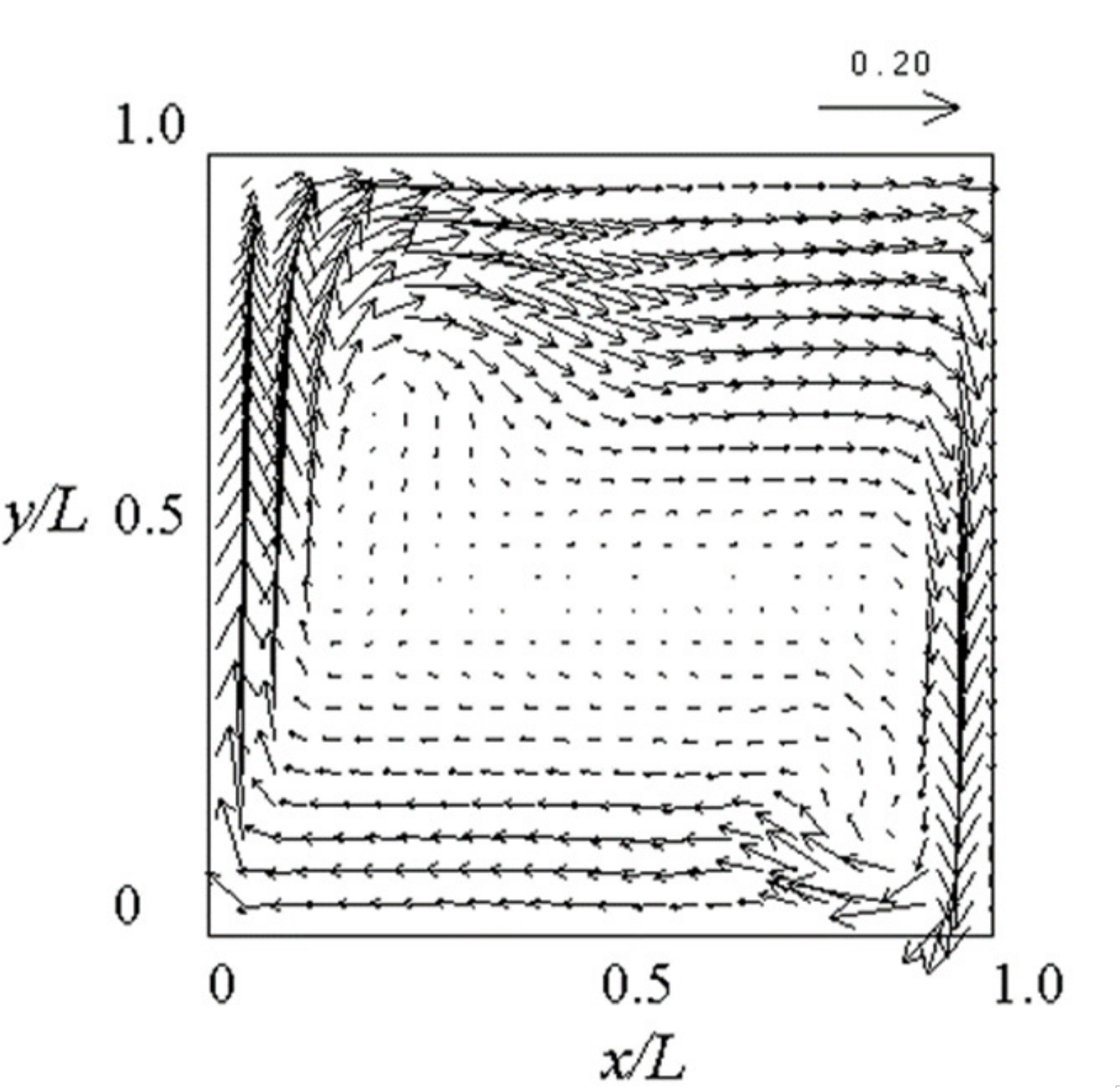} \\
{\small (a) Velocity vectors}
\end{center}
\end{minipage}
\begin{minipage}[t]{0.24\hsize}
\begin{center}
\includegraphics[trim=0mm 0mm 0mm 0mm, clip, width=41mm]{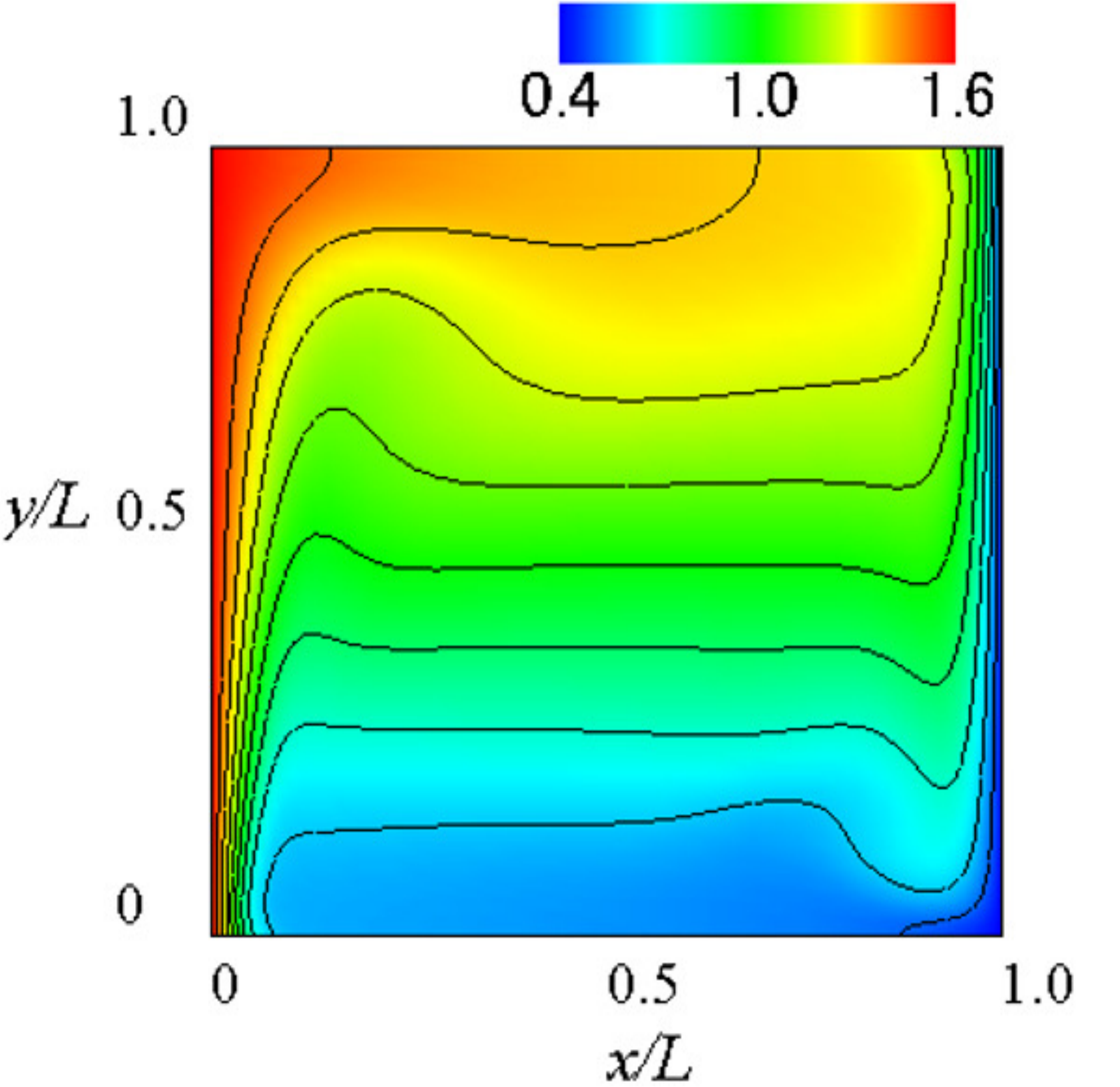} \\
{\small (b) Temperature contours}
\end{center}
\end{minipage}
\begin{minipage}[t]{0.24\hsize}
\begin{center}
\includegraphics[trim=0mm 0mm 0mm 0mm, clip, width=41mm]{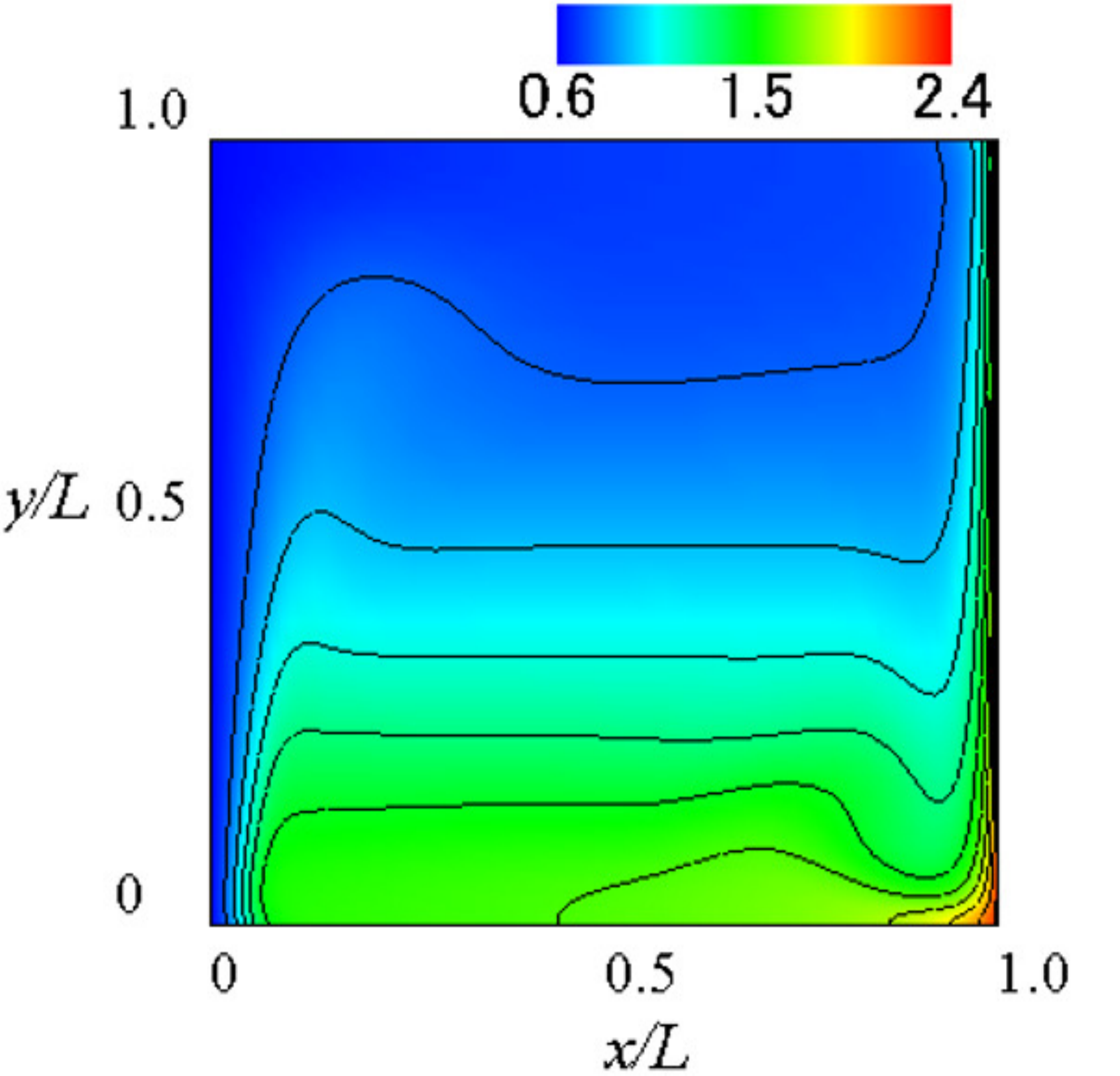} \\
{\small (c) Density contours}
\end{center}
\end{minipage}
\begin{minipage}[t]{0.24\hsize}
\begin{center}
\includegraphics[trim=0mm 0mm 0mm 0mm, clip, width=41mm]{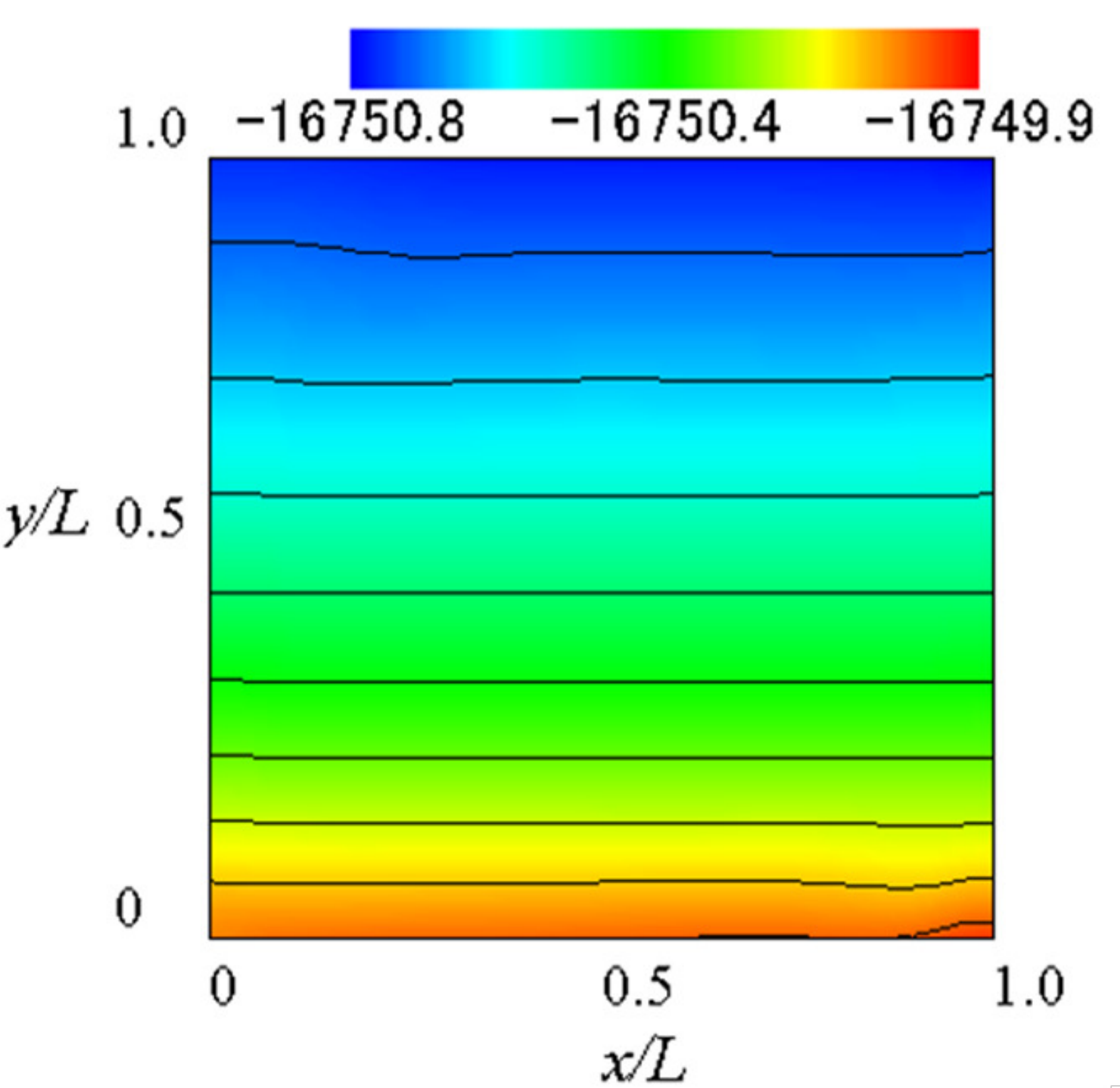} \\
{\small (d) Pressure contours}
\end{center}
\end{minipage}
\caption{Velocity, temperature, density, and pressure fields 
for $\Delta T = 720^{\circ}$C and $Ra = 10^6$}
\label{cavity_fig07Ra6}
\end{figure}

Figure \ref{cavity_fig07Ra6} shows the velocity, temperature, density, 
and pressure distributions at $Ra = 10^6$. 
Fluid heated near the hot wall is pushed up by buoyancy and transported to the cold wall. 
On the other hand, the transported high-temperature fluid descends 
while being cooled by the cold wall and flows into the high-temperature side again. 
A series of such movements of the fluid generates a clockwise heat convection. 
At $Ra = 10^2$, thermal conduction is dominant. 
As the Rayleigh number increases, convective heat transfer becomes dominant. 
The flow and temperature fields obtained using this computational method are qualitatively 
similar to the existing result \citep{Vierendeel_et_al_2003}. 
The spatial variation in the pressure at each Rayleigh number is low, 
and this numerical method can capture minute variations. 
The density has a distribution corresponding to the temperature distribution 
and changes in the range $0.6< \rho/\rho _0<2.4$. 
As the Boussinesq approximation assumes the density to be constant, 
such changes in density cannot be captured. 
It is considered that the spatial variation of the density affects 
the heat transfer characteristics. 
Hence, the results using the Boussinesq approximation and this computational method are compared later.

\begin{figure}[!t]
\centering
\includegraphics[trim=0mm 0mm 0mm 0mm, clip, width=75mm]{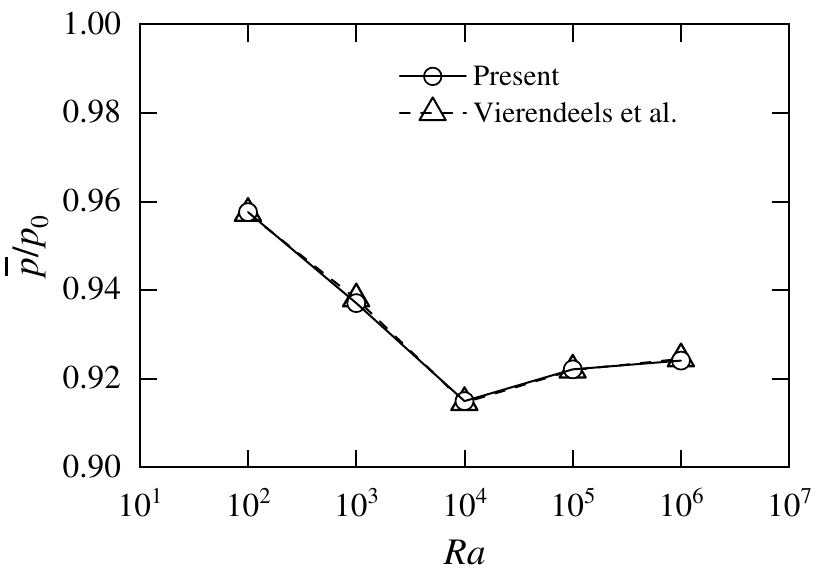}
\vspace*{-0.5\baselineskip}
\caption{Average pressure for $\Delta T = 720^{\circ}$C}
\label{cavity_fig05}
\end{figure}

To verify the validity of the pressure obtained using this numerical method, 
we compare the average pressure shown below with the previous result 
\citep{Vierendeel_et_al_2003}:
\begin{equation}
   \overline{p}/p_0 = \int_V \frac{p}{(\kappa -1) \rho _0 c_v T_0} dV, 
\end{equation}
where $V$ represents the computational domain. 
Figure \ref{cavity_fig05} shows the variation in the average pressure $\overline{p}/p_0$ 
with the Rayleigh number. 
For all Rayleigh numbers, 
the present results agree well with the existing results \citep{Vierendeel_et_al_2003}.

\begin{figure}[t]
\begin{minipage}[t]{0.49\hsize}
\begin{center}
\includegraphics[trim=0mm 0mm 0mm 0mm, clip, width=75mm]{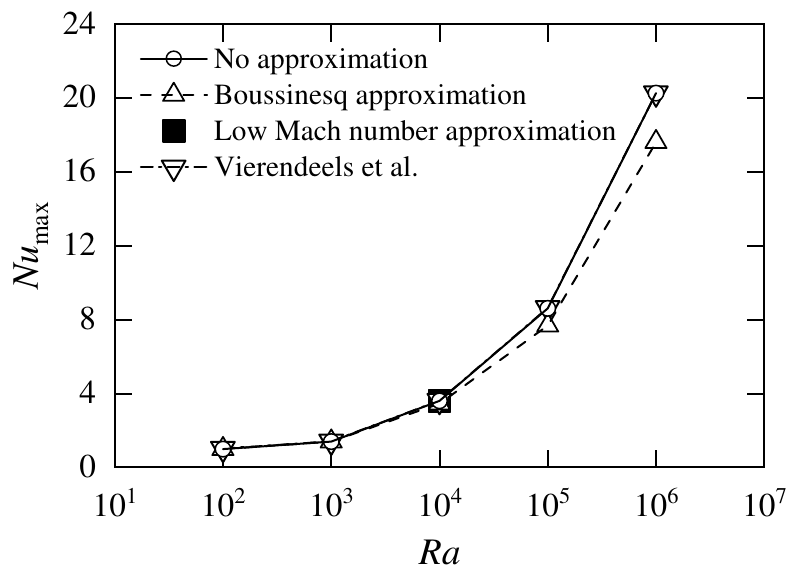} \\
{\small (a) Maximum Nusselt number}
\end{center}
\end{minipage}
\begin{minipage}[t]{0.49\hsize}
\begin{center}
\includegraphics[trim=0mm 0mm 0mm 0mm, clip, width=75mm]{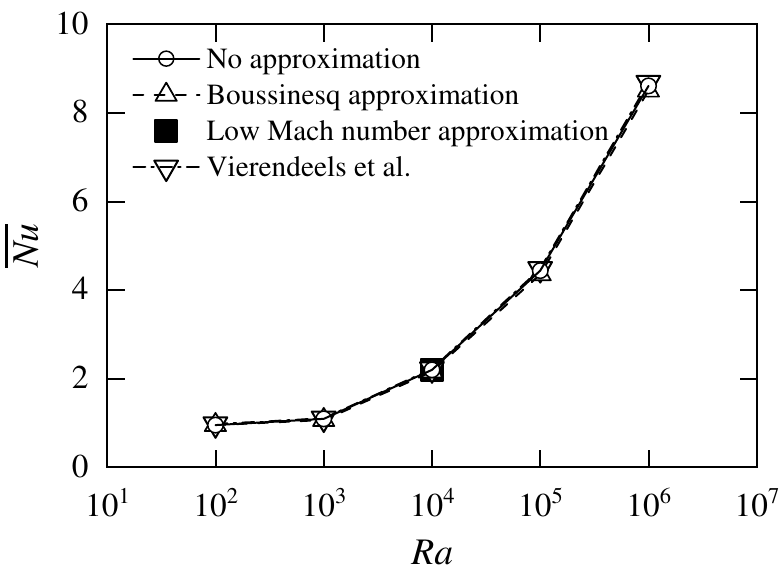} \\
{\small (b) Average Nusselt number}
\end{center}
\end{minipage}
\caption{Nusselt number distributions without approximation 
and with Boussinesq approximation for $\Delta T = 720^{\circ}$C}
\label{cavity_fig04}
\end{figure}

Subsequently, the heat transfer coefficient is compared with the previous result \citep{Vierendeel_et_al_2003}. 
Figure \ref{cavity_fig04} shows the result of the present numerical method 
without approximation to the fundamental equation 
and the result using the Boussinesq approximation. 
Figure \ref{cavity_fig04} (a) shows the maximum Nusselt number $Nu_\mathrm{max}$ 
at the heated wall for each Rayleigh number. 
$Nu$ is defined as $Nu = h_l L/k$, where $h_l$ is the local heat transfer coefficient. 
For $Ra \leq 10^4$, $Nu_\mathrm{max}$ with the Boussinesq approximation agrees well with 
that without the approximation. 
However, when $Ra > 10^4$, 
$Nu_\mathrm{max}$ with Boussinesq approximation is lower that without approximation. 
At $Ra = 10^5$ and $10^6$, heat transfer by convection is dominant over heat conduction. 
In the Boussinesq approximation, 
a change in buoyancy due to density is converted to a change due to temperature difference. 
Hence, when convection becomes dominant, 
the buoyancy cannot be approximated only by the temperature difference, 
and it is considered that a difference 
between the results of no approximation and the Boussinesq approximation appears.

Figure \ref{cavity_fig04}(b) shows the area-averaged Nusselt number $\overline{Nu}$ 
at the heated wall for each Rayleigh number. 
Unlike the trend of $Nu_\mathrm{max}$, $\overline{Nu}$ with the Boussinesq approximation 
agrees with the result without the approximation. 
It can be seen from this that the heat transfer characteristics 
in the calculations without approximation and with the Boussinesq approximation are the same 
on average but differ locally. 
When the low Mach number approximation is applied, 
it was confirmed that the calculation becomes unstable and diverges 
under the conditions of $Ra = 10^2$, $10^3$, $10^5$, and $10^6$. 
It is considered that the calculation became unstable 
due to the decrease of $Ma$ at $Ra = 10^2$ and $10^3$ 
and due to the increase in the local Courant number at $Ra = 10^5$ and $10^6$.

In Fig. \ref{cavity_fig05}, the average pressure is lowest at $Ra = 10^4$. 
As the mean pressure increases, 
the calculation becomes unstable for the low Mach number approximation. 
The computation at $Ra = 10^4$ is stable, 
and $Nu_\mathrm{max}$ and $\overline{Nu}$ agree with the results without approximation. 
It was confirmed from this result that the calculation is likely to be unstable 
when the low Mach number approximation is applied 
and that this computational method without approximation is effective. 
Moreover, for all Rayleigh numbers, 
the present results without approximation agree well with the existing results \citep{Vierendeel_et_al_2003}. 
This study obtained valid results, even with fewer grid points than in the earlier work. 
From the above results, it can be seen that this computational method is effective 
even in the analysis of incompressible flow considering the density variation.

\begin{figure}[!t]
\centering
\includegraphics[trim=0mm 0mm 0mm 0mm, clip, width=75mm]{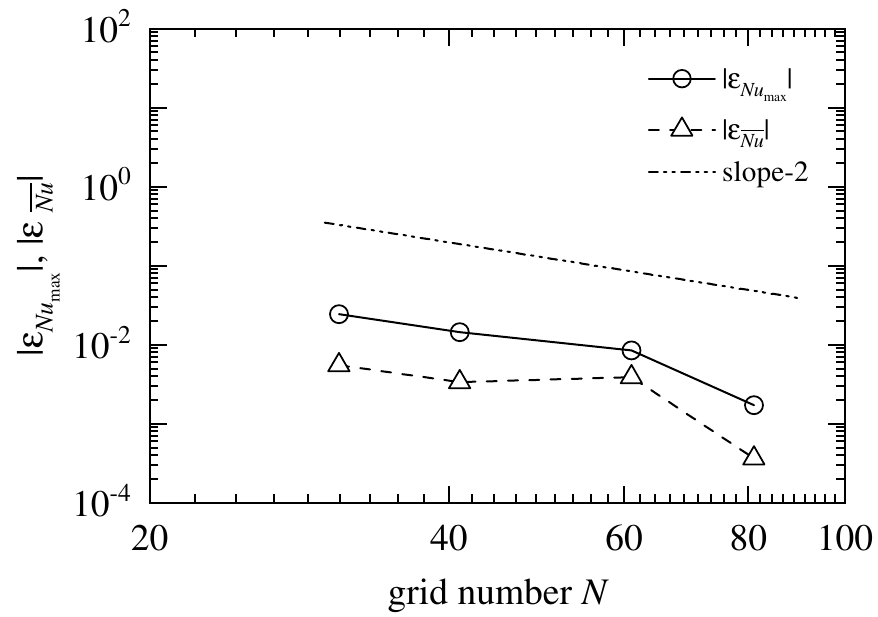} \\
\vspace*{-0.5\baselineskip}
\caption{Relative errors of maximamu and average Nusselt numbers 
for $\Delta T = 720^{\circ}$C and $Ra = 10^6$: No approximation}
\label{cavity_fig08}
\end{figure}

To investigate the convergence of the calculation results, 
we calculated the condition of $\Delta T = 720^\circ$C, $Ra = 10^6$, 
and $Pr = 0.71$ using the grids $N \times N \times 2$ ($N = 31$, 41, 81, 121). 
For the maximum Nusselt number $Nu_\mathrm{max}$ and the average Nusselt number $\overline{Nu}$, 
Fig. \ref{cavity_fig08} shows the relative differences, 
$|\varepsilon_{Nu_\mathrm{max}}|$ and $|\varepsilon_{\overline{Nu}}|$, 
between the result obtained using the grid with $N = 121$ 
and the results for each grid. 
The relative difference decreases as the number of grid points increases 
and approximately changes at a slope of $-2$. 
From the above results, it can be seen that 
the computational method used in this analysis has good convergence to the number of grid points.

\subsection{Three-dimensional Taylor decaying vortex}

We analyzed a three-dimensional Taylor decaying vortex 
to investigate the validity of the present numerical method 
for the three-dimensional analysis considering viscosity. 
The calculated result was compared with the exact solution. 
In this calculation, the wavelength of the vortex is $\lambda$, 
the maximum velocity of the vortex is $U$, 
and uniform initial values of density and temperature are $\rho_0$ and $T_0$, respectively. 
The reference values used in this calculation are as follows: 
the length, velocity, time, density, pressure, temperature, and internal energy are 
$l_\mathrm{ref} = \lambda$, $u_\mathrm{ref} = U$, 
$t_\mathrm{ref} = \lambda/U $, $\rho_\mathrm{ref} = \rho_0$, 
$p_\mathrm{ref} = (\kappa -1) \rho_\mathrm{ref} e_\mathrm{ref}$, 
$T_\mathrm{ref} = T_0$, and $e_\mathrm{ref} = c_v T_\mathrm{ref}$, respectively. 
Antuono \citep{Antuono_2020} obtained a periodic three-dimensional analytical solution 
using the method of Ethier and Steinman \citep{Ethier&Steinman_1994}. 
The nondimensionalized exact solution by the reference values is expressed as
\begin{align}
   u_{1,2} &= \alpha \left[ 
     \sin(k x + \theta_{1,2}) \cos(k y + \phi_{1,2}) \sin(k z + \psi_{1,2}) 
   \right. \nonumber \\
   & \, \left. 
   - \cos(k z + \theta_{1,2}) \sin(k x + \phi_{1,2}) \sin(k y + \psi_{1,2}) 
   \right] e^{-3 \frac{k^2}{Re} t}, \\
   v_{1,2} &= \alpha \left[ 
     \sin(k y + \theta_{1,2}) \cos(k z + \phi_{1,2}) \sin(k x + \psi_{1,2}) 
   \right. \nonumber \\
   & \, \left. 
   - \cos(k x + \theta_{1,2}) \sin(k y + \phi_{1,2}) \sin(k z + \psi_{1,2}) 
   \right] e^{-3 \frac{k^2}{Re} t}, \\
   w_{1,2} &= \alpha \left[ 
     \sin(k z + \theta_{1,2}) \cos(k x + \phi_{1,2}) \sin(k y + \psi_{1,2}) 
   \right. \nonumber \\
   & \, \left. 
   - \cos(k y + \theta_{1,2}) \sin(k z + \phi_{1,2}) \sin(k x + \psi_{1,2}) 
   \right] e^{-3 \frac{k^2}{Re} t}, \\
   p_{1,2} &= - \frac{|\bm{u}_{1,2}|^2}{2},
   \label{3D_Taylor_vortex}
\end{align}
where $\alpha = 4 \sqrt{2}/(3 \sqrt{3})$. 
Subscripts 1 and 2 denote two solutions which have similar distributions. 
$k = 2\pi$ is the dimensionless wavenumber, 
and the Reynolds number is defined as $Re = U \lambda/\nu$. 
The phases $\theta$, $\phi$, and $\psi$ are given as
\begin{equation}
   \theta_1 = \psi_1 - \frac{5 \pi}{6}, \quad 
   \phi_1 = \psi_1 - \frac{\pi}{6}, \quad 
   \psi_1 = \cos^{-1} \left( \frac{R}{\sqrt{1 + R^2}} \right),
\end{equation}
\begin{equation}
   \theta_2 = \phi_1, \quad 
   \phi_2 = \theta_1, \quad 
   \psi_2 = \psi_1,
\end{equation}
where $R$ is a parameter set to a value 
excluding the singular point value $R = \pm 1/\sqrt{3}$. 
In this research, we set $R = 0$ as with the existing research \citep{Antuono_2020}. 
The kinetic energy $K$ and the volume-integrated kinetic energy $\langle K \rangle$ are given as follows:
\begin{equation}
   K = \frac{1}{2} | \bm{u} |^2,
\end{equation}
\begin{equation}
   \langle K \rangle = \int_V K dV = \frac{1}{2} e^{-\frac{6 k^2}{Re}t}.
\end{equation}

The computational domain is a cube with a side $\lambda$. 
As the initial conditions, exact solutions $\bm{u} _1$ and $p_1$ are given 
for the velocity and pressure, respectively. 
Periodic boundary conditions are imposed at all boundaries. 
For this analysis, we use a uniform grid of $N \times N \times N$ and set $N = 41$. 
The grid has almost the same resolution as the existing research \citep{Antuono_2020}. 
The Mach number in this calculation was set to $Ma = U/c_0 = 0.001$, 
where $c_0 = \sqrt{\kappa(\kappa-1) c_v T_0}$. 
Similarly to the previous study \citep{Antuono_2020}, 
under the conditions of $Re = 50$, $10^3$ and time step $\Delta t/(\lambda/U) = 0.01$, 
the calculations were performed until time $t/( \lambda/U) = 10$, 
and this calculation result was compared with the exact solution. 
The Courant number defined as $\mathrm{CFL} = \Delta t U/\Delta x$ is CFL = 0.4, 
and the Courant number considering the speed of sound is $\mathrm{CFL} = \Delta t (U +c_0)/\Delta x = 400.0$, 
where $\Delta x$ is a grid width. 
Additionally, to verify the conservation of kinetic energy, 
we perform the calculation under the condition $Re = \infty$.

\begin{figure}[!t]
\begin{minipage}[t]{0.49\hsize}
\begin{center}
\includegraphics[trim=0mm 0mm 0mm 0mm, clip, width=75mm]{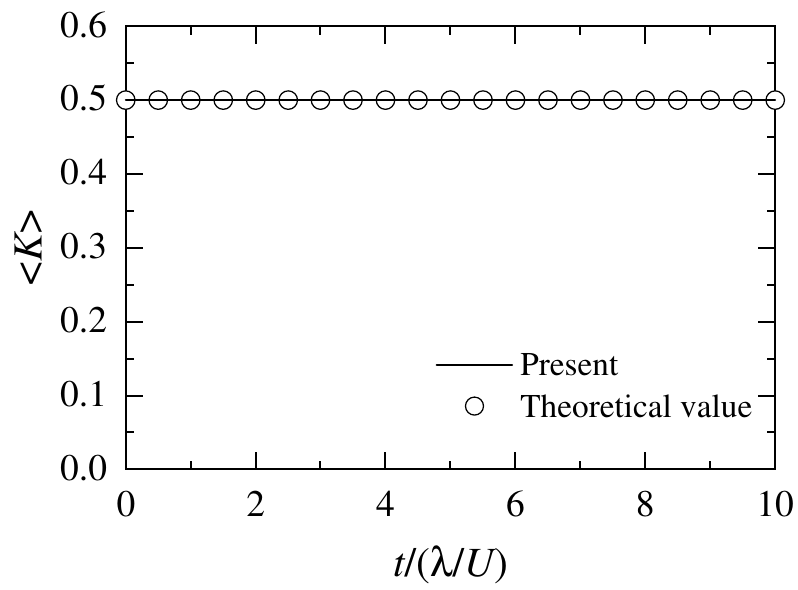} \\
{\small (a) $Re = \infty$}
\end{center}
\end{minipage}
\begin{minipage}[t]{0.49\hsize}
\begin{center}
\includegraphics[trim=0mm 0mm 0mm 0mm, clip, width=75mm]{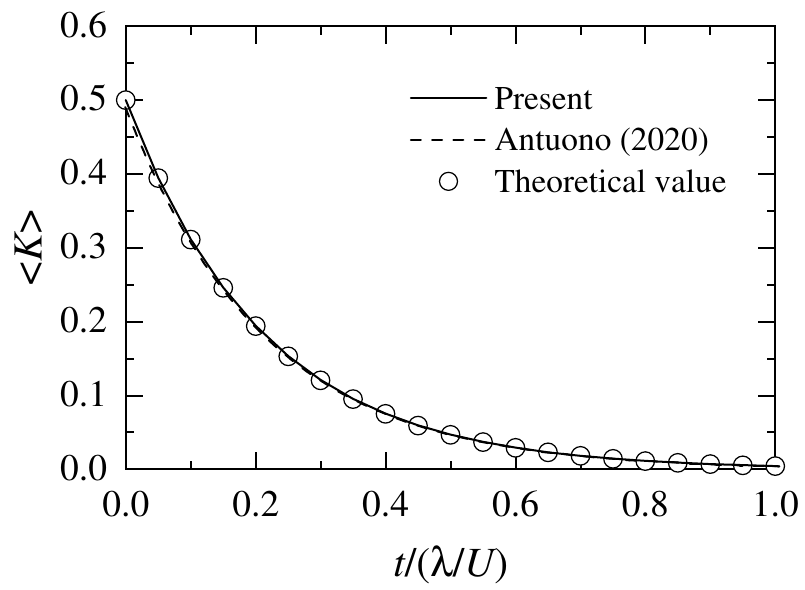} \\
{\small (b) $Re = 50$}
\end{center}
\end{minipage}

\vspace*{0.5\baselineskip}
\begin{minipage}[t]{0.49\hsize}
\begin{center}
\includegraphics[trim=0mm 0mm 0mm 0mm, clip, width=75mm]{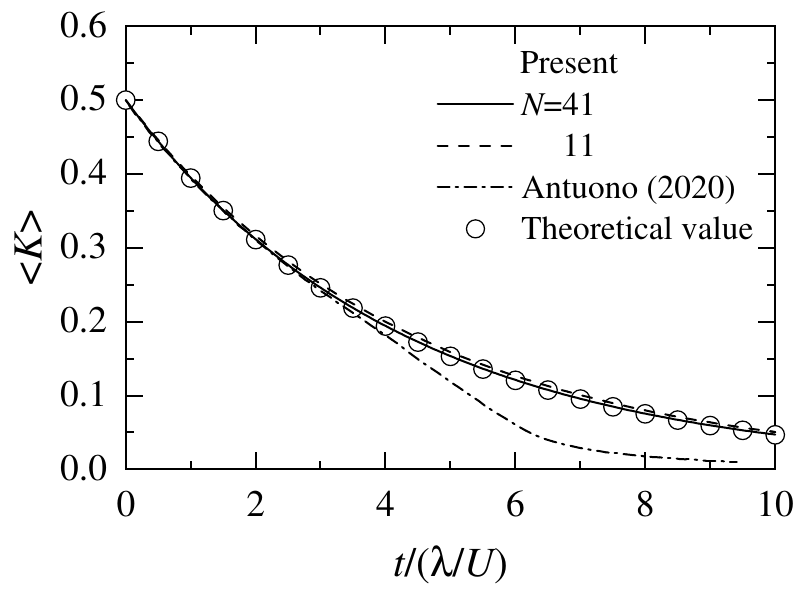} \\
{\small (c) $Re = 1000$ without disturbance}
\end{center}
\end{minipage}
\begin{minipage}[t]{0.49\hsize}
\begin{center}
\includegraphics[trim=0mm 0mm 0mm 0mm, clip, width=75mm]{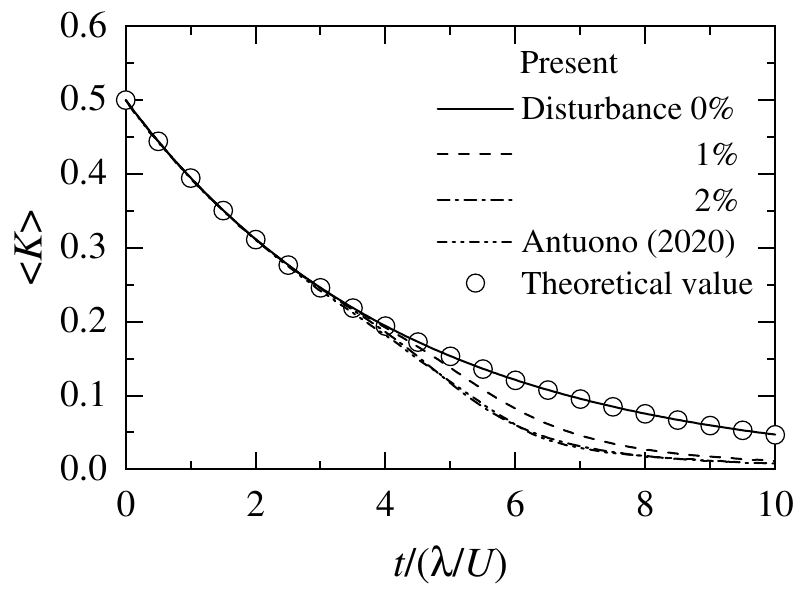} \\
{\small (d) $Re = 1000$ with disturbance}
\end{center}
\end{minipage}
\caption{Total amount of kinetic energy}
\label{decayfig1}
\end{figure}

Figure \ref{decayfig1} shows the time variation of the total amount 
of the kinetic energy $\langle K \rangle$ up to time $t/(\lambda/U) = 10$. 
Our calculation results at $Re = \infty$, 50, and 1000 agree well with the exact solution. 
As shown in Fig. \ref{decayfig1}(a), for $Re = \infty$, 
the kinetic energy does not decay with time, 
and the excellent energy conservation property is achieved. 
For $Re = 50$ in Fig. \ref{decayfig1}(b), 
the kinetic energy decays to almost zero at $t/(\lambda/U) = 1.0$. 
At $Re = 1000$ in Fig. \ref{decayfig1}(c), 
the attenuation of the kinetic energy is suppressed. 
The result for $N = 11$ is also included to confirm the grid dependence on the calculated result. 
Moreover, for $Re = 1000$, Fig. \ref{decayfig1}(d) shows the results 
when the initial disturbance is added to the initial velocity. 
The initial turbulence levels correspond to $1\%$ and $2\%$ of the initial average kinetic energies. 
In the earlier study, at $Re = 1000$, 
the difference between the calculated result and exact solution increased with time, 
and a transition to turbulent flow was observed. 
Without initial disturbance, no such transition to turbulent flow is observed in this study, 
and the calculated result agrees with the exact solution. 
Figure \ref{decayfig1}(c) shows that the difference from the existing value is not caused by grid dependence. 
When the initial disturbance is added, 
the calculated result and the exact solution start to differ from around $t/(\lambda/U) = 5$, 
and this trend is similar to that in the existing research \citep{Antuono_2020}. 
Increasing the initial disturbance level advances the transition 
and gives a distribution similar to the existing result \citep{Antuono_2020}. 
From the above results, we can find that this numerical method can also be applied to incompressible flows.

\section{Conclusion}
\label{sec5}

In this study, we proposed a method for simultaneous relaxation of velocity, pressure, 
density, and internal energy 
to analyze low Mach number compressible flows. 
This method uses a conservative finite difference scheme 
with excellent energy conservation properties. 
We investigated the accuracy, convergence, and energy conservation properties 
of the present computational scheme 
and thus obtained the following findings.

In the analysis for sound wave propagation in an inviscid compressible flow, 
the amplitude amplification ratio and frequency of sound wave obtained 
by this numerical method agree well with the theoretical values. 
Moreover, even when the Courant number is high, 
the calculation accuracy, convergence, and energy conservation are excellent.

In the analysis of a three-dimensional periodic inviscid compressible flow, 
the present numerical method conserves each total amount of the momentum, total energy, and entropy in time, 
even in discrete equations. 
In this method, which discretely calculates the kinetic and internal energies at the same temporallevel, 
the total energy is discretely preserved at the rounding error level. 
When no approximation is applied to the fundamental equations, 
the excellent conservation properties of momentum, total energy, and entropy are achieved. 

For compressible isotropic turbulence, 
the momentum and total energy are discretely conserved in an inviscid fluid. 
Furthermore, this computational method can capture fluctuations such as turbulence kinetic energy 
and has high calculation accuracy in the analysis considering compressibility and viscosity.

Similarly, for the Taylor--Green decaying vortex problem, 
the momentum and total energy are discretely conserved in an inviscid flow. 
The time variation in the dissipation rate of kinetic energy agrees well with the previous results. 
We confirmed the validity of this computational scheme for compressible viscous flows.

We analyzed natural convection in a cavity 
and confirmed the validity of this numerical method 
even in incompressible flows considering density variation. 
When the temperature difference and Rayleigh number are high, 
the maximum Nusselt number with the Boussinesq approximation is lower than without the approximation, 
and the results differ depending on the computational method.

In the analysis for a three-dimensional Taylor decaying vortex, 
the kinetic energy is discretely preserved in an inviscid flow. 
In the viscous analysis, 
the time variation of the kinetic energy agrees well with the exact solution. 
We found that this numerical method can be applied to incompressible flows 
and have high calculation accuracy.

This study clarified the accuracy and validity of the present numerical method 
by analyzing various flow models 
and demonstrated the possibility of applying this method to complex flow fields.

\section*{Acknowledgment}

The numerical results in this research were obtained using the supercomputing resources 
at Cyberscience Center, Tohoku University. 
We would like to express our gratitude to Associate Professor Yosuke Suenaga 
of Iwate University for his support of our laboratory. 
The authors wish to acknowledge the time and effort of everyone involved in this study.


\bibliography{reference_low_mach_bibfile}

\end{document}